\documentclass[a4paper,11pt]{article}
\pdfoutput=1 

\usepackage{jheppub} 

\usepackage[T1]{fontenc} 

\usepackage{tabularx}
\usepackage{verbatim}
\usepackage{cancel}
\usepackage{mathtools}
\usepackage{floatrow}
\usepackage{xcolor}
\usepackage{subcaption}
\usepackage{xfrac}
\usepackage{multirow}
\usepackage{standalone}
\usepackage{pdflscape}
\usepackage{float}
\restylefloat{table}

\usepackage{tikz}
\usepackage[compat=1.1.0]{tikz-feynman}
\tikzfeynmanset{warn luatex=false}
\tikzfeynmanset{/tikzfeynman/warn luatex = false}

\DeclareMathOperator{\sgn}{sgn}
\DeclareMathOperator{\Res}{Res}
\DeclareMathOperator{\CT}{CT}
\DeclareMathOperator{\PV}{PV}
\DeclareMathOperator{\arsinh}{arsinh}
\newcommand{\dd}{\mathrm{d}}
\newcommand{\ii}{\mathrm{i}}
\newcommand{\order}{\mathcal{O}}
\renewcommand{\Re}{\operatorname{Re}}
\renewcommand{\Im}{\operatorname{Im}}
\definecolor{myBlue}{rgb}{0.000,0.3155,0.50}
\definecolor{myLightBlue}{rgb}{0.5,0.8155,1.000}
\definecolor{myRed}{rgb}{0.972,0.6536,0.6208}
\definecolor{myDarkRed}{rgb}{0.651,0.0938,0.0364}

\chardef\MyArticleWithColor=\pdfcolorstackinit page direct{0 g}

\title{\boldmath Numerical integration of loop integrals through local cancellation of threshold singularities
}

\author{D. Kermanschah}

\affiliation{ETH Z\"urich,\\R\"amistrasse 101, 8092 Z\"urich, Switzerland}

\emailAdd{d.kermanschah@gmail.com}

\abstract{
We propose a new approach that allows for the separate numerical calculation of the real and imaginary parts of finite loop integrals.
We find that at one-loop the real part is given by the Loop-Tree Duality integral supplemented with suitable counterterms and the imaginary part is a sum of two-body phase space integrals, constituting a locally finite representation of the generalised optical theorem.
These expressions are integrals in momentum space, whose integrands were specially designed to feature local cancellations of threshold singularities.
Such a representation is well suited for Monte Carlo integration and avoids the drawbacks of a numerical contour deformation around remaining singularities.
Our method is directly applicable to a range integrals with certain geometric properties but not yet fully generalised for arbitrary one-loop integrals.
We demonstrate the computational performance with examples of one-loop integrals with various kinematic configurations, which gives promising prospects for an extension to multi-loop integrals.
}

\begin{document} 
\maketitle
\flushbottom

\section{Introduction}
\label{sec:introduction}

The endeavor of automating perturbative calculations in Quantum Field Theories beyond next-to-leading order has lead to a continuous development of new methods and computational techniques.
Unfortunately, increasing numbers of loops, legs and mass scales pose a challenge to many established approaches.
However, some promise better scaling with prospects to automation.
In this regard, numerical methods have proven to be very powerful.
In this article, we focus on methods based on the numerical integration of loop integrals.
The development of such methods may enable a uniform treatment of loop and phase space integrals, where the latter are already commonly performed numerically with Monte~Carlo event generators \cite{Papadopoulos_2001,Ohl_2001,Krauss_2002,Maltoni_2003,Gleisberg_2004,Pukhov:2004ca}.
This would allow for a change in methodology in how we treat real and virtual corrections.
In fact, the cancellation of infrared singularities between real radiation and virtual loops \cite{PhysRev.52.54,Kinoshita:1962ur,PhysRev.133.B1549} can be realised locally, sidestepping the need for dedicated infrared subtraction schemes.
This programme has been pursued by D. Soper with foundational work in \cite{Soper_1998,Soper_2000,Soper_2001,Kr_mer_2002,Kr_mer_2004}, based on local cancellations among interferences,
and has recently been extended to all orders in \cite{Capatti_2021}.
Other approaches, based on mappings between real and virtual momentum configurations, were studied in \cite{Buchta_2014,Sborlini_2016,Sborlini_2016_2,Hern_ndez_Pinto_2016,Seth_2016,Runkel_2020}.
\par

However, the numerical integration of loop integrals with traditional methods such as Monte~Carlo integration \cite{nla.cat-vn740522,James:1980yn} is challenging.
Not only does this apply to ultraviolet or infrared divergent integrals but also to finite integrals in $D=4$ dimensions.
The reason is the presence of threshold singularities in the integrand.
Fortunately, ultraviolet and infrared singular regions can be removed with local counterterms, which puts them on the same footing as finite integrals.
The construction of such counterterms at one loop has been first pursued in \cite{Nagy_2003,Assadsolimani_2010,Becker_2010,Becker_2012_3}.
Examples of local ultraviolet subtraction at two-loops were studied in \cite{Driencourt_Mangin_2019} that can be generalised using a local representation of the BPHZ forest formula treatment \cite{Bogoliubov:1957gp,Hepp:1966eg,Zimmermann:1969jj}.
Infrared singularities at higher loop orders can be removed at amplitude level satisfying universal factorisation properties, as shown in \cite{Anastasiou_2019,Anastasiou_2021}.
\par

Threshold singularities arise for particular values of loop momenta for which virtual particles are on their mass shell, i.e. $q_i^2=m_i^2$.
In loop integrands, they appear as the poles of Feynman propagators and are formally regulated with the $\ii \epsilon$ causal prescription, which slightly shifts the singularity away from the integration domain.
Numerically, threshold singularities can be regulated according to Cauchy’s theorem by a deformation of the integration contour into the complex plane.
Indeed, shifting the threshold singularity into the upper (lower) complex half-plane is equivalent to deforming the contour in the complex plane to pass the pole from below (above).
On the deformed contour, the integrand is free of poles and Monte~Carlo methods can be employed.
Such contour deformations can be constructed directly in four-dimensional loop momentum space, first at one loop \cite{Gong_2009, Becker_2012,Becker_2012_2,Becker_2012_3} and later extended to more loops \cite{Becker_2013}.
It can however be beneficial to analytically manipulate the integrand first before constructing a contour deformation.
For example, one finds that in Feynman parameter space some configurations of external momenta do not even introduce threshold singularities.
Others require a contour deformation like the ones constructed in \cite{Nagy_2006,Binoth_2005,Lazopoulos_2007,Anastasiou_2007,Anastasiou_2008}, which were successfully employed in combination with sector decomposition \cite{Binoth_2000,Heinrich:2008si,CARTER20111566,SMIRNOV2009735,Borowka_2018}.
\par

Moreover, for loop integrals in momentum space, it was discovered that analytically integrating out the loop energies, resulting in the Loop-Tree Duality (LTD) representation \cite{Catani_2008,Capatti_2019,Runkel_2019,Aguilera_Verdugo_2020,Aguilera_Verdugo_2021_2,Ram_rez_Uribe_2021}, simplifies the singularity structure drastically: In spatial momentum space threshold singularities are confined to compact regions and can be avoided altogether for certain kinematics.
A corresponding contour deformation was first studied at one-loop for selected momentum configurations \cite{buchta2015numerical} and recently constructed from geometrical considerations for arbitrary multi-loop integrals \cite{Capatti_2020}.
\par

Contour deformation has shown to be a powerful tool that enabled the precise calculation of challenging multi-scale multi-loop integrals. 
However, the construction of a valid deformation field as in \cite{Capatti_2020} is computationally costly.
In fact, it relies on numerically solving a convex optimisation problem that is increasingly hard to solve the more threshold singularities are present.
This is especially cumbersome because every single phase space point requires its own deformation.
Moreover, a contour deformation is controlled by many parameters.
Finding their numerically optimal setting usually requires a careful study of the contour and the resulting integrand.
Even if the contour is mathematically correct, it may still be in the close vicinity of a pole, or introduce a complicated Jacobian.
This may result in a large variance of the integrand, which directly impairs the convergence of the Monte Carlo integration.
\par

In this article, we intend to improve on the drawbacks of a deformation-based approach and instead investigate an alternative way of regulating threshold singularities based on subtraction.
First efforts with such a strategy were already made in \cite{kilian2009numerical}.
However, the described method inflicted a systematic error in the presence of intersecting threshold singularities that could not be estimated.
For this reason, the method presented in this article follows a new construction that does not compromise on intersecting thresholds.
We propose a residue based scheme that subtracts threshold singularities of one-loop integrals with local counterterms.
The subtracted integral will correspond to the real (dispersive) part of the integral.
It is free of poles and can be directly evaluated by Monte Carlo numerical integration.
The counterterms are constructed such that they trivially integrate and contribute exclusively to the imaginary (absorptive) part of the integral.
Eventually, we find a local representation of the optical theorem that connects the imaginary part of a one-loop integral to the two-body phase space integral of tree-level interferences.
The resulting expressions can be represented in terms of cut diagrams \cite{doi:10.1063/1.1703676} with modified causal prescriptions.
We apply our numerical method to a range of finite one-loop integrals with up to 30 external legs and various kinematic configurations, and assess the computational performance.
Furthermore, we identify the method's limitations by carefully investigating the threshold singularity structure.
In fact, we find that the boundary conditions for our results cannot always be met for certain complicated threshold overlap structures.
\par

The outline of the article is as follows:
In sect.~\ref{sec:resolving_causal_prescription}, we sketch our subtraction method in some generality.
First, in sect.~\ref{sec:generic_case}, we consider integrands with single poles only, where later in sect.~\ref{sec:degenerate_case} we interpret our results in the presence of higher-order poles, which gives rise to local cancellations among residues.
We apply our method to a simple example in sect.~\ref{sec:simple_example}.
In sect.~\ref{sec:feynman_integrals}, we target one-loop integrals.
First, in sect.~\ref{sec:one_loop_ltd}, we set the stage by recalling the LTD formalism.
We then apply our method to the LTD representation in sect.~\ref{sec:integrating_magnitude} arriving at the main result of this article in sect.~\ref{sec:local_unitarity}.
In sect.~\ref{sec:divergent_loop_integrals} we address divergent loop integrals before presenting results obtained with our method in sect.~\ref{sec:results} for a range of one-loop configurations with 
threshold singularity structures of various complexity.
A detailed connection to the optical theorem is made in sect.~\ref{sec:unitarity_raised_props_ot}.
We make final remarks and discuss future directions in sect.~\ref{sec:conclusion}.
In appendices~\ref{app:residue_cancellations} and \ref{app:dual_propagator} we show the local cancellations for second order poles explicitly and give a summary of the geometry of threshold singularities within LTD.

\section{Resolving the causal prescription}
\label{sec:resolving_causal_prescription}
Let us start by recalling a useful distributional identity from complex analysis, the Sokhotski-Plemelj theorem on the real line,
\begin{align}
\label{eq:sokhotski_plemelj}
    \lim_{\epsilon\to0}\frac{1}{x\pm\ii\epsilon}
    =
    \PV
    \frac{1}{x}
    \mp
    \ii \pi \delta(x),
\end{align}
for $x \in\mathbb{R}$, where $\PV$ denotes the Cauchy principal value, $\delta$ the Dirac delta function and $\epsilon>0$ the regulating \emph{causal prescription}, named after its role in physics.
It states that a pole of an integrand, located on the complex plane just above or below the real integration interval, will contribute to an imaginary part of the integral.
In fact, the real and imaginary parts can be a priori separated into two summands without explicitly solving the integral.
The real part is then given by the Cauchy principal value, by symmetrically evaluating the bounds of the improper integral.
The imaginary part is the fractional residue at the pole.
\par

In this section, we will consider a general integral over the real line regulated by a causal prescription $\epsilon>0$.
By making use of the Sokhotski-Plemelj theorem we will resolve the integral's $\epsilon$-dependence and separate its real and imaginary part.
Furthermore, we remove the Cauchy principal value prescription with suitable counterterms to make its evaluation applicable for Monte~Carlo numerical integration.
\par

First, we will discuss the \emph{generic case} in sect.~\ref{sec:generic_case}, where we consider single poles only, at pairwise distinct locations (for all values of $\epsilon>0$ including the limit $\epsilon\to0$).
We will derive a set of simple counterterms as well as explicit expressions for both real and imaginary parts of the integral.
As a consequence of the counterterms, the remaining integral expression for the real part is free of poles in the integration space.
\par

Next, we will consider the \emph{degenerate case} in sect.~\ref{sec:degenerate_case}, where multiple single poles merge into higher-order poles (in the limit $\epsilon\to0$).
These poles can be divided into two categories, \emph{pinched} and \emph{non-pinched} poles.
In the presence of non-pinched poles the expressions in sect.~\ref{sec:generic_case} remain valid and feature local cancellations among residues and separately among counterterms.
In the presence of pinched poles however, these cancellations are broken rendering the integral divergent.
This divergent behaviour is reminiscent of infrared singular loop integrals in $D=4$ dimensions, whose soft and collinear infrared singularities define \emph{pinch surfaces} in the integration space (cf. \cite{Collins:1989gx}).
However, the singularities we will encounter are at worst \emph{locally} pinched in a proper subspace of the integration domain.
In this case, as we discuss in sect.~\ref{sec:multidimensional_integrals}, multi-dimensional integrals remain finite but the regulator $\epsilon>0$ cannot be directly removed as opposed to integrals in one dimension.
\par

Last but not least, we will apply these results to a simple example in sect.~\ref{sec:simple_example}.

\subsection{Generic case}
\label{sec:generic_case}

We introduce the integral
\begin{align}
\label{eq:generic_integral}
    I
    =
    \lim_{\epsilon \to 0}
    \int_{\mathbb{R}}
    \dd x\;
    \mathcal{I}^\epsilon(x),
\end{align}
where the integrand $\mathcal{I}^\epsilon(x)$ has single poles at $n$ pairwise distinct locations ${x_1^\epsilon,\dots,x_n^\epsilon}$ for $\epsilon>0$ as well as in the limit $\epsilon\to 0$, with the property that each point $x_i^\epsilon$ has a non-zero imaginary part that vanishes for $\epsilon\to 0$.
This means that for $\epsilon\to0$, the integrand has poles on the real axis.
We further assume that 
$\mathcal{I}^0(x)$ is real $\forall x\in\mathbb{R}\setminus\{x_1^0,\dots,x_n^0\}$ and that $I$ is finite, meaning that the causal prescription $\epsilon>0$ is a sufficient regulator of the integral.
However, the integral is ill-defined if the limit $\epsilon\to0$ is taken prior to integration.
\par

According to the Sokhotski-Plemelj identity in eq.~\eqref{eq:sokhotski_plemelj}, we a priori know that whenever the integration contour is close to a single pole, the integral $I$ gets an imaginary contribution.
To apply this identity systematically, we first subtract each pole at $x_i^\epsilon$ with a local counterterm $\CT_i^\epsilon$ and then add it back by directly employing eq.~\eqref{eq:sokhotski_plemelj}.
We now study the conditions that such a counterterm has to satisfy.
\par

Around each single pole at $x_i^\epsilon$ in the complex plane the integrand behaves as
\begin{align}
\label{eq:expansion_around_pole}
    \mathcal{I}^\epsilon(x)
    = \frac{
        \Res[\mathcal{I}^\epsilon(y),y=x_i^\epsilon]
        }{x-x_i^\epsilon}
    + \order((x-x_i^\epsilon)^0).
\end{align}
Therefore, any valid counterterm $\CT_{i}^\epsilon$ has to remove this leading order contribution of the integrand $\mathcal{I}^\epsilon$ in order to cancel the pole at $x_i^\epsilon$.
In particular, any choice of function with identical leading order term like the above can serve as a counterterm $\CT_{i}^\epsilon$.
Counterterms are therefore not unique.
To determine what a good counterterm is, we go one step ahead and consider the obstacles, when integrating it using the Sokhotski-Plemelj identity in eq.~\eqref{eq:sokhotski_plemelj}.
\par

According to eq.~\eqref{eq:expansion_around_pole}, any valid counterterm must satisfy
\begin{align}
\label{eq:counterterm_integral}
    \lim_{\epsilon \to 0}
    \int_\mathbb{R} \CT_{i}^\epsilon(x)
    \dd x
    =
    \PV \int_\mathbb{R} \CT_{i}^0(x)
    \dd x
    +
    \ii
    \pi
    \sgn(\Im x_i^{+0})
    \Res[\mathcal{I}^0(y),y=x_i^0],
\end{align}
where the winding number $\sgn(\Im x_i^{+0})\coloneqq\lim_{\epsilon \to 0} \sgn(\Im x_i^\epsilon)$ is independent of $\epsilon>0$.
The above equation can be simplified by choosing the counterterm to be anti-symmetric in $x$ around $x_i^0$, i.e. $\CT_{ij}^0 (x+x_i^0)=-\CT_{ij}^0 (-x+x_i^0)$, such that the Cauchy principal value over $\mathbb{R}$ vanishes.
Furthermore, $\CT_i^\epsilon$ must not introduce new poles apart from the one at $x_i^\epsilon$ and the sum of counterterms must be such that the integral over the subtracted integrand $\mathcal{I}^0-\sum_i\CT_i^0$ is ultraviolet (UV) finite.
\par

Notice that the leading order term in eq.~\eqref{eq:expansion_around_pole} is already anti-symmetric and its Cauchy principal value integral vanishes.
It might however introduce a UV singularity to the subtracted integral\footnote{
    Convergence of the integral
    $\displaystyle\lim_{R\to\infty} \int_{-R}^R f(x)\, \dd x$
    does not imply convergence of the improper integral
    $\int_{-\infty}^\infty f(x)\, \dd x
    = \displaystyle\lim_{a\to-\infty}\displaystyle\lim_{b\to\infty}
    \int_{a}^b f(x)\,\dd x$,
    as e.g. for $f(x)=x$.
}.
In general, it is therefore necessary to introduce a (symmetric) UV suppression to the counterterm in order to guarantee convergence of the subtracted integral, as
\begin{align}
    \label{eq:counterterm}
    \CT_{i}^\epsilon(x)
    =
    \frac{
        \Res[\mathcal{I}^\epsilon(y),y=x_i^\epsilon]
        }{x-x_i^\epsilon}
    \chi_i(x-x_i^\epsilon),
\end{align}
where $\chi_i$ is an arbitrary \emph{symmetric} function with $\chi_i(0)=1$ and $\lim_{x\to \infty} \chi_i(x)=0$.\footnote{
If 
the integrand $\mathcal{I}^\epsilon$ is a rational function, we can choose $\chi_i\equiv1$ and still get a UV convergent subtracted integral (which in fact vanishes), since in this case the counterterms correspond to the partial fraction decomposition of the integrand, which algebraically cancel all poles (and the integrand).
}
A simple suppression function is $\chi_i(x)=\Theta(|x|-c_i)$, where $\Theta$ is the Heaviside step function and $c_i>0$ is an arbitrary UV cutoff.
A smooth alternative is the Gaussian suppression
\begin{align}
    \chi_i(x)
    = \exp\left(
        -\frac{x^2}{2 a_i^2}
    \right)
\end{align}
with width $a_i>0$.
We will see in sect.~\ref{sec:degenerate_case} that if two single poles at $x_i^\epsilon$ and $x_j^\epsilon$ merge into a double pole without pinching the integration contour, i.e. $\sgn(\Im x_i^{+0})=\sgn(\Im x_j^{+0})$, the two suppression functions $\chi_i$ and $\chi_j$ must be identical, such that their corresponding counterterms do not introduce additional singularities.
\par
Given the counterterms in eq.~\eqref{eq:counterterm}, whose Cauchy principal value integral vanishes by construction, we can now use the identity in eq.~\eqref{eq:counterterm_integral} to separate the real and imaginary part of the integral $I$, as
\begin{align}
\label{eq:real_integral}
    \Re I
    &=
    \int_\mathbb{R}
    \left(
        \mathcal{I}^0(x)-\sum_{i=1}^n \CT_i^0(x)
    \right)
    \dd x, \\
\label{eq:imag_integral}
    \Im I
    &=
    \pi
    \sum_{i=1}^n
    \sgn(\Im x_i^{+0})
    \Res[\mathcal{I}^0(y),y=x_i^0],
\end{align}
where we use the shorthand $\sgn(\Im x_i^{+0})\coloneqq\lim_{\epsilon\to0}(\Im x_i^\epsilon)$.
It holds that $\Re I = \PV I$.
\par

This is the main result of this section.
It states that the real part of $I$ is given by the integral over the \emph{subtracted integrand} evaluated at $\epsilon=0$.
It is free of poles on the real line.
The imaginary part of $I$ is given by the sum of fractional residues at $\epsilon=0$.
Keep in mind that the winding numbers implicitly depend on the sign of the causal prescription $\epsilon>0$.
Also, note that whereas the counterterms are summed over with equal prefactor, the sum of residues is weighted with the corresponding winding numbers.
\par

We recall that the separation of real and imaginary part as above relies on the assumption that $\mathcal{I}^0(x)$ is real away from the poles at locations $x_1^0,\dots,x_n^0$.
Furthermore, we emphasize that in this derivation the residue and its corresponding counterterm in eq.~\eqref{eq:counterterm} concern only \emph{single} poles.
We will see in sect.~\ref{sec:degenerate_case} that eqs.~\eqref{eq:real_integral} and \eqref{eq:imag_integral} hold more generally in the presence of \emph{non-pinched} higher-order poles, where (at least) two single poles $x_i^\epsilon$ and $x_j^\epsilon$ merge into a double pole from the same side of the integration contour, here $\sgn(\Im x_i^{+0})=\sgn(\Im x_j^{+0})$.
In this case, the expressions for $\Re I$ and $\Im I$ feature local cancellations in the sum of residues as well as in the sum of counterterms.
In the presence of \emph{pinched} poles however, where $\sgn(\Im x_i^{+0})=-\sgn(\Im x_j^{+0})$, these cancellations are broken in the imaginary part of $I$, rendering eq.~\eqref{eq:imag_integral} divergent.
Our assumption that the particular integral $I$ in eq.~\eqref{eq:generic_integral} is finite rules out the appearance of such pinched poles.
This is consistent, since it is a priori clear that in this case of pinched poles, the integral $I$ is ill-defined, as any possible contour deformation is constrained to go \emph{through} the pole\footnote{
Despite the appearance of pinched poles in one integration dimension, the integral can still be finite if there remain more integrations in extra dimensions to perform (cf. sect.~\ref{sec:multidimensional_integrals}).
In this case, we are dealing with \emph{local} pinches, whose appearance depends on the parameterisation of the integration space.
}.
\par

The formulae in eqs.~\eqref{eq:real_integral} and \eqref{eq:imag_integral} constitute a an alternative representation of eq.~\eqref{eq:generic_integral}.
This representation for $\Re I$ in eq.~\eqref{eq:real_integral} is now suitable for direct numerical integration since the integrand is free of poles in the integration space. 
Note that $\Im I$ in eq.~\eqref{eq:imag_integral} is already solved analytically\footnote{
If $I$ were a multi-dimensional integral over $n$ integration variables (see sect.~\ref{sec:multidimensional_integrals}) the corresponding expressions for eqs.~\eqref{eq:real_integral} and \eqref{eq:imag_integral} would be $n$- and $n-1$-dimensional integrals, respectively.
}.
In sect.~\ref{sec:feynman_integrals}, we apply the results of this section to one-loop integrals.

\subsection{Degenerate case}
\label{sec:degenerate_case}

The results in the previous section, namely the expressions in eq.~\eqref{eq:real_integral} and eq.~\eqref{eq:imag_integral}, were derived under the assumption that the integrand $\mathcal{I}^\epsilon$ has single poles only.
In this section, we lift this assumption and investigate the \emph{degenerate} case, where the integrand in eq.~\eqref{eq:generic_integral} has higher-order poles in the limit $\epsilon\to0$.
We can assume that for non-vanishing $\epsilon>0$ the integrand still only has single poles, since we only want higher-order poles for $\epsilon\to0$.
\par

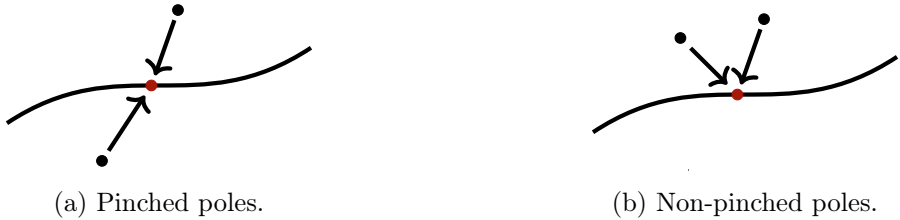
\begin{figure}[H]
	\begin{subfigure}[t]{0.49\textwidth}
		\centering
		\begin{tikzpicture}
			\coordinate (u) at (0.25,1.5);
			\coordinate (d) at (-0.75,-0.5);
			\coordinate (m) at (-0.1,0.5);
			\node (U) at (u) {};
			\node (D) at (d) {};
			\node (M) at (m) {};
			\filldraw [black] (u) circle (2pt);
			\filldraw [black] (d) circle (2pt);
			\draw [ultra thick] (-2,0) .. controls (-0.5,1) and (0.5,0) .. (2,1);
			\draw [ultra thick,->] (U) edge (M);
			\draw [ultra thick,->] (D) edge (M);
			\filldraw [myDarkRed] (m) circle (2pt);
		\end{tikzpicture}
		\caption{\label{fig:pinched}Pinched poles.}
	\end{subfigure}\hfill
	\begin{subfigure}[t]{0.49\textwidth}
		\centering
		\begin{tikzpicture}
			\coordinate (d) at (-0.75,-0.5);
			\coordinate (u1) at (0.25,1.5);
			\coordinate (u2) at (-0.85,1.25);
			\coordinate (m) at (-0.1,0.5);
			\node (U1) at (u1) {};
			\node (U2) at (u2) {};
			\node (M) at (m) {};
			\filldraw [black] (u1) circle (2pt);
			\filldraw [black] (u2) circle (2pt);
			\draw [ultra thick] (-2,0) .. controls (-0.5,1) and (0.5,0) .. (2,1);
			\draw [ultra thick,->] (U1) edge (M);
			\draw [ultra thick,->] (U2) edge (M);
			\draw (d) -- (d);
			\filldraw [myDarkRed] (m) circle (2pt);
		\end{tikzpicture}
		\caption{\label{fig:non-pinched}Non-pinched poles.}
	\end{subfigure}
	\caption{\label{fig:(non)pinch-poles}Two ways that poles merge onto the integration contour for $\epsilon\to0$.}
\end{figure}

Let us start with two single poles $x_i^\epsilon$ and $x_j^\epsilon$ (i.e. $x_i^\epsilon\neq x_j^\epsilon$) for $\epsilon>0$.
For $\epsilon\to 0$, we assume that these two poles merge into a double pole on the integration contour, which implies that\footnote{
Note that the condition $x_i^0=x_j^0$ alone does not guarantee a double pole.
For example the function $f(x)=\frac{1}{x-x_i^\epsilon}+\frac{1}{x-x_j^\epsilon}$ only has single poles, also for $x_i^0\to x_j^0$.
}
\begin{align}
    x_i^0=x_j^0.
\end{align}
As depicted in fig.~\ref{fig:(non)pinch-poles}, we distinguish between two ways the poles can merge for $\epsilon\to 0$ (cf. \cite{Cohen2007}):
If the two poles approach the integration contour from opposite sides, we call them \emph{pinched}, since they pinch the contour for $\epsilon\to0$.
On the other hand, if the two poles approach the contour from the same side, we call them \emph{non-pinched}.
\par

In our case, where the contour is the real line, the poles at $x_i^\epsilon$ and $x_j^\epsilon$ are pinched if 
\begin{align}
\label{eq:pinched_definition}
    \sgn(\Im x_i^{+0})=-\sgn(\Im x_j^{+0}).
\end{align}
They are non-pinched if
\begin{align}
\label{eq:non_pinched_definition}
    \sgn(\Im x_i^{+0})=\sgn(\Im x_j^{+0}).
\end{align}
The generalisation to poles of order higher than two is analogous.
The single poles $x_1^\epsilon,\dots,x_n^\epsilon$ that merge into a pole of order $n$ are pinched if at least two out of $n$
are pinched.
Otherwise they are non-pinched.
We can also expand this terminology to include higher order poles for non-vanishing $\epsilon>0$.
In fact, a pole of order $n$ can be understood as $n$ overlapping single poles, which are non-pinched since they follow the same trajectory for $\epsilon\to0$.
\par

We can now investigate the integral in eq.~\eqref{eq:generic_integral} in the degenerate limit.
On first sight, our expressions for its real and imaginary part in eqs.~\eqref{eq:real_integral} and \eqref{eq:imag_integral} seem to break down:
The residue $R_i^\epsilon$ of a single pole at $x_i^\epsilon$ diverges if the pole merges with another one at $x_j^\epsilon$.
So does the counterterm $\CT_i^\epsilon\propto R_i^\epsilon/(x-x_i^\epsilon)$.
Note, however, that the same holds for the other pole at $x_j^\epsilon$ i.e. $R_j^\epsilon$ and $\CT_j^\epsilon$ diverge as well.
Despite their individual divergences, the \emph{sum} of residues and counterterms remains finite, as we will explicitly demonstrate below in sect.~\ref{sec:local_cancellations}.
As a result, we will find that for non-pinched poles the expression for $\Im I$ in eq.~\eqref{eq:imag_integral} as well as the for $\Re I$ in eq.~\eqref{eq:real_integral} will feature local cancellations among residues and counterterms, respectively.
In the presence of pinched poles however $\Im I$ is singular due to the poles' opposite winding numbers $\sgn(\Im x_i^{+0})$.
The expression for $\Re I$ is always free of singularities, also in the pinched case.
\par

\subsubsection{Origin and mechanism of local cancellations}
\label{sec:local_cancellations}

In this section we demonstrate the mechanism of local cancellations by considering the simple case, where two single poles merge and combine into a double pole.
It is evident that the combination of $n$ single poles into a pole of order $n$ can be studied analogously.
We will show that this type of cancellations is a simple property of calculus with residues.
\par
To characterise the scenario of two merging poles, we introduce a function
\begin{align}
\label{eq:integrand_local_cancellations}
    \mathcal{I}(x,y,z) = \frac{f(x)}{g(x,y,z)},
\end{align}
which is a ratio of two functions $f$ and $g$ without poles, where $g$ is assumed to have single zeros at $x=y$ and $x=z$ if $y\neq z$ or a double zero if $y=z$.
This implies that
\begin{align}
    g^{(1,0,0)}(y,y,z)&\neq0,
    \qquad
    g^{(1,0,0)}(z,y,z)\neq 0,\\
    g^{(1,0,0)}(y,y,y)&=0,
    \qquad
    g^{(2,0,0)}(y,y,y)\neq0,
\end{align}
for $y\neq z$, where $g^{(a,b,c)}(x,y,z)\coloneqq \partial_{x}^a\, \partial_{y}^b\, \partial_{z}^c\, g(x,y,z)$ for integers $a,b, c$ refers to the respective partial derivative.
Under the assumption that $y\neq z$, the function $\mathcal{I}$ only features single poles like the integrand in eq.~\eqref{eq:generic_integral}.
Then, the residues at the two locations $x=y$ and $x=z$ are given by
\begin{align}
    \begin{split}
    \label{eq:residues_local_cancellations}
    R_y(y,z)
    &\coloneqq \Res[\mathcal{I}(x,y,z),x=y]
    = \frac{f(y)}{g^{(1,0,0)}(y,y,z)},
    \\
    R_z(y,z)
    &\coloneqq \Res[\mathcal{I}(x,y,z),x=z]
    = \frac{f(z)}{g^{(1,0,0)}(z,y,z)}.
    \end{split}
\end{align}
Note that the residues are well defined, since their denominators are non-zero by assumption.
Their corresponding counterterms, following the construction in eq.~\eqref{eq:counterterm}, are
\begin{align}
    \label{eq:counterterms_local_cancellations}
    \CT_y(x,y,z)
    \coloneqq \frac{R_y(y,z)}{x-y} \chi_y(x-y),
    \qquad
    \CT_z(x,y,z)
    \coloneqq \frac{R_z(y,z)}{x-z} \chi_z(x-z),
\end{align}
where $\chi\coloneqq\chi_y\equiv\chi_z$ is a symmetric suppression function.
For merging poles, it is necessary that the suppression functions are chosen to be equal for all corresponding counterterms.
\par

\begin{figure}[t]
	\centering
	\begin{tikzpicture}
		\coordinate (l) at (-0.8,0.8);
		\coordinate (r) at (0.7,-0.7);
		\coordinate (m) at (-0.15,-0.1);
		\node (L) at (l) {};
		\node (R) at (r) {};
		\node (M) at (m) {};
		\draw [ultra thick, ->]  plot[smooth cycle, tension=1] coordinates {(-1,1.25) (1,1) (1,-1) (-1,-1.25) };
		\draw [ultra thick, ->] (1-0.0001,1) -- (1-0.001,1+0.001);
		\draw [ultra thick, ->] (-1+0.00015,-1.25) -- (-1+0.001,-1.25-0.001);
		\filldraw [black] (l) circle (2pt);
		\filldraw [black] (r) circle (2pt);
		\draw [ultra thick,->] (L) edge (M);
		\draw [ultra thick,->] (R) edge (M);
		\filldraw [myDarkRed] (m) circle (2pt);
	\end{tikzpicture}
	\caption{\label{fig:double_residue} Two single poles that merge into a double pole for $\epsilon\to0$ within a closed integration contour.
	The sum of their residues converges to the residue around the double pole.
	}
\end{figure}
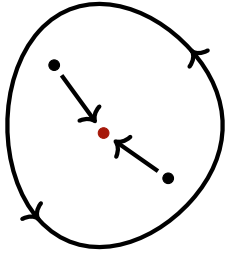

We now present three key results that demonstrate how the residues, counterterms and the integrand combine in the limit $z\to y$.
The detailed derivation can be found in app.~\ref{app:residue_cancellations}.
First of all, notice that both residues $R_y(y,z)$ and $R_z(y,z)$ diverge in the limit $z \to y$.
However, their sum converges to the residue around a double pole, i.e.
\begin{align}
\label{eq:limit_sum_residues}
    \lim_{z\to y}
    \left(
        R_{y}(y,z) + R_{z}(y,z)
    \right)
    = \Res[\mathcal{I}(x,y,y),x=y]
    \eqqcolon R(y)
    .
\end{align}
This is a simple consequence of the residue theorem, as depicted in fig.~\ref{fig:double_residue}.
\par

It then follows that for $z\to y$ the sum of the two counterterms not only converges despite being individually divergent, i.e.
\begin{multline}
\label{eq:limit_sum_counterterms}
    \lim_{z\to y} \left(\CT_y(x,y,z) + \CT_z(x,y,z)\right)
    =
    \frac{\chi(x-y)}{x-y}
    R(y)
    +
    \left(
        \frac{\chi(x-y)}{(x-y)^2}
        -
        \frac{\chi'(x-y)}{(x-y)}
    \right)
    \frac{2f(y)}{\tilde g''(y)},
\end{multline}
where $\tilde g(x) \coloneqq g(x,y,y)$, but also serves as a valid subtraction term for the double pole of $\mathcal{I}$, i.e. the subtracted integrand converges to
\begin{multline}
\label{eq:limit_subtracted_integrand}
    \lim_{x\to y}
    \lim_{z\to y}
    \left(
        \mathcal{I}(x,y,z)-\CT_y(x,y,z)-\CT_z(x,y,z)
    \right)
    =
    C(y)
    - \chi'(0) R(y)
    + \chi''(0) \frac{f(y)}{\tilde g''(y)}
\end{multline}
where $C$ is a constant in $x$ and $z$ (see app.~\ref{app:residue_cancellations}).
\par
The above results show that the residues and counterterms of two single poles combine to the residue and to a valid counterterm of a double pole in the limit where the two poles merge.
The property of residues in eq.~\eqref{eq:limit_sum_residues} is what eventually allows us to construct a single counterterm per each single pole, which generalises naturally to higher-order poles as the limit in which single poles merge.
\par
This finding can now be applied to the problem of computing the integral $I$ in eq.~\eqref{eq:generic_integral}.
We will see that the fractional residues in eq.~\eqref{eq:imag_integral} do not always allow for the cancellation mechanism discussed above.
\par
First of all, consider the real part of the integral $I$ in eq.~\eqref{eq:real_integral}.
It is the integral over the subtracted integrand.
Note that all counterterms contribute to the integrand with the same sign.
Therefore, two counterterms corresponding to two degenerate poles $x_i^\epsilon$ and $x_j^\epsilon$ locally cancel each other's singularities as well as the double pole of the integrand $I$, as in eqs.~\eqref{eq:limit_sum_counterterms} and \eqref{eq:limit_subtracted_integrand}.
It therefore follows that the subtracted integral is finite and amounts to the Cauchy principle value.
This holds true independently of whether or not the poles $x_i^\epsilon$ and $x_j^\epsilon$ are pinched.
\par

Secondly, we discuss the imaginary part of $I$ in eq.~\eqref{eq:imag_integral} in the degenerate limit.
It is a sum of single pole residues, weighted with their winding number.
The winding numbers of two residues are equal if two degenerate poles $x_i^\epsilon$ and $x_j^\epsilon$ are non-pinched.
In this case, the two residues exhibit local cancellations and combine into the residue of a double pole in the limit $\epsilon\to 0$, according to eq.~\eqref{eq:limit_sum_residues}.
It follows that eq.~\eqref{eq:imag_integral} is still valid in the presence of non-pinched poles.
\par

The winding numbers are opposite however, if $x_i^\epsilon$ and $x_j^\epsilon$ are pinched.
In this case, we are left with a difference of single pole residues that diverges for $\epsilon\to0$.
The two residues do not exhibit local cancellation but instead contribute a pole to the imaginary part of $I$.
In the presence of pinched poles, the limit $\epsilon\to0$ of eq.~\eqref{eq:imag_integral} therefore does not exist and we cannot remove the causal prescription $\epsilon>0$.
Note that this divergent behaviour is expected since every possible integration contour is constrained to go \emph{through} the pole, in which case the integral $I$ is ill-defined to begin with.
This argument is however only valid for one-dimensional integrals.
We will discuss the multi-dimensional case in the next section.
\par

\subsubsection{Multi-dimensional integrals}
\label{sec:multidimensional_integrals}

The above discussion applies to integrals over a single integration dimension.
Although our results can simply be generalised to the multi-dimensional case by iteratively performing one-dimensional integrals, some notions may need clarification.
\par

For our purposes, we consider a simple extension to a multi-dimensional integral of the form 
\begin{align}
\label{eq:multi_dim_integral_start}
    I =
    \lim_{\epsilon\to0}
    \int_{\mathbb{R}\times V}
    \dd (x,z) \, \mathcal{I}^\epsilon(x,z),
\end{align}
where $x\in \mathbb{R}$ and $z\in V$ for some region $V$.
We assume the integrand $\mathcal{I}^\epsilon$ to have the same properties in the one-dimensional case, in particular, that all its poles in the integration domain can be parameterised by $x(z)$.
\par

By virtue of Fubini's theorem one can perform the integral over $x$ first, allowing us to apply our subtraction method in one dimension.
However, subtleties arise in the treatment of the causal regulator $\epsilon$.
This can already be seen in a simple scenario, where two single poles pinch the $x$-contour for some value of $z$.
Their residues in $x$ then each have a remaining single pole in $V$ that must be regulated for the $z$-integration to be convergent.
In general, this can be achieved by deforming the $z$-contour around the remaining poles.
Therefore, the regulator $\epsilon$ cannot yet be removed in the residues as it dictates the orientation of the contour deformation.
As a consequence the integrals of the residues generally do not have a pure phase.
Moreover, the Cauchy principal value cannot be identified with the subtracted integral alone but also gets contributions from the residue integrals.
\par

We shall therefore use the expression 
\begin{equation}
\label{eq:multi_dim_integral}
    \begin{split}
        I &=
        \int_V \dd z
        \int_{\mathbb{R}} \dd x
        \left(
            \mathcal{I}^0(x,z)-\sum_{i=1}^n \CT_i^0(x,z)
        \right)
        \dd x \\
        &\phantom{{}=}+
        \ii \pi
        \lim_{\epsilon\to0}
        \int_{V}\dd z
        \sum_{i=1}^n
        \sgn(\Im x_i^{+0})
        \Res[\mathcal{I}^\epsilon(y,z),y=x_i^\epsilon],
    \end{split}
\end{equation}
when using our subtraction method with the simple multi-dimensional integrals in eq.~\eqref{eq:multi_dim_integral_start}.
Note that the subtracted integral is still evaluated at $\epsilon=0$.
The residues however have a remaining $\epsilon$ dependence.
\par

As we have seen in sect.~\ref{sec:degenerate_case}, integrals with a single integration dimension are always singular in the presence of pinched poles.
However, in multi-dimensional integrals poles have to pinch \emph{all} integration dimensions for the integral to be singular.
Such \emph{pinch surfaces} in the integration domain cannot be avoided by any deformation of the contour.
They appear for example in loop integrals as infrared singularities.
On the other hand, singularities that only pinch the contour in a proper subspace may be avoided by deforming the contour in the remaining integration subspace.
We refer to such poles as \emph{local pinches}.
Their presence is an artifact of the choice of parameterisation of the integration domain.
The pinched poles we encounter in the following, unless explicitly referred to as infrared soft or collinear singularities, are all local as they do not result in a singularity of the integral.
\par

In general, the expression in eq.~\eqref{eq:multi_dim_integral} is not directly suited for Monte~Carlo numerical integration as the integral over the residues still depends on the causal prescription $\epsilon$.
Numerically, the residue integral can then be regulated with a contour deformation in $z$ or by iterating our subtraction method.
However, the regulator $\epsilon$ can be removed if the integrand in eq.~\eqref{eq:multi_dim_integral_start} has no poles that pinch the $x$-contour.
In this case, we can separate real and imaginary part of the integral analogous to eqs.~\eqref{eq:real_integral} and \eqref{eq:imag_integral}, which can both be directly performed with Monte~Carlo integration.
In particular, this applies to all one-loop integrals we consider in sect.~\ref{sec:results}.

\subsection{Example}
\label{sec:simple_example}
We now apply our expressions in eqs.~\eqref{eq:real_integral} and \eqref{eq:imag_integral} and illustrate the mechanism of local cancellations in a simple example.
\par

Consider an integral as in eq.~\eqref{eq:generic_integral} with integrand
\begin{align}
    \label{eq:example_integrand}
    \mathcal{I}^\epsilon(x)
    &=
    \frac{1}{\vphantom{\sqrt{x^2}}x-(z-\ii \epsilon)
        }
    \frac{1}{
        \sqrt{x^2-(y^2+\ii \epsilon)+2^2}-2
        },
\end{align}
where $y\in[-2,2]$ and $z\in\mathbb{R}$.
The integrand has three poles at $x_0^\epsilon=z-\ii\epsilon$, $x_\pm^\epsilon=\pm\sqrt{y^2+\ii\epsilon}$.
The integral in the limit $\epsilon\to0$ can be evaluated analytically for values of $z$ and $y$ by e.g. rationalising the square root.
It is however not a priori clear how to evaluate this integral numerically as the limit $\epsilon\to0$ is numerically unstable, due to cancellations of theoretically infinitely large numbers.
Of course, Monte Carlo integration with vanishing $\epsilon$ is bound to fail due to the poles in the integration domain.
However, according to eq.~\eqref{eq:real_integral}, we can remove the regulator if we subtract the poles with local counterterms. 
The subtracted integrand can then be evaluated by Monte Carlo numerical integration and directly yields the real part, or equivalently the Cauchy principle value, of the integral.
Its imaginary part is given by the residues at the poles as in eq.~\eqref{eq:real_integral}.
\par

The poles are at pairwise different locations if $z\neq \pm y \neq 0$.
In this case, we can compute three residues for single poles, each multiplied by their respective winding number, as
\begin{align}
    \label{eq:example_residue0}
    \sgn(\Im x_0^{+0}) \Res[\mathcal{I}^\epsilon(y),y=x_0^\epsilon]
    &=
    \frac{-1}{\sqrt{(z-\ii \epsilon)^2-(y^2+\ii \epsilon)+2^2}-2}, \\
    \label{eq:example_residuePM}
    \sgn(\Im x_\pm^{+0}) \Res[\mathcal{I}^\epsilon(y),y=x_\pm^\epsilon]
    &=
    \frac{1}{\pm \sqrt{y^2+\ii \epsilon}-(z-\ii \epsilon)}
    \frac{2}{\sqrt{y^2+\ii\epsilon}},
\end{align}
According to eq.~\eqref{eq:imag_integral} the imaginary part of the integral for $\epsilon\to 0$ is given by the sum of the above fractional residues.
After some algebraic manipulations, we find
\begin{align}
\label{eq:example_imag}
    \frac{\Im I}{\pi}
    =
    \frac{
        |y|
        \left(2+\sqrt{z^2-y^2+2^2}\right)
        +4 z
    }{
        |y|
        \left(y^2-z^2\right)
    }
    .
\end{align}
The subtracted integrand is given by
\begin{align}
    \label{eq:example_real}
    \mathcal{I}^0-\CT_0^0-\CT_+^0-\CT_-^0,
\end{align}
where the counterterms $\CT_i^\epsilon$ for each of the three poles $x_i^\epsilon$ for $i\in\{0,\pm\}$ are constructed from the above residues as in eq.~\eqref{eq:counterterm}.
\par
According to sect.~\ref{sec:degenerate_case}, our discussion has to be slightly modified if the integrand has higher-order poles.
Indeed, the integrand in eq.~\eqref{eq:example_integrand} can feature higher-order poles if $y=0$ or if $z=\pm y$.
\par
If $z=-|y|< 0$ the poles $x_0^\epsilon$ and $x_-^\epsilon$ merge into a non-pinched double pole for $\epsilon\to 0$.
Note that the residue at $x_0^\epsilon$ as well as the residue at $x_-^\epsilon$ individually diverge for $z\to -|y|$.
However, the poles in these two residues are spurious as they locally cancel in their sum.
Therefore, the limit of eq.~\eqref{eq:example_imag} for $z\to -|y|$ and $\epsilon \to 0$ exists and is given by
\begin{align}
    \frac{\Im I}{\pi}
    =
    \frac{2}{y^2}-\frac{1}{4}.
\end{align}
Also our initial counterterms remain valid as their limit for $z\to -|y|$ exists (see eq.~\eqref{eq:appendix_counterterms_local_cancellations} in the appendix).
\par

If $y=0$ for arbitrary $z$ or $z=|y|\geq0$, the integrand in eq.~\eqref{eq:example_integrand} has pinched double or triple poles.
In this case, the divergences in the corresponding residues for $z\to |y|$ or for $y\to 0$ do not locally cancel in the sum of the (weighted) residues in eqs.~\eqref{eq:example_residue0} and \eqref{eq:example_residuePM}.
Therefore, the imaginary part is singular in these limits.
If we were to perform an additional integration, e.g. over $z$, we would find that we can still deform around the pole $z=|y|>0$, therefore defining only a local pinch in the starting integrand in eq.~\eqref{eq:example_integrand}. 
However, for $y=0$ even the multi-dimensional integral is divergent.
\par
Note that the same counterterms still cancel the poles of the integrand, also in the pinched case.
Therefore, the subtracted integral still converges to the real part of the original integral, i.e. the Cauchy principal value integral.
\par

\begin{table}[H]
\begin{center}
\begin{tabular}{l|c|r|c|c}
pole struct.
    & configuration
    & analytic $\Re I$ ~~~~~~~~~~~
    & num. $\Re I$
    & $\Im I$ \\
    \hline\hline
single only
    & $z=+4, y=\phantom{-}2$
    & $-\frac{2}{3} \log(2)\approx -0.462098$
    & $-0.462097(2)$
    & $-\frac{7}{6}\pi$ \\ 
    \hline
non-pinched
    & $z=\:-|y| \,=-1$
    & $1-\frac{3}{4}\log(3)\approx \phantom{-}0.176041$
    & $\phantom{-}0.176042(5)$
    & $\phantom{-}\frac{7}{4}\pi$ \\
    \hline
pinched
    & $z=\:+|y| \,=\phantom{-}2$
    & $\frac{1}{2}=-0.5\phantom{00000}$
    & $-0.500000(3)$
    & $\phantom{-}\infty$ \\
    \hline
pinched
    & $z=-6, y=\phantom{-}0$
    & $\frac{\sqrt{10}}{9}\arsinh(3)-\frac{1}{3}
    \approx \phantom{-}0.305604$
    & $\phantom{-}0.305603(2)$
    & $\phantom{-}\infty$ \\
    \hline
pinched
    & $z=\phantom{-}\:y\phantom{|}\phantom{|} \,=\phantom{-}0$
    & $\phantom{.}0\phantom{.000000}$
    & $-0.000003(6)$
    & $\phantom{-}\infty$ \\
    \hline
\end{tabular}
\end{center}
\caption{\label{tab:example} The integral of eq.~\eqref{eq:example_integrand} for five different configurations of $y$ and $z$.
In the first configuration the integrand has only single poles, in the next three a single and a double pole and in the last configuration a triple pole.
}
\end{table}

In tab.~\ref{tab:example} we present the values of our example integral for different pole structures.
The numerical result for the real part, or equivalently the Cauchy principal value, was obtained with our subtraction method and Monte Carlo numerical integration.
For all counterterms $\CT_i^\epsilon$ we chose the Gaussian UV suppression $\chi_i(x) = \exp(-x^2)$.
The numerical integration was performed with the \texttt{AdaptiveMonteCarlo} integration strategy of \textsc{Mathematica}'s \texttt{NIntegrate} with $10^6$ samples.
\par
Note that as an alternative to our method, the Cauchy principal value of one-dimensional integrals can also be numerically evaluated by symmetrising the integrand in the neighbourhood of each of its poles.
However, such a symmetrisation cannot be easily generalised for higher-order poles and multi-dimensional integrals with complicated intersecting singular surfaces in the integration domain.

\section{Loop integrals}
\label{sec:feynman_integrals}

Integrals with the same properties as the one in eq.~\eqref{eq:generic_integral} appear in perturbative quantum field theory as loop integrals.
In fact, a generic one-loop diagram with $n$ external legs can be written as
\begin{align}
\label{eq:one_loop_integral}
    \ii I =
	\begin{gathered}
	    \vcenter{\hbox{\includegraphics[page=2,scale=1]{img/diagrams.pdf}}}
	\end{gathered}
    =
	\int 
	\frac{\dd^4 k}{(2\pi)^4}
	\mathcal{I},
	\quad
	\mathcal{I}
	= \frac{N}{\prod_{i=1}^n D_i},
	\quad
	D_i = q_i^2 - m_i^2 +\ii\epsilon,
\end{align}
where $N$ is a polynomial in the loop energy $k^0$ and regular in $\vec{k}$.
The Feynman propagator $1/D_i$ depends on the four-momentum $q_i = k+p_i$, with $p_i$ a linear combination of external momenta and an arbitrary constant shift, the mass $m_i$ and the causal prescription $\epsilon>0$.
The limit $\epsilon\to0$ is left implicit.
The results derived in this section hold for pairwise distinct propagators but can easily be extended for propagators raised to higher powers than one.
\par

In general, the one-loop integral $\ii I$ is plagued by infrared and ultraviolet divergences.
They have to be regulated to make sense of the integral.
However, these divergences do only occur in the integration over the \emph{spatial} momentum $\vec{k}$, which is why the energy integration over $k^0$ can be performed without having to worry about infrared or ultraviolet divergences yet.
This is what we will do now.
\par

\subsection{Loop-Tree Duality}
\label{sec:one_loop_ltd}

After integrating out the loop energy of a Feynman integral one arrives at what is referred to as the Loop-Tree Duality (LTD) expression.
Inspired by the Feynman Tree Theorem \cite{1972mwm..book..355F,Brandhuber_2006,kilian2009numerical,Bierenbaum_2010}, the LTD expression was originally presented at one-loop in \cite{Catani_2008} and later generalised to $n$-loops in \cite{Capatti_2019,Runkel_2019, Aguilera_Verdugo_2020}.
In principle however, such a representation of loop integrals goes back to Time-Ordered Perturbation Theory (TOPT) \cite{Sterman:1993hfp,Bourjaily_2021}.
An explicit connection between LTD and TOPT was recently made in \cite{capatti2020manifestly}.
\par

For the purpose of this work, only the one-loop LTD expression is needed.
Although it is well established and may be stated here without comment, we believe it is instructive to present an alternative derivation with the method we outlined in the previous section.
This amounts to applying the result in eq.~\eqref{eq:multi_dim_integral} to the $k^0$-integration.
\par

\subsubsection{Derivation}
\label{sec:ltd_derivation}
As a first step, we notice that the denominator $D_i$ in eq.~\eqref{eq:one_loop_integral} is a polynomial in the integration variable $k^0$ and factorises as
\begin{align}
    D_i =
    \left(k^0+p_i^0-E_i\right)
    \left(k^0+p_i^0+E_i\right),
\end{align}
where $E_i = \sqrt{\vec{q}_i^2+m_i^2-\ii\delta}$ is the positive on-shell energy.
This implies that the integrand in eq.~\eqref{eq:one_loop_integral}
has $2n$ poles at $x_{i_\sigma}^\epsilon \coloneqq \sigma E_i-p_i^0$ for $i\in\{1,\dots,n\}$ and $\sigma\in\{+1,-1\}$ that approach the real line for $\epsilon\to 0$.
According to the result in eq.~\eqref{eq:multi_dim_integral} the integral can then be expressed in terms of the residues
\begin{align}
\label{eq:dual_integrand}
    R_{i_{\sigma}}(\vec{k})
	=
    \Res[\mathcal{I}(k^0),k^0=\sigma E_i-p_i^0]
    = 
    \frac{\sigma}{2E_i}
    \frac{N\vert_{i_\sigma}}{\prod_{j\neq i}D_j\vert_{i_\sigma}},
\end{align}
in this context commonly referred to as \emph{dual integrands}, where we use the shorthand notation $f\vert_{i_\sigma}$ denoting the evaluation of a function 
$f(k)$ at the on-shell loop momentum $k=(\sigma E_i-p_i^0,\vec{k})$.
Note that the residue $R_{i_\sigma}$ is a function of the spatial loop momentum $\vec{k}$ alone.
In terms of cutting rules, where one replaces $D_i=(q_i^2-m_i^2+\ii\epsilon)^{-1}$ with $-(2\pi\ii)\delta(q_i^2-m_i^2)\Theta(\sigma q_i^0)$ the dual integrand can be identified with the cut diagram
\begin{align}
    \!\!\!\!\!
    \begin{gathered}
	    \vcenter{\hbox{\includegraphics[page=3,scale=1]{img/diagrams.pdf}}}
	\end{gathered}
	=
    -\ii
    \int \frac{\dd^3 \vec{k}}{(2\pi)^3}
    \sigma R_{i_{\sigma}}
    \approx
	\int \frac{\dd^4 k}{(2\pi)^4}
	(-2\pi\ii)\delta(q_i^2-m_i^2)\Theta(\sigma q_i^0)
    \frac{N}{\prod_{j\neq i} D_j}
	,
\end{align}
where $\approx$ stands for equality up to $\order(\epsilon)$ in the causal prescription $\epsilon$.
This originates from the fact that in contrast to the residue in eq.~\eqref{eq:dual_integrand} a naive replacement of a propagator with a Dirac delta function does not propagate the causal prescription correctly\footnote{
\label{foot:causal_prescription}
To clarify compare
$\Res[\frac{f(x)}{x-(x^*+\ii \epsilon)},x=x^*]=f(x^*+\ii \epsilon)$
and
$\int_\mathbb{R} \dd x\, \delta(x-x^*)f(x)=f(x^*)$, where $f$ is analytic and $x^*\in \mathbb{R}$, which differ at order $\order(\epsilon)$.
}.
Furthermore, following sect.~\ref{sec:resolving_causal_prescription}, one then constructs the subtracted integrand.
In our case, where the integrand is a rational function, the counterterms in eq.~\eqref{eq:counterterm} can be chosen such that the subtracted integrand vanishes exactly with partial fraction decomposition.
Therefore, the sum of fractional residues alone directly yields the one-loop LTD expression
\begin{align}
\label{eq:one_loop_ltd_averaged}
    \ii I
    =
    -\ii 
    \int \frac{\dd^3 \vec{k}}{(2\pi)^3}
    \mathcal{I}_{\rm LTD}
    =
    -\frac{\ii}{2}
    \int \frac{\dd^3 \vec{k}}{(2\pi)^3}
    \sum_{\sigma\in\{\pm1\}}
    \sum_{i=1}^n
    \sigma R_{i_{\sigma}},
\end{align}
where we applied eq.~\eqref{eq:multi_dim_integral} and $\sgn \Im(\sigma E_i-p_i^0)=-\sigma$.
\par

Readers familiar with the traditional derivation of the LTD expression \cite{Catani_2008}, where one directly applies residue theorem closing the contour on either the lower or the upper complex half-plane, may recognise this expression as the average over both contour closures.
The expression in eq.~\eqref{eq:one_loop_ltd_averaged} therefore features twice as many summands as necessary.
This choice of contour closure, reflecting time-reversal symmetry, implies that $\mathcal{I}_{\rm LTD} = \mathcal{I}_{\rm LTD}^+ \equiv \mathcal{I}_{\rm LTD}^{-} \equiv \frac{1}{2}(\mathcal{I}_{\rm LTD}^++\mathcal{I}_{\rm LTD}^-)$, where $\mathcal{I}_{\rm LTD}^\sigma \coloneqq \sum_{i=1}^n \sigma R_{i_\sigma}$ are all locally identical.
Diagrammatically, we can write
\begin{align}
	\begin{gathered}
	    \vcenter{\hbox{\includegraphics[page=2,scale=1]{img/diagrams.pdf}}}
	\end{gathered}
	=
	\sum_{i=1}^n
    \begin{gathered}
	    \vcenter{\hbox{\includegraphics[page=3,scale=1]{img/diagrams.pdf}}}
	\end{gathered}
\end{align}
for $\sigma$ either $+1$ or $-1$.
\par

\subsubsection{Higher-order poles: Threshold singularities}
\label{sec:threshold_singularities}

According to the discussion in sect.~\ref{sec:degenerate_case}, the appearance of higher-order poles in the integrand in eq.~\eqref{eq:one_loop_integral} can lead to local (mis-)cancellations in the LTD expression in eq.~\eqref{eq:one_loop_ltd_averaged}.
These higher-order poles appear locally for specific configurations in the spatial loop momentum $\vec{k}$ but also as raised propagators.
Let us explicitly study second order poles and apply the terminology of sect.~\ref{sec:degenerate_case}.
Poles of order higher than two can then be studied analogously.
\par

The single poles of the one-loop integrand of eq.~\eqref{eq:one_loop_integral} are given by $x_{i_\sigma}^\epsilon \coloneqq \sigma E_i-p_i^0$.
It has a double pole exactly when $x_{i_\sigma}^0=x_{j_\sigma}^0$, i.e. when two propagators go on-shell simultaneously, satisfying
\begin{align}
    \label{eq:singular_surface}
    \sigma E_i - p_i^0 = \rho E_j - p_j^0
\end{align}
for $\epsilon\to 0$ and real solutions of the spatial loop momenta $\vec{k}$ (see existence conditions in app.~\ref{app:dual_propagator}).
\par
According to eqs.~\eqref{eq:pinched_definition} and \eqref{eq:non_pinched_definition} and using that $\Im E_i<0$ and $\Im E_j<0$, one then finds that this double pole pinches the $k^0$-integration contour if $\sigma=-\rho$.
The corresponding singular surface we call \emph{causal threshold} (as in \cite{aguileraverdugo2019causality}) or \emph{E-surface} (as in \cite{Capatti_2019})
\begin{align}
    E_{ij}^\sigma \coloneqq
    \sigma (E_i+E_j) -p_i^0 + p_j^0
    = 0.
\end{align}
It does \emph{not} locally cancel in the sum of dual integrands and defines a single pole in the integration domain of the LTD expression in eq.~\eqref{eq:one_loop_ltd_averaged}.
Therefore, one \emph{cannot} remove the causal prescription $\epsilon>0$, which guarantees that the remaining spatial integrals stay well defined.
\par
On the other hand, the double pole is non-pinched if $\sigma=\rho$.
The corresponding singular surface we call \emph{non-causal threshold} (as in \cite{aguileraverdugo2019causality}) or \emph{H-surface} (as in \cite{Capatti_2019})
\begin{align}
    H_{ij}^\sigma \coloneqq
    \sigma(E_i-E_j)-p_i^0+p_j^0
    = 0.
\end{align}
It locally cancels between the dual integrands $R_{i_\sigma}$ and $R_{j_\rho}$ and therefore defines a \emph{spurious} single pole.
This implies that the regulator $\epsilon>0$ can be removed (i.e. set to zero) without consequences.
\par

This cancellation mechanism of H-surfaces between dual integrands is usually referred to as \emph{dual cancellation} \cite{Capatti_2019,Catani_2008,kilian2009numerical}.
Since H-surfaces are the non-pinched double poles of the loop integrand, it is a priori clear why these cancellations are realised.
\par

In this context, it is interesting to see that the \emph{dual propagator} $1/D_{i_\sigma}$ factorises exactly into one H-surface and one
%
%
E-surface\footnote{
We may refer to the equation $E_{ij}^\sigma=0$, its solutions, as well as to $E_{ij}^\sigma$ as E-surface (and analogous for H-surfaces).
}, i.e.
\begin{align}
    D_j\vert_{i_\sigma}
    = H_{ij}^\sigma E_{ij}^\sigma.
\end{align}
A more detailed geometrical study of the dual propagator is given in app.~\ref{app:dual_propagator}.
\par

Furthermore, the H-surface has the symmetry $H_{ij}^\sigma=-H_{ji}^\sigma$, in accordance with its pairwise appearance in both in the dual integrand $R_{i_\sigma}$ and in $R_{j_\sigma}$. 
Similarly, an E-surface satisfies $E_{ij}^\sigma=-E_{ji}^{-\sigma}$.
This implies that the E-surface also appears pairwise, in $R_{i_\sigma}$ and $R_{j_{-\sigma}}$.
However, in contrast to the H-surface, only one of the two residues contributes to the LTD expression (or both, in the average over both contour closures, but with opposite winding numbers).
\par

Moreover, note that the E-surface defines the kinematic regime, where two particles, one incoming, the other outgoing, are on-shell.
More precisely, we lie on the E-surface $E_{ij}^\sigma$ if the propagator momentum $q_i$ is on its mass shell with \emph{positive} energy ($q_i^0>0$) and $q_j$ is on-shell with \emph{negative} energy ($q_j^0<0$).
The E-surface therefore defines exactly the two-body phase space.
\par

In the next section, we attempt to resolve the causal prescription $\epsilon$ for the surviving E-surface poles by tackling the three dimensional integration over the spatial momenta $\vec{k}$.
More specifically, we will integrate out one more dimension.

\subsection{Integrating out a spatial dimension}
\label{sec:integrating_magnitude}
Analogous to the previous section, where we integrated out the loop energy $k^0$, we now proceed with the integral over a spatial dimension of the loop momentum, again following the general strategy in sect.~\ref{sec:resolving_causal_prescription}.
Recall that the $k^0$-integrand is a rational function in the integration variable, which simplified the calculation drastically.
For the $\vec{k}$-integrand (i.e. the LTD integrand), however, this is not the case anymore.
This is why we now elaborate the application of the result in eq.~\eqref{eq:multi_dim_integral} more carefully.
In the next paragraph, we first set the basis and sketch the procedure.
\par

As we have seen in the previous sect.~\ref{sec:one_loop_ltd}, the LTD expression can still feature poles on E-surfaces that render the integral ill-defined if not regulated with the causal prescription.
To remove these singularities following the method described in sect.~\ref{sec:resolving_causal_prescription}, we have to cast the remaining integrations in the LTD integral in eq.~\eqref{eq:one_loop_ltd_averaged} into a form compatible with eq.~\eqref{eq:multi_dim_integral_start}.
More precisely, this requires us to identify an integration dimension that is the whole real line.
One might therefore choose Cartesian coordinates.
However, we choose a spherical coordinate system. This allows to avoid the appearance of locally pinched higher-order poles (see sect.~\ref{sec:multidimensional_integrals}) in many cases and to realise more local cancellations among residues and counterterms.
\par

We will use slightly modified spherical coordinates, such that the radius can be negative with angles spanning a single hemisphere only\footnote{
Equivalently, one can use the usual spherical coordinates, which however obscures a symmetry of the final expressions.
}.
More concretely, we parameterise the spatial loop momentum as
\begin{align}
	\vec{k} = r \hat k = r \left(\sin\theta \cos\phi, \sin\theta\sin\phi,\cos\theta \right)
\end{align}
where $r\in\mathbb{R}$ is a real number 
%
%
and $\hat k\in H^2$ lies in the unit hemisphere, i.e. $\cos\theta \in [-1,1]$, $\phi \in [0,\pi]$.
In these coordinates the one-loop integral in eq.~\eqref{eq:one_loop_ltd_averaged} reads
\begin{align}
\label{eq:ltd_integral_starting}
	\ii I = -\ii 
	\int_{H^2} \frac{\dd^2\hat k}{(2\pi)^3}
	\int_{\mathbb{R}} \dd r r^2 \mathcal{I}_{\rm LTD}(r \hat{k}).
\end{align}
As a next step, one has to check if the integrand $r^2\mathcal{I}_{\rm LTD}(r\hat{k})$ satisfies the assumptions for the starting expression eq.~\eqref{eq:multi_dim_integral_start}.
This includes the identification of the poles in $r$ for $\epsilon\to0$.
From sect.~\ref{sec:threshold_singularities}, we already know that these poles are given by the E-surfaces of the LTD integrand $\mathcal{I}_{\rm LTD}$.
Given the poles in $r$, we can then write the integral $\ii I$ in eq.~\eqref{eq:ltd_integral_starting} in the form of eq.~\eqref{eq:multi_dim_integral} in terms of their residues and respective counterterms.

\subsubsection{Parameterisation of E-surfaces}
\label{sec:poles_in_r}

The poles of the one-loop LTD integrand are given by the E-surfaces of sect.~\ref{sec:threshold_singularities}.
They are defined by $E_{ij}^\sigma \coloneqq \sigma(E_i+ E_j) - p_i^0+ p_j^0=0$ for $\epsilon\to0$, where $E_l=\sqrt{(\vec{k}+\vec{p}_l)^2+m_l^2-\ii \epsilon}$, $l\in\{i,j\}$.
The E-surface $E_{ij}^\sigma$ only has real solutions in $\vec{k}$ if\footnote{
If $p_{ij}^2=(m_i+m_j)^2>0$ the E-surface collapses into a single point, an integrable singularity.
If however $p_{ij}^2=(m_i+m_j)^2=0$, the E-surface squeezes into the line segment between $-\vec{p}_i$ and $-\vec{p}_j$, whose endpoints correspond to two soft singularities connected by a collinear singularity of an infrared divergent integral (cf. app.~\ref{app:dual_propagator}).
We assume pinched (squeezed) E-surfaces to be absent in our loop integrand.
}
\begin{align}
\label{eq:existence_condition_e_surface}
    p_{ij}^2>(m_i+m_j)^2, \quad \sigma p_{ij}^0 >0,
\end{align}
for $\epsilon\to 0$, real parameters $p_l$, $m_l$, $l\in\{i,j\}$ and $p_{ij}= p_i-p_j$ (cf. app.~\ref{app:dual_propagator} or \cite{kilian2009numerical, buchta2015numerical, Capatti_2019}).
\par

To find the local\footnote{
It is crucial that all poles are parameterised in the same variable $r$.
A separate parameterisation of each E-surface $E_{ij}^\sigma$, as e.g. of the usual two-body phase space in the rest frame of $p_{ij}$ will in general break local cancellations of E-surface intersections.
} parameterisation $r(\hat k)$ of the E-surface in the integration space, we apply a few tricks.
First, we note that if the E-surface $E_{ij}^\sigma$ exists, the H-surface $H_{ij}^\sigma$ does not.
Instead of the E-surface directly, we can therefore study the zeros of the dual propagator $D_j\vert_{i_\sigma} = H_{ij}^\sigma E_{ij}^\sigma$ in the regime, where the existence conditions eq.~\eqref{eq:existence_condition_e_surface} is satisfied.
Furthermore, we exploit the invariance of the existence condition as well as the dual propagator under orthochronous Lorentz boosts
\begin{align}
	\Lambda_{\vec{\beta}} = 
	\begin{pmatrix}
		\gamma & -\gamma \vec{\beta}^T \\
	-\gamma \vec{\beta} \phantom{~~} & \mathbb{I}_3 + (\gamma-1) \vec{\beta} \cdot \vec{\beta}^T \\
	\end{pmatrix},
	\qquad
	\gamma=\frac{1}{\sqrt{1-\vec{\beta}^2}},
\end{align}
to first find a parameterisation in a simpler Lorentz frame.
\par

\begin{figure}
\begin{subfigure}[t]{.49\textwidth}
    \centering
    \includegraphics[scale=0.7]{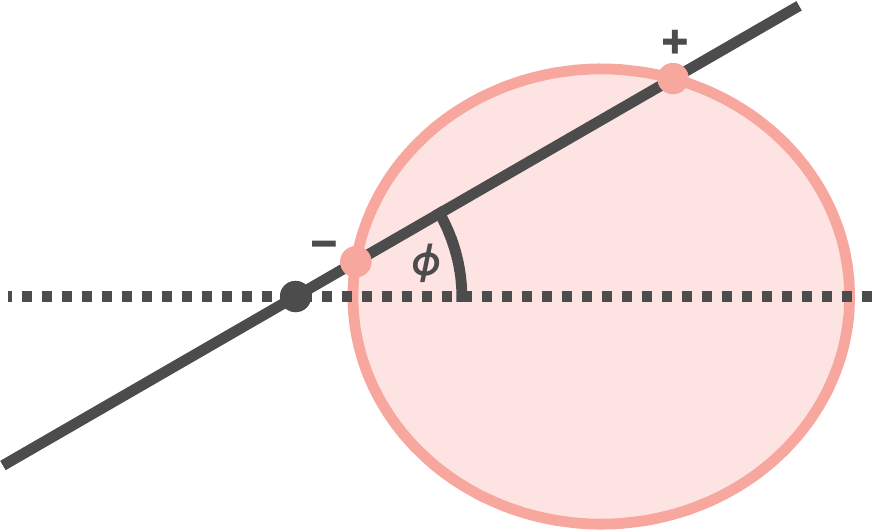}
    \caption{
    Two intersections at $\vec{k}=r_{i_\sigma j_{-\sigma}}^{\pm}(\hat k) \hat k$ in some reference frame $K$.
    Along the direction $\hat k(\phi)$, both solutions are positive, $r_{i_\sigma j_{-\sigma}}^{+}>r_{i_\sigma j_{-\sigma}}^{-}>0$.
    }
    \label{subfig:ellipse}
\end{subfigure}\hfill
\begin{subfigure}[t]{.49\textwidth}
    \centering
    \includegraphics[scale=0.7]{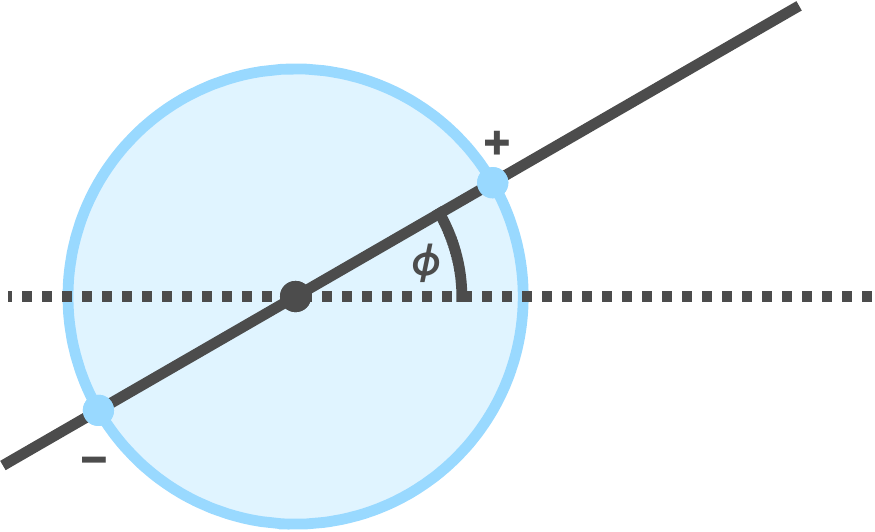}
    \caption{
    The intersections at $\vec{k}
    =\pm|\vec{q}_1^{\, \prime}|\hat k$ (see eq.~\eqref{eq:q_rad_on_shell}) in the rest frame $K^\prime$ of $p_{ij}$.
    Along the direction $\hat k(\phi)$, we find a positive and a negative solution $r_{i_\sigma j_{-\sigma}}^{\pm}=\pm|\vec{q}_1^{\, \prime}|$. 
    }
    \label{subfig:circle}
\end{subfigure}
\caption{
\label{fig:surface_ib_os}
An E-surface $E_{ij}^\sigma$ 
in two different reference frames $K$ and $K^\prime$ in the spatial loop momentum space of (\subref{subfig:ellipse}) $\vec{k}=\vec{q}_i$ and (\subref{subfig:circle}) $\vec{k}=\vec{q}_i^{\,\prime}$, where $K^\prime$ is the rest frame of $p_{ij}$.
Note that in general, the origin $\vec{q}_i=\vec{0}$ does not lie within the ellipse.
However, a Lorentz boost transforms the ellipse into a circle with center at $\vec{q}_i^{\, \prime}=\vec{q}_j^{\, \prime}=\vec{0}$.
As a consequence, the general solutions $r_{i_\sigma j_{-\sigma}}^\pm(\hat k)$ may both be positive or may not even be real for all directions $\hat k$.
}
\end{figure}

Since the momentum $p_{ij}$ is timelike, we can evaluate the dual propagator $D_j\vert_{i_\sigma}$ in its rest frame, where
%
%
\begin{align}
    D_j\vert_{i_\sigma}
    = q_j^2\vert_{i_\sigma}-m_j^2+\ii\epsilon
    = (q_i^\prime\vert_{i_\sigma} - p_{ij}^\prime)^2 -m_j^2+\ii\epsilon
    = m_i^2 + p_{ij}^2 - 2E_i^\prime \sqrt{p_{ij}^2} - m_j^2,
\end{align}
with $E_i^\prime=\sqrt{|\vec{q}_i^{\,\prime}|^2+m_i^2-\ii\epsilon}$ and $p_{ij}^\prime=\Lambda_{\vec{\beta}_{ij}} p_{ij}=(\sigma \sqrt{p_{ij}^2},\vec{0})$, where $\vec{\beta}_{ij}=\vec{p}_{ij}/p_{ij}^0$.
On the E-surface, where $D_j\vert_{i_\sigma}=0$, we therefore have that
\begin{align}
\label{eq:q_rad_on_shell}
    |\vec{q}_i^{\,\prime}|=\sqrt{\frac{\lambda\left(p_{ij}^2,m_j^2,m_i^2\right)}{4 p_{ij}^2}+ \ii\epsilon},
\end{align}
which means that the E-surface in $\vec{q}_i^{\,\prime}$-space is a sphere (see fig.~\ref{fig:surface_ib_os}).
\par

As a next step, we go back to the loop momentum frame by expressing the energy component of $q_i^\prime = \Lambda_{\vec{\beta}_{ij}} (k+p_i)$ with both on-shell conditions $k^0 + p_i^0=\sigma\sqrt{(\vec{k}+{p}_i)+m_i^2-\ii\epsilon}$ and eq.~\eqref{eq:q_rad_on_shell}, such that
\begin{align}
\label{eq:lorentz_trsf_energy}
    \sigma E_i^{\prime}
    =
    \gamma_{ij} \left( \sqrt{(\vec{k}+\vec{p}_i)^2+m_i^2-\ii\delta}-\vec{\beta}_{ij}\cdot (\vec{k}+\vec{p}_i)\right),
\end{align}
where $\gamma_{ij}=1/\sqrt{1-\vec{\beta}_{ij}^2}$.
After appropriate squaring of eq.~\eqref{eq:lorentz_trsf_energy} we find a quadratic equation in $r$ with coefficients
\begin{align}
    2a_{ij}(\hat k)
    &= 
    1-\left(\beta_{ij}\cdot\hat k\right)^2
    , \\
    b_{ij}^\sigma(\hat k)
    &= \hat k \cdot
    \left(\vec{p}_i
    - \vec{\beta}_{ij} \left(
        \frac{\sigma E_i^\prime}{\gamma_{ij}}
        + \vec{\beta}_{ij}\cdot \vec{p}_i
    \right)
    \right), \\
    2c_{ij}^\sigma
    &= |\vec{p}_i|^2+m_i^2-\ii\epsilon-\left(
        \frac{\sigma E_i^\prime}{\gamma_{ij}}
        + \vec{\beta}_{ij}\cdot \vec{p}_i
    \right)^2,
\end{align}
resulting in the E-surface parameterisation
\begin{align}
\label{eq:r_pole}
    r_{i_\sigma j_{-\sigma}}^{\pm}(\hat k)
    &\coloneqq
    \frac{-b_{ij}^\sigma(\hat k)\pm\sqrt{\Delta_{ij}^\sigma(\hat k)}}{2a_{ij}(\hat k)},
    \qquad
    \Delta_{ij}^\sigma(\hat k)
    \coloneqq
    b_{ij}^\sigma(\hat k)^2 - 4a_{ij}(\hat k)c_{ij}^\sigma,
\end{align}
where we require that the discriminant (for vanishing $\epsilon$) is positive, $\Delta_{ij}^\sigma(\hat k)\geq0$.
If the E-surface $E_{ij}^\sigma$ satisfies the existence condition in eq.~\eqref{eq:existence_condition_e_surface}, we are guaranteed to find a direction $\hat k\in H^2$ in the unit hemisphere for which the discriminant is positive.
\par

Geometrically speaking, the points $r=r_{i_\sigma j_{-\sigma}}^{\pm}(\hat k)$ are the intersections of the E-surface\footnote{
The line is a tangent of the E-surface if $\Delta_{ij}^\sigma(\hat k)=0$ or a secant $\Delta_{ij}^\sigma(\hat k)>0$.
} with the line $\vec{k}=r \hat k$ for fixed $\hat k$ in the spatial loop momentum space.
Moreover, if the origin $\vec{0}$ lies within the E-surface, the discriminant is positive for all directions $\hat k\in H^2$ and yields exactly one positive $r_{i_\sigma j_{-\sigma}}^{+}(\hat k)$ and one negative solution $r_{i_\sigma j_{-\sigma}}^{-}(\hat k)$ (see fig.~\ref{fig:surface_ib_os}).
\par

Last but not least, we note that the pole $r_{i_\sigma j_{-\sigma}}^\rho(\hat k)$ lies in the upper ($\rho=+1$) or lower ($\rho=-1$) complex half-plane and approaches the real line for $\epsilon\to0$, satisfying
\begin{align}
\label{eq:sign_r_pole}
    \lim_{\epsilon\to 0} \sgn \Im r_{i_\sigma j_{-\sigma}}^{\rho}(\hat k) = \rho.
\end{align}
Other useful identities are $r_{i_\sigma j_{-\sigma}}^{\pm}(\hat k) = -r_{i_\sigma j_{-\sigma}}^{\mp}(-\hat k)$,
which reflects the spherical nature of our parameterisation, and
\begin{align}
\label{eq:r_symmetry}
    r_{i_\sigma j_{-\sigma}}^{\pm}(\hat k) =
    r_{j_{-\sigma} i_{\sigma}}^{\pm}(\hat k),
\end{align}
which follows from the symmetry $E_{ij}^\sigma = -E_{ji}^{-\sigma}$.

\subsubsection{E-surface residues}
\label{sec:e_surface_residues}

The residue at the pole $r=r_{i_\sigma j_{-\sigma}}^{\rho}$ of the LTD expression in eq.~\eqref{eq:ltd_integral_starting} can directly be computed from the dual integrand $R_{i_\sigma}$\footnote{
Or equivalently from $R_{j_{-\sigma}}$. Eq.~\eqref{eq:r_symmetry} then implies that $R_{i_\sigma j_{-\sigma}}^\rho=R_{j_{-\sigma} i_\sigma}^\rho$.
The residues also satisfy $R_{i_\sigma j_{-\sigma}}^\pm(\hat k)=-R_{i_\sigma j_{-\sigma}}^\mp(-\hat k)$.
}:
We simply replace the respective E-surface $E_{ij}^\sigma$
by its derivative with respect to $r$ and evaluate the integrand at $r=r_{i_\sigma j_{-\sigma}}^\rho$, i.e.
\begin{align}
\label{eq:e_surface_residue}
	R_{i_\sigma j_{-\sigma}}^{\rho}(\hat k)
	&= \Res[r^2\mathcal{I}_{\rm LTD}(r \hat k), r= {r_{i_\sigma j_{-\sigma}}^{\rho}(\hat k)}]
	= \Res[r^2 R_{i_\sigma}(r \hat k), r= {r_{i_\sigma j_{-\sigma}}^{\rho}(\hat k)}] \\
	&=
	\Theta\left(r_{i_\sigma j_{-\sigma}}^{\rho}\in\mathbb{R}\right)
	\frac{1}{-4E_iE_j}
	\frac{r^2}{
	\left(
	    \frac{\vec{q}_i}{E_i} + \frac{\vec{q}_j}{E_j}
	\right) \cdot \hat k}
	\Bigg\vert_{r= r_{i_\sigma j_{-\sigma}}^{\rho}}
	\left.
	\frac{N}{\prod\limits_{l\neq i,j}D_l}
	\right\vert_{
	   \substack{\rho \ \ \quad \\ i_\sigma j_{-\sigma}}},
\end{align}
where the shorthand notation $f\vert_{i_\sigma j_{-\sigma}}^\rho$ denotes the evaluation of $f(k)$
at the loop momentum
\begin{align}
    k(\hat k) =
    \begin{pmatrix}
    \sigma\sqrt{
        \left(r_{i_\sigma j_{-\sigma}}^\rho(\hat k) \hat k+\vec{p}_i\right)^2
        + m_i^2 -\ii\epsilon}
        -p_i^0
    \\
    r_{i_\sigma j_{-\sigma}}^\rho(\hat k) \hat k
    \end{pmatrix}.
\end{align}
The Heaviside function appears since the residue is non-zero only if the E-surface has a real solution in $r$ (i.e. if $\Delta(\hat k)\geq 0$ in eq.~\eqref{eq:r_pole}).
\par

In terms of cutting rules, where one replaces $D_i=(q_i^2-m_i^2+\ii\epsilon)^{-1}$ with $(-2\pi\ii)\delta(q_i^2-m_i^2)\Theta(\sigma q_i^0)$ and $D_j=(q_j^2-m_j^2+\ii\epsilon)^{-1}$ with $(-2\pi\ii)\delta(q_j^2-m_j^2)\Theta(-\sigma q_j^0)$ the E-surface residue can be identified with the double cut diagram\footnote{
With $\approx$ we mean equality up to terms $\order(\epsilon)$ (cf. footnote~\ref{foot:causal_prescription}).
%
}
\begin{align}
\label{eq:e_surface_residue_diagram}
    \begin{gathered}
	    \vcenter{\hbox{\includegraphics[page=4,scale=1]{img/diagrams.pdf}}}
	\end{gathered}
	&=
	\int_{H^2} \frac{\dd^2\hat k}{(2\pi)^2}
    \sum_{\rho\in\{\pm1\}}
    \rho
    R_{i_\sigma j_{-\sigma}}^{\rho}
    \approx
	-\int \dd \Pi_2
	\frac{N}{\prod_{l\neq i,j} D_l},
\end{align}
where $\dd \Pi_2$ is the Lorentz invariant phase space measure\footnote{
It satisfies
\begin{align}
    \dd \Pi_2
    &\coloneqq
    -\frac{\dd^4 k}{(2\pi)^4}
    (-2\pi\ii)\delta(q_i^2-m_i^2)\Theta(\sigma q_i^0)
    (-2\pi\ii)\delta(q_j^2-m_j^2)\Theta(-\sigma q_j^0)
    =
    \frac{\dd^3 \vec{k}}{(2\pi)^3}
	\frac{1}{2E_i} \frac{1}{2E_j}
	(2\pi)
	\delta(E_{ij}^\sigma)
\end{align}
}
for the two particles with momenta $q_i$ and $q_j$ with opposite energy flow, i.e. $q_i^0=\sigma E_i$ and $q_j^0=-\sigma E_j$.
\par

We now consider the residue prefactor in eq.~\eqref{eq:e_surface_residue} in more detail.
First, note that the vector 
\begin{align}
    \vec{n}_{ij}^\rho \coloneqq
    \left.\left(
        \frac{\vec{q}_i}{E_i} + \frac{\vec{q}_j}{E_j}
    \right)\right\vert_{r= {r_{i_\sigma j_{-\sigma}}^\rho}}
\end{align}
is the outward pointing normal of E-surface $E_{ij}^\sigma$.
From geometrical considerations, and since $r_{i_\sigma j_{-\sigma}}^+ \geq r_{i_\sigma j_{-\sigma}}^-$, we can infer that $ (\vec{n}_{ij}^\rho\cdot \hat k)\rho>0$.
This implies that the two residues for different $\rho$ have opposite signs.
\par

Furthermore, we observe that the residue has integrable singularities whenever $\vec{n}_{ij}^\rho \perp \hat{k}$, i.e. at the tangents of the E-surface $E_{ij}^\sigma$.
By placing the origin of the three-dimensional integration space in the interior of the E-surface, we are guaranteed that $\vec{n}_{ij}^\rho \cdot \hat{k}$ cannot vanish.
If this can be achieved for all E-surfaces of the LTD integrand depends on the external momenta.
\par


\subsubsection{E-surface counterterms}
\label{sec:counterterms}
We can now subtract the E-surface poles from the LTD integrand with counterterms constructed as in eq.~\eqref{eq:counterterm} with a Gaussian suppression function
\begin{align}
\label{eq:ltd_counterterm}
	\CT_{i_\sigma j_{-\sigma}}^{\rho}(r,\hat k) = \frac{ R_{i_\sigma j_{-\sigma}}^{\rho}(\hat k)}{r-r_{i_\sigma j_{-\sigma}}^{\rho}(\hat k)} \exp\left(-\frac{(r-r_{i_\sigma j_{-\sigma}}^{\rho}(\hat k))^2}{(a_{ij}E_\mathrm{cm})^2}\right),
\end{align}
where $a_{ij}$ is a dimensionless coefficient and $E_{\rm cm}$ is the center of mass energy defined by the external momenta of the Feynman diagram.
\par

Note that we have the symmetry $\CT_{i_\sigma j_{-\sigma}}^\rho(-r,-\hat k)=\CT_{i_\sigma j_{-\sigma}}^{-\rho}(r,\hat k)$, which implies that the sum $\dd^3\vec{k} \CT_{i_\sigma j_{-\sigma}}(\vec{k})\coloneqq \dd r \dd^2 \hat k \sum_{\rho\in\{\pm1\}} \CT^\rho_{i_\sigma j_{-\sigma}}(r,\hat k)$ with $\vec{k}=r\hat k=(-r)(-\hat k)$, indeed behaves like a function in Euclidean space.
\par

\begin{figure}
\begin{subfigure}[t]{.49\textwidth}
    \centering
    \includegraphics[scale=0.7]{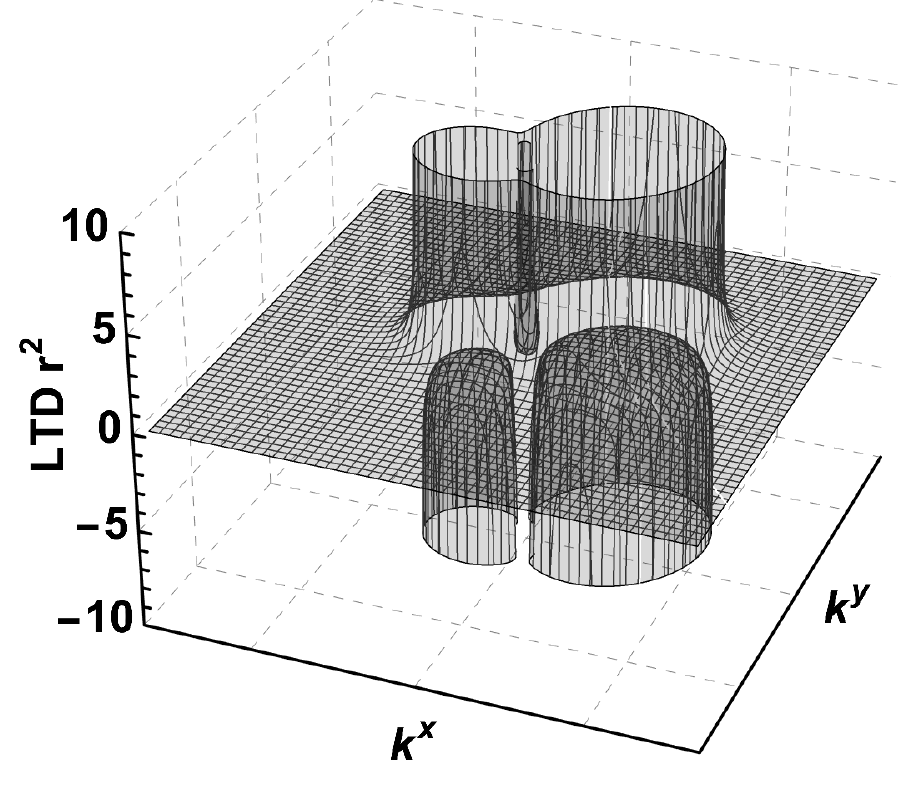}
    \caption{LTD integrand with intersecting threshold singularities.}
    \label{subfig:ltd}
\end{subfigure}\hfill
\begin{subfigure}[t]{.49\textwidth}
    \centering
    \includegraphics[scale=0.7]{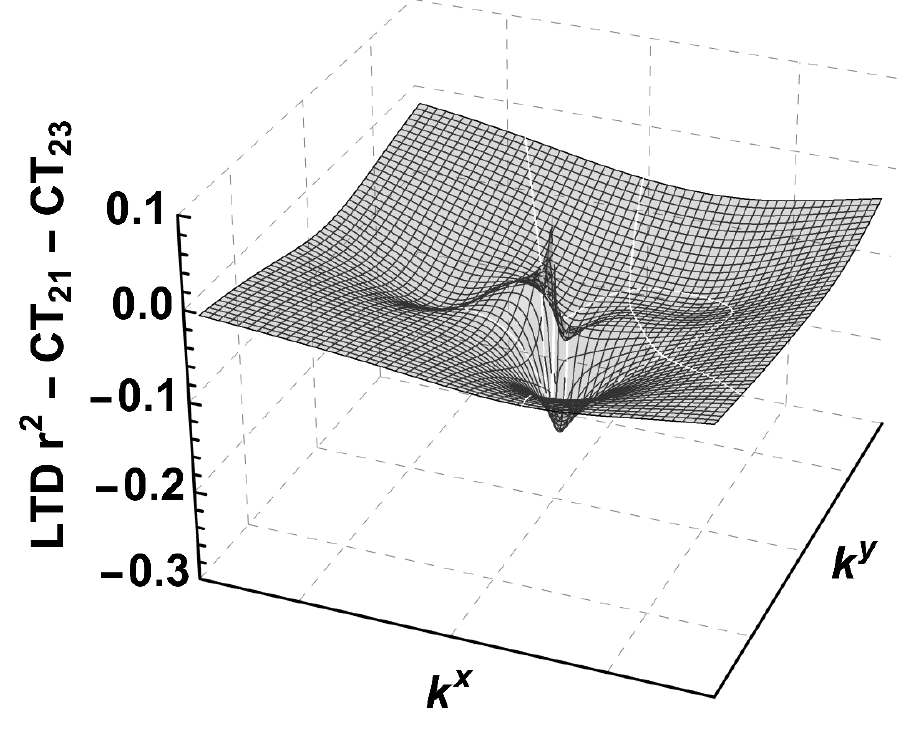}
    \caption{Threshold subtracted integrand with counterterms $\CT_{ij}\coloneqq\CT_{i_\sigma j_{-\sigma}}^++\CT_{i_\sigma j_{-\sigma}}^-$.}
    \label{subfig:subtracted_ltd}
\end{subfigure}
\caption{\label{fig:subtracted_integrand}
The LTD integrand of the scalar triangle integral in two dimensional momentum space featuring two intersecting E-surfaces in (\subref{subfig:ltd}) and the corresponding subtracted integrand according to eq.~\eqref{eq:subtracted_ltd} in (\subref{subfig:subtracted_ltd}).
The subtracted integrand is free of poles and can be directly evaluated with Monte~Carlo numerical integration.
The fine elliptic and hyperbolic lines are artefacts of numerical instabilities in the local cancellations between the LTD integrand and the E-surface counterterms and in the dual cancellations of H-surfaces.
}
\end{figure}

\subsubsection{Local optical theorem}
\label{sec:local_unitarity}

We have come far in our endeavour of following sect.~\ref{sec:resolving_causal_prescription} and integrating out another spatial dimension from the LTD integral in eq.~\eqref{eq:ltd_integral_starting}:
We identified the poles of the integrand in eq.~\eqref{eq:r_pole}, computed their residues in eq.~\eqref{eq:e_surface_residue} and constructed their counterterms in eq.~\eqref{eq:ltd_counterterm}.
Finally, following eq.~\eqref{eq:multi_dim_integral}, we put everything together.
\par

We can further simplify the expression eq.~\eqref{eq:multi_dim_integral} in the case, where no locally pinched poles (with respect to the integration contour of $r$) are present in the LTD integrand of eq.~\eqref{eq:ltd_integral_starting}.
In this case, all intersections of E-surfaces and their corresponding poles in residues and counterterms will locally cancel amongst themselves, such that the sum of fractional residues is free of poles in the integration domain.
The causal prescription can then be removed and the integral can be explicitly split up into real and imaginary part\footnote{
We use the slightly confusing identities $\Re I=\Im \ii I$ and $\Im \ii (\ii I)=-\Im I = \Re \ii I$.
}
\begin{align}
\label{eq:subtracted_ltd}
	\Re I
	&= -\int_{H^2} \frac{\dd^2\hat k}{(2\pi)^3}
	\int_\mathbb{R}\dd r
	\Bigg(r^2 \mathcal{I}_{\rm LTD}(r \hat k) - 
	\sum_{(i,j)\in\mathcal{E}} \sum_{\rho\in\{\pm1\}}
	\CT_{i_\sigma j_{-\sigma}}^{\rho}(r,\hat k)
	\Bigg) \\
\label{eq:fract_residues_ltd}
	-\Im I
	&= \frac12 \int_{H^2} \frac{\dd^2\hat k}{(2\pi)^2}
	\sum_{(i,j)\in\mathcal{E}} \sum_{\rho\in\{\pm1\}}
	\rho R_{i_\sigma j_{-\sigma}}^{\rho}(\hat k),
\end{align}
where the existing E-surfaces are labelled by
\begin{align}
\label{eq:exisiting_e_surfaces}
    \mathcal{E}=\left\{(i_\sigma,j_{-\sigma})\in\{1,\dots,n\}^2\middle\vert (p_i-p_j)^2>(m_i+m_j)^2 \land \sigma (p_i-p_j)^0>0\right\}.
\end{align}
The sign $\sigma \in \{+1,-1\}$ can be chosen arbitrarily, a reflection of time-reversal symmetry of the one-loop integral, and does not affect local representation of the above result.
\par

We can express the imaginary part in eq.~\eqref{eq:fract_residues_ltd} diagrammatically using our definitions of the double cut diagram in eq.~\eqref{eq:e_surface_residue_diagram} and the loop diagram in eq.~\eqref{eq:one_loop_integral} as
\begin{align}
\label{eq:fract_residues_ltd_diag}
    2\Im \ii
    \begin{gathered}
	    \vcenter{\hbox{\includegraphics[page=2,scale=1]{img/diagrams.pdf}}}
	\end{gathered}
	&=
    \sum_{(i,j)\in\mathcal{E}_\sigma}
    \begin{gathered}
	    \vcenter{\hbox{\includegraphics[page=4,scale=1]{img/diagrams.pdf}}}
	\end{gathered}
\end{align}
which shows that our eq.~\eqref{eq:fract_residues_ltd} is a local\footnote{
The integrand is locally finite as a consequence of local cancellations of threshold singularities.
} one-loop representation\footnote{
Up to terms $\order(\epsilon)$ (cf. footnote \ref{foot:causal_prescription}).
}
of the generalised optical theorem, a direct implication of the unitarity of the $S$-matrix,
\begin{align}
\label{eq:generalised_optical_theorem}
    2 \Im \mathcal{M}(i\to f) = \sum_x \int \dd \Pi_x \mathcal{M}^*(f\to x) \mathcal{M}(i\to x),
\end{align}
for matrix elements $\mathcal{M}$ of initial and finial states $i$ and $f$.
The intermediate cut states are denoted by $x$.
The equivalence of E-surface residues and two-body phase space integrals can be made manifest by expressing each E-surface residues $R_{i_\sigma j_{-\sigma}}^\rho$ in the rest frame of $p_{ij}$.
However, keep in mind that such a different parameterisation for each residue definitively breaks the local cancellation mechanism between the remaining threshold singularities.
\par

The expressions in eqs.~\eqref{eq:subtracted_ltd} and \eqref{eq:fract_residues_ltd} for the real and imaginary part of one-loop integrals set the basis for the numerical results presented in sect.~\ref{sec:results}.
In the absence of locally pinched singularities, the integrands become free of poles and the two- and three-dimensional integrals can be directly numerically evaluated with Monte~Carlo without a further need of regulation.
However, the expressions for real and imaginary part are not valid if the LTD integrand has poles that pinch the $r$-integration contour.
In this case, the integrand of residues in eq.~\eqref{eq:fract_residues_ltd} has poles in the integration domain.
In general, these poles have to be regulated with, e.g. the causal prescription $\epsilon>0$, a contour deformation or subtraction.
In any case, this integral will then give rise to an absorptive part which in turn contributes to the real part of the integral\footnote{
This is analogous to the fact that the LTD integral only has E-surfaces if there are pinched poles in $k^0$-integration of the loop integrand in eq.~\eqref{eq:one_loop_integral}.
Without E-surfaces, the LTD integrand has no poles, can be directly numerically integrated and the loop integral has a pure phase.
}.
At this stage, real and imaginary part cannot be formally separated and the integral should be represented as in eq.~\eqref{eq:multi_dim_integral}.
This implies that the subtracted integral, although perfectly convergent also in the presence of locally pinched poles, does not yield the real part of the loop integral.
\par

The occurrence of locally pinched poles depends on the external kinematics but also on the choice of E-surface parameterisation.
We will elaborate on their existence in general in sect.~\ref{sec:e_surface_intersections}.
Fortunately, we can avoid them in many cases by simply moving the origin of our parameterisation frame.
This is discussed in sect.~\ref{sec:e_surface_grouping}.
\par


\subsubsection{Higher-order poles: E-surface intersections}
\label{sec:e_surface_intersections}

Before we apply our strategy of sect.~\ref{sec:degenerate_case} for identifying poles in the sum of E-surface residues, let us take a step back and think about how poles in E-surface residues come about.
Here, we only discuss E-surface residues, and ignore their corresponding counterterms, since the latter, as we have seen in sect.~\ref{sec:degenerate_case}, 
always exhibit local cancellations in the case of higher-order poles.
\par

\paragraph{Poles of the residues}
First, one notices that the residue $R^\rho_{i_\sigma j_{-\sigma}}$ at E-surface $E_{ij}^\sigma$ in eq.~\eqref{eq:e_surface_residue} has a pole if one of its propagators $D_l\vert_{i_\sigma j_{-\sigma}}=H_{il}^\sigma E_{il}^\sigma$ vanishes.
This is exactly the case if E-surface $E_{ij}^\sigma$ intersects with \emph{either} E-surface $E_{il}^\sigma$ \emph{or} with H-surface $H_{il}^\sigma$, i.e.
\begin{align}
\label{eq:relevant_intersections}
    E_{ij}^\sigma \cap E_{il}^\sigma \neq \emptyset
    \quad \lor \quad
    E_{ij}^\sigma \cap H_{il}^\sigma \neq \emptyset.
\end{align}
It cannot intersect with both as only one of the two can have their existence condition satisfied.
We will now argue that the poles of E-surface residues come in pairs with corresponding poles in another E-surface residues.
Furthermore, the residues exhibit pairwise local cancellation if \emph{summed} together.
If instead, we consider their difference, the cancellation mechanism is broken.
\par

\begin{figure}
    \centering
    {\includegraphics[scale=1]{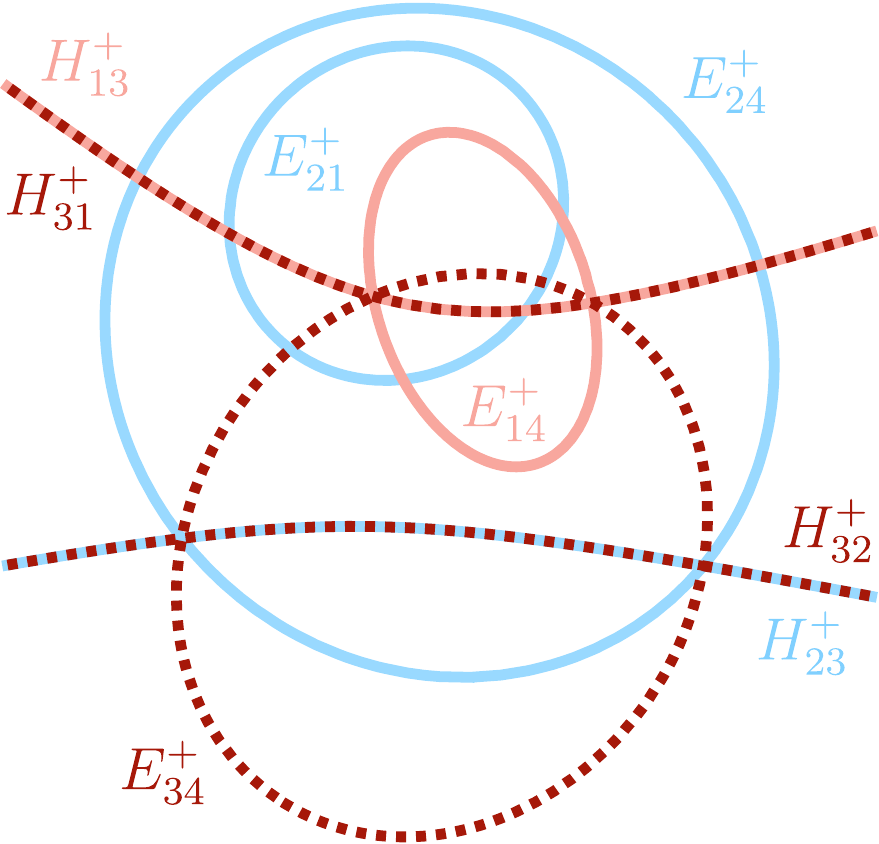}}
    {\caption{The singular surfaces of a box integral in two-dimensional spatial momentum space.
    The dual integrand ${\color{myRed}R_{1_+}}$ has poles on the surfaces
    ${\color{myRed}E_{14}^+}$ and 
    ${\color{myRed}H_{13}^+}$,
    ${\color{myLightBlue}R_{2_+}}$ has poles on
    {\color{myLightBlue}$E_{21}^+$}, {\color{myLightBlue}$E_{24}^+$} and {\color{myLightBlue}$H_{23}^+$},
    and ${\color{myDarkRed}R_{3_+}}$ on
    {\color{myDarkRed}$E_{34}^+$}, ${\color{myDarkRed}H_{31}^+}=-{\color{myRed}H_{13}^+}$ and ${\color{myDarkRed}H_{32}^+}=-{\color{myLightBlue}H_{23}^+}$.
    For this particular configuration, the dual integrand $R_{4_+}$ has no poles.
    First, we observe that H-surfaces appear pairwise.
    They locally cancel between the dual integrands.
    Secondly, we see that an intersection between an E- and an H-surface of the \emph{same} dual integrand is identical to the one between two E-surfaces of \emph{different} dual intagrands. Here, ${\color{myRed}E_{14}^+}\, \cap\, {\color{myRed}H_{13}^+} = {\color{myRed}E_{14}^+}\, \cap\, {\color{myDarkRed}E_{34}^+}$ and ${\color{myLightBlue}E_{24}^+}\, \cap\, {\color{myLightBlue}H_{23}^+} = {\color{myLightBlue}E_{24}^+}\, \cap\, {\color{myDarkRed}E_{34}^+}$.
    This implies that all higher-order poles of dual integrands are E-surface intersections, which is why all poles of E-surfaces residue can be locally cancelled between E-surface (and not H-surface) residues only.
    The remaining intersections of E- and H-surfaces are irrelevant as they do not correspond higher-order poles in the dual integrands and therefore do not introduce poles to E-surface residues.
    }\label{fig:box_singular_surfaces}}
\end{figure}

From symmetry, it is clear that if E-surface $E_{ij}^\sigma$ intersects with E-surface $E_{il}^\sigma$, then the residue $R^\rho_{i_\sigma l_{-\sigma}}$ of E-surface $E_{il}^\sigma$ also has a pole from the propagator $D_j\vert_{i_\sigma l_{-\sigma}}=E_{ij}^\sigma H_{ij}^\sigma\vert_{i_\sigma l_{-\sigma}}$.
Note that we can not make the same argument for the intersection with H-surfaces, since we do not consider H-surface residues.
However, if we are on the intersection of $E_{ij}^\sigma$ and $H_{il}^\sigma$, then we also lie on the E-surface $E_{lj}^\sigma=E_{ij}^\sigma-H_{il}^\sigma$, i.e.
\begin{align}
     E_{ij}^\sigma \cap H_{il}^\sigma \neq \emptyset
     \quad \Leftrightarrow \quad
     E_{ij}^\sigma \cap E_{lj}^\sigma \neq \emptyset.
\end{align}
We therefore find the corresponding pole in the 
residue $R^\rho_{l_\sigma j_{-\sigma}}$ of E-surface $E_{lj}^\sigma$ at the intersection with H-surface $H_{li}^\sigma=-H_{il}^\sigma$.
\par

The above description of pairwise appearances of intersections can be visualised by plotting the singular surfaces of the dual integrands (including the dual cancelling ones) in spatial loop momentum space, as in fig.~\ref{fig:box_singular_surfaces}.
At this point we emphasise that \emph{not} all intersections of singular surfaces in momentum space correspond to poles in E-surfaces residues.
Whereas two E-surfaces $E_{ij}^\sigma$ and $E_{ab}^\sigma$ may share solutions, their intersection only results in a pole in their corresponding residues if either $a=i$ or $b=j$, as we have seen from the discussion above.
As a consequence, H-surfaces and their intersections can then be generally ignored (unless equivalent to intersections among E-surfaces).
\par

We can formally parameterise a pole of the E-surface residue with a coordinate $\lambda$ at $\lambda=0$.
The behavior of the residue around the pole can then be found by expanding $R^\rho_{i_\sigma j_{-\sigma}}(\hat k(\lambda))$ around $\lambda=0$.
We find that the E-surface residue $R^\rho_{i_\sigma j_{-\sigma}}$ and its locally cancelling partner $R^\rho_{a_\sigma b_{-\sigma}}$ indeed share the same singular behaviour but with opposite signs, i.e.
\begin{align}
\label{eq:e_surface_residue_expansion}
    R^\rho_{i_\sigma j_{-\sigma}}
	&=
	\frac{1}{\lambda}
	\left[
	\frac{r^2}{8E_iE_jE_l}
	\frac{1}{\vec{n}_{ij}\times \vec{n}_{ab}}
	\frac{1}{\vec k \times \frac{\dd\hat k}{\dd \lambda}}
	\frac{N}{\prod_{m\neq i,j,l} D_m}
	\right]_{\lambda=0}
	+\order(\lambda^{-2})
	=
	-R^\rho_{a_\sigma b_{-\sigma}},
\end{align}
where $(a,b)=(i,l)$ for the intersection of $E_{ij}^\sigma$ and $E_{il}^\sigma$, and $(a,b)=(l,j)$ for the intersection of $E_{ij}^\sigma$ and $E_{lj}^\sigma$ (or equivalently $H_{il}^\sigma$).
The quantities above in eq.~\ref{eq:e_surface_residue_expansion} are understood to be evaluated at the intersection of said surfaces.
\par

There is a simple criterion to determine if poles exist in the E-surface residue, or equivalently, if two E-surfaces $E_{ij}^\sigma$ and $E_{ab}^\sigma$ with $(a,b)=(i,l)$ or $(a,b)=(l,j)$ intersect.
It boils down to checking whether there are solutions to the equations $D_i=D_j=D_l=0$ with fixed energy flow $\sgn q_i^0=\sigma$, $\sgn q_j^0=-\sigma$, $\sgn q_l^0=-\sigma$ or $\sgn q_l^0=\sigma$, respectively.
Therefore, we evaluate $D_l$ at the momentum $q_i^\prime=(\sigma E_i^\prime, |\vec{q}_i^{\,\prime}|\hat q_i^\prime)$ in the rest frame of $p_{ij}=p_i-p_j$, where $E_i^\prime$ and $|\vec{q}_i^{\,\prime}|$ are constants determined by eq.~\eqref{eq:q_rad_on_shell}, that puts propagators $i$ and $j$ on-shell with energies $\sgn q_i^0=\sigma$ and $\sgn q_j^0=-\sigma$.
The equation $D_l=0$ then has a solutions for real momenta, meaning that $E_{ij}^\sigma$ intersects with either $E_{il}^\sigma$ or $E_{lj}^\sigma$, if $|\cos\theta^\prime|\leq 1$, where
\begin{align}
\label{eq:intersection_condition}
    \cos\theta^\prime
    =
    \frac{1}{|\vec{p}_{il}^{\,\prime}||\vec{q}_i^{\,\prime}|}
    \left(
        \sigma E_i^\prime p_{il}^{\prime0}
        - \frac{m_i^2+p_{il}^2-m_l^2}{2}
    \right),
\end{align}
with $p_{il}=p_i-p_l$.
Which one of the two intersection types 
we are dealing with can e.g. be checked with the E-surface existence conditions.
Eq.~\eqref{eq:intersection_condition} is equivalent to 
eq.~(82) in \cite{kilian2009numerical}.
\par

Finally, note that eq.~\eqref{eq:e_surface_residue_expansion} does not imply that the poles of E-surface residues always locally cancel in the imaginary part in eq.~\eqref{eq:fract_residues_ltd}.
This is because the residues are not always guaranteed to contribute with equal signs $\rho=\lim_{\epsilon\to0} \sgn\Im r_{i_\sigma j_{-\sigma}}^\rho$.
This subtlety is best discussed in the framework we laid out in sect.~\ref{sec:degenerate_case}.


\paragraph{Local cancellations}
In the above discussion, we managed to give a geometrical description of how poles arise in E-surface residues.
The important aspect, of whether or not these poles locally cancel in the integrand of eq.~\eqref{eq:fract_residues_ltd} may not have become completely clear.
We are devoting ourselves to that now.
\par

Recall from sect.~\ref{sec:degenerate_case} that the local appearance of higher-order poles in the LTD integrand in eq.~\eqref{eq:ltd_integral_starting} for specific values of $\hat{k}$ leads to poles in the E-surface residues in eq.~\eqref{eq:e_surface_residue} and their counterterms eq.~\eqref{eq:ltd_counterterm}.
Only if the higher-order poles do not pinch the $r$-integration contour for $\epsilon\to 0$, we find local cancellation in the imaginary part in eq.~\eqref{eq:fract_residues_ltd}.
If the poles are pinched however, eq.~\eqref{eq:fract_residues_ltd} will be singular for $\epsilon\to0$.
\par

In the discussion above, we found that certain intersections between E-surfaces lead to local double poles.
At such an intersection, following the E-surface parameterisation in sect.~\ref{sec:poles_in_r}, it holds that
\begin{align}
\label{eq:pole_r_higher_order}
    r_{i_\sigma j_{-\sigma}}^\rho(\hat k^*) = 
    r_{a_\sigma b_{-\sigma}}^\tau(\hat k^*),
\end{align}
for $\epsilon\to0$, some particular values $\hat k^*$ and $(a,b)=(i,l)$ or $(a,b)=(l,j)$.
According to eqs.~\eqref{eq:pinched_definition} and \eqref{eq:non_pinched_definition} and using eq.~\eqref{eq:sign_r_pole}, we find that this double pole is pinched if $\rho=-\tau$.
The solutions $\hat k^* \in H^2$ of
\begin{align}
    r_{i_\sigma j_{-\sigma}}^\rho(\hat k^*) = r_{a_\sigma b_{-\sigma}}^{-\rho}(\hat k^*)
\end{align}
for $\epsilon\to0$ are therefore poles in the integration domain of the integrand in eq.~\eqref{eq:fract_residues_ltd}.
If $\rho=\tau$ the double pole is non-pinched, such that the surface defined by
\begin{align}
    r_{i_\sigma j_{-\sigma}}^\rho(\hat k^*) = r_{a_\sigma b_{-\sigma}}^{\rho}(\hat k^*)
\end{align}
for $\epsilon\to0$ defines spurious poles as they locally cancel among the E-surface residues.
\par

\begin{figure}
\begin{subfigure}[t]{.37\textwidth}
    \centering
    \includegraphics[scale=0.6]{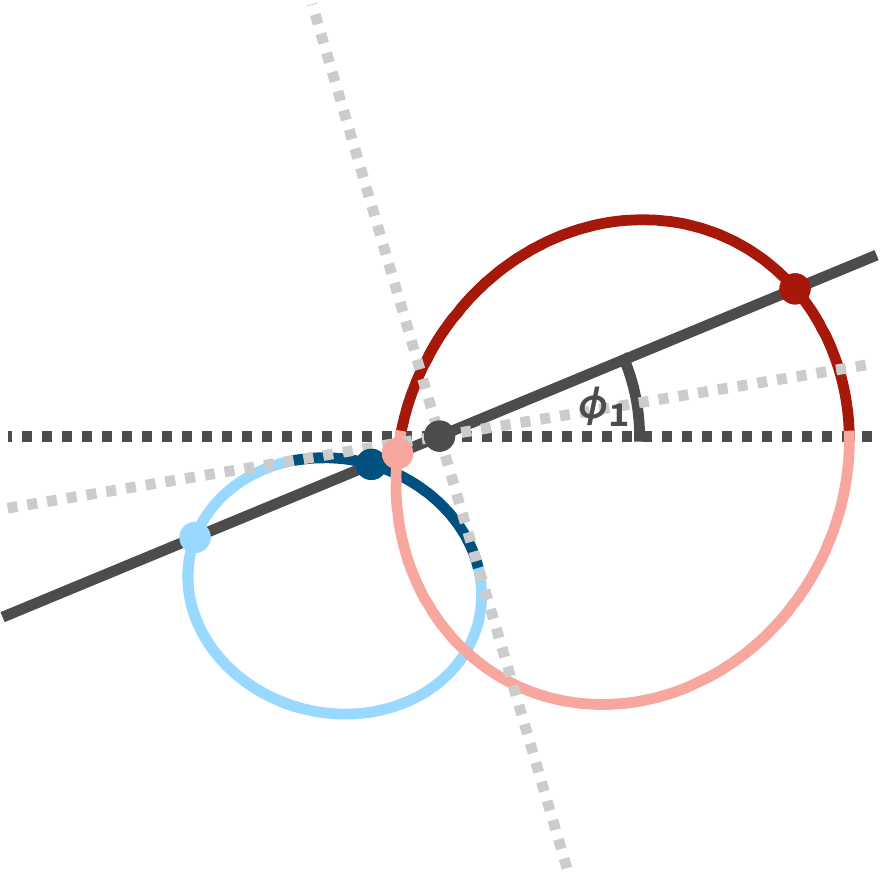}
    \caption{E-surfaces $E_{21}^+$ (blue) and $E_{23}^+$ (red) with highlighted ellipse segments parameterised by $r_{i_+j_-}^+$ (dark) and $r_{i_+j_-}^-$ (light), respectively.}
    \label{subfig:ellipse_miscancellations}
\end{subfigure}\hfill
\begin{subfigure}[t]{.61\textwidth}
    \centering
    \includegraphics[scale=0.95]{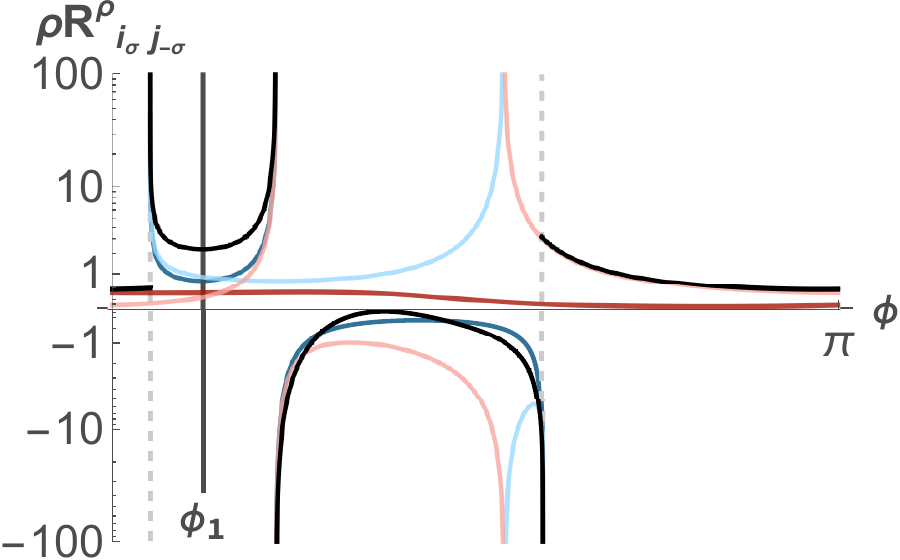}
    \caption{The E-surface residues $+R_{2_+1_-}^+$ (dark blue), $-R_{2_+1_-}^-$ (light blue), $+R_{2_+3_-}^+$ (dark red), $-R_{2_+3_-}^-$ (light red) and their sum (black).
    The singularities at the gray dashed lines are integrable and one in between is a single pole.
    }
    \label{subfig:residues_miscancellations}
\end{subfigure}
\caption{
\label{fig:triangle_miscancellations}
Local (mis)cancellations in the imaginary part of a scalar triangle according to a naive application of eq.~\eqref{eq:fract_residues_ltd}.
The E-surfaces $E_{ij}^+$ in two-dimensional spatial momentum space are shown in (\subref{subfig:ellipse_miscancellations}) and their corresponding residues $\rho R_{i_+j_-}^\rho$ in (\subref{subfig:residues_miscancellations}).
They are parameterised by the radial solutions $r_{i_+j_-}^\rho$.
For angles $\phi\in[0,\pi)$ the E-surfaces are scanned at two points simultaneously.
The upper intersection, where $r_{21}^+=r_{23}^-$, results in the miscancellation between the poles in $+R_{2_+1-}^+$ and $-R_{2_+3-}^-$.
This is due to the opposite orientation of the parameterisation of the intersecting ellipse segments (dark blue, light red).
The lower intersection however, where $r_{21}^-=r_{23}^-$, corresponds to spurious, locally cancelling poles in $-R_{2_+1-}^-$ and $-R_{2_+3-}^-$.
In this case, the intersecting ellipse segments (light blue, light red) are parameterised with the same (positive) orientation.
Another notable feature are the integrable singularities of the residues $\pm R_{2_+1_-}^\pm$ at the tangents (gray) of the ellipse $E_{21}^+$.
}
\end{figure}

\begin{figure}
\begin{subfigure}[t]{.37\textwidth}
    \centering
    \raisebox{0.4\height}{\includegraphics[scale=0.6]{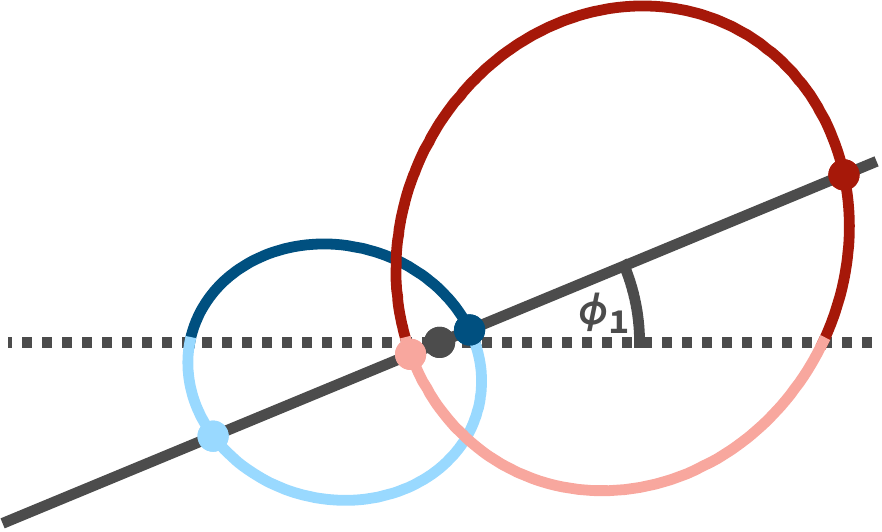}}
    \caption{E-surfaces $E_{21}^+$ (blue) and $E_{23}^+$ (red) with highlighted ellipse segments parameterised by $r_{i_+j_-}^+$ (dark) and $r_{i_+j_-}^-$ (light), respectively.}
    \label{subfig:ellipse_cancellations}
\end{subfigure}\hfill
\begin{subfigure}[t]{.61\textwidth}
    \centering
    \includegraphics[scale=0.95]{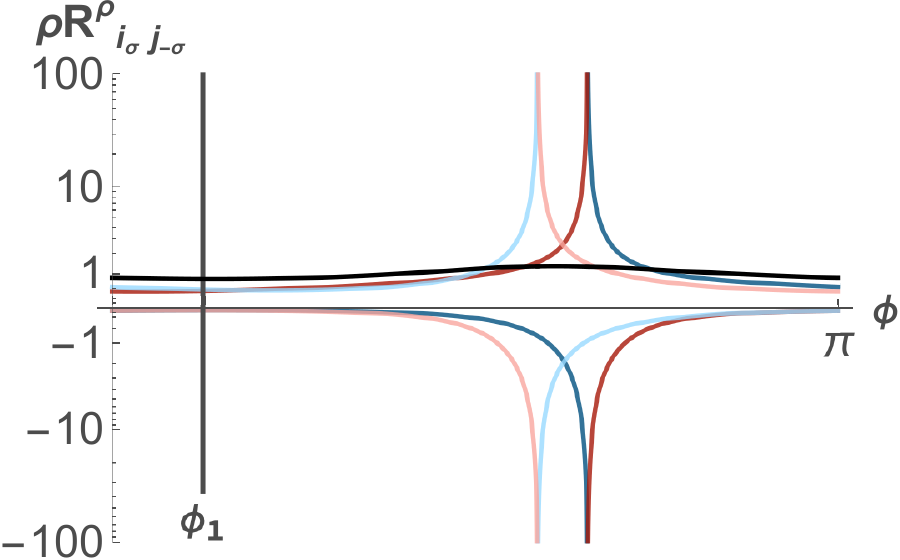}
    \caption{The E-surface residues $+R_{2_+1_-}^+$ (dark blue), $-R_{2_+1_-}^-$ (light blue), $+R_{2_+3_-}^+$ (dark red), $-R_{2_+3_-}^-$ (light red) and their sum (black).}
    \label{subfig:residues_cancellations}
\end{subfigure}
\caption{
\label{fig:triangle_cancellations}
Local cancellations in the imaginary part of a scalar triangle according to eq.~\eqref{eq:fract_residues_ltd}.
The E-surfaces $E_{ij}^+$ in two-dimensional spatial momentum space are shown in (\subref{subfig:ellipse_cancellations}) and their corresponding residues $\rho R_{i_+j_-}^\rho$ in (\subref{subfig:residues_cancellations}).
They are parameterised by the radial solutions $r_{i_+j_-}^\rho$.
For angles $\phi\in[0,\pi)$ the E-surfaces are scanned at two points simultaneously.
Both intersections, where $r_{21}^+=r_{23}^+$ and $r_{21}^-=r_{23}^-$, correspond to spurious, locally cancelling poles in the residues $+R_{2_+1-}^+$ and $+R_{2_+3-}^+$, and $-R_{2_+1-}^-$ and $-R_{2_+3-}^-$, respectively.
In fact, placing the origin within the ellipses' overlap ensures local cancellations and avoids the introduction of integrable singularities (cf. fig.~\ref{fig:triangle_miscancellations}).
}
\end{figure}

Pinched and non-pinched poles can be visualised in spatial loop momentum space by plotting the E-surfaces with respect to the origin at $\vec{k}=0$ (see figs.~\ref{fig:triangle_miscancellations} and \ref{fig:triangle_cancellations}).
We find that non-pinched, cancelling intersections are parameterised with equal orientation, whereas
pinched, miscancelling intersections are parameterised with opposite orientation.
The comparison of figs.~\ref{fig:triangle_miscancellations} and \ref{fig:triangle_cancellations}, where the latter in contrast to the former has no pinched poles, illustrates that the placing of the origin $\vec{k}=0$ can be crucial for local cancellations to be realised.
However, the conclusion that we can only avoid miscancellations by placing the origin within the overlap of E-surfaces is wrong (e.g. see sect.~\ref{sec:exterior_sources}).
Nevertheless, it may be impossible to find a single parameterisation for all E-surfaces in which all poles locally cancel.
In this case, one shall check if certain E-surfaces can be integrated independently of others.
How this property can be exploited is discussed in the next section.

\bigbreak
\par
Finally, we emphasize that only the intersections that result in higher-order poles, i.e. the type in eq.~\eqref{eq:relevant_intersections}, or equivalently in eq.~\eqref{eq:pole_r_higher_order}, transform covariantly under Lorentz transformations, as opposed to the other intersections of E-surfaces and H-surfaces in spatial momentum space.
More precisely, these intersections are associated with the loop four-momentum
\begin{align}
    k = \begin{pmatrix}
        \sigma E_i - p_i^0 \\
        r_{i_\sigma j_{-\sigma}}^\rho \hat k^*
    \end{pmatrix},
\end{align}
which simultaneously satisfies the three on-shell constraints $D_i=0, D_j=0, D_l = 0$ with fixed energy flow $\sgn q_i^0=\sigma, \sgn q_j^0=-\sigma, \sgn q_l^0=-\sigma$ or $\sgn q_l^0=\sigma$ for $(a,b)=(i,l)$ or $(a,b)=(l,j)$, respectively.
The other intersections do not have a single associated four-momentum for which three propagators go on-shell.
This implies that E-surface intersections of the covariant type exist in every Lorentz frame, whereas the other intersections are frame dependent (see fig.~\ref{fig:hexagon_groups}).
\par


\subsubsection{Local cancellations within groups of residues}
\label{sec:e_surface_grouping}

\begin{figure}
    \centering
    \hfill
    \begin{subfigure}[t]{0.39\textwidth}
        \centering
        \includegraphics[scale=0.95]{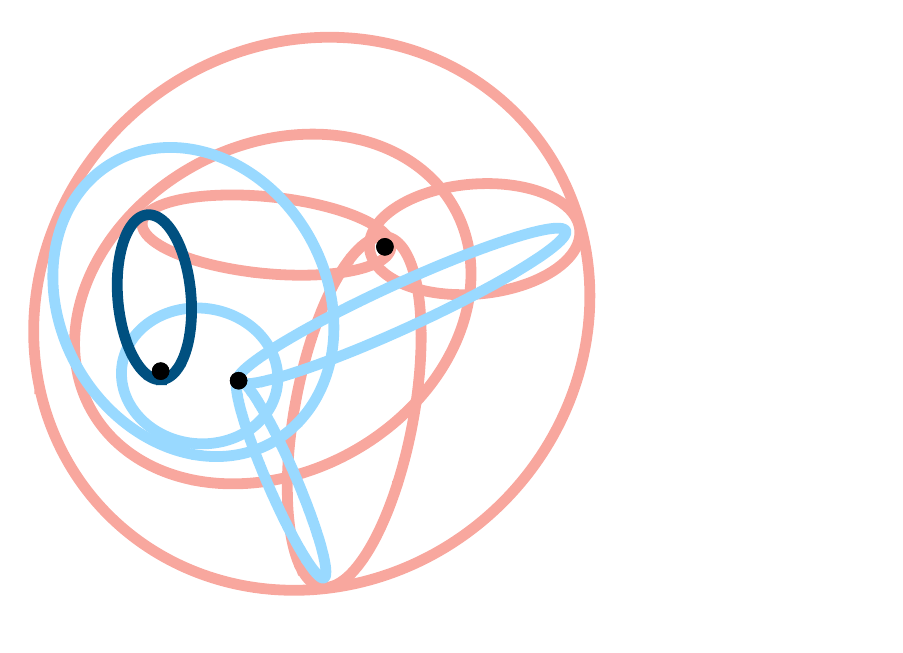}
        \caption{\label{subfig:hexagon_groups}
    }
    \end{subfigure}%
    \begin{subfigure}[t]{0.59\textwidth}
        \centering
        \includegraphics[scale=0.95]{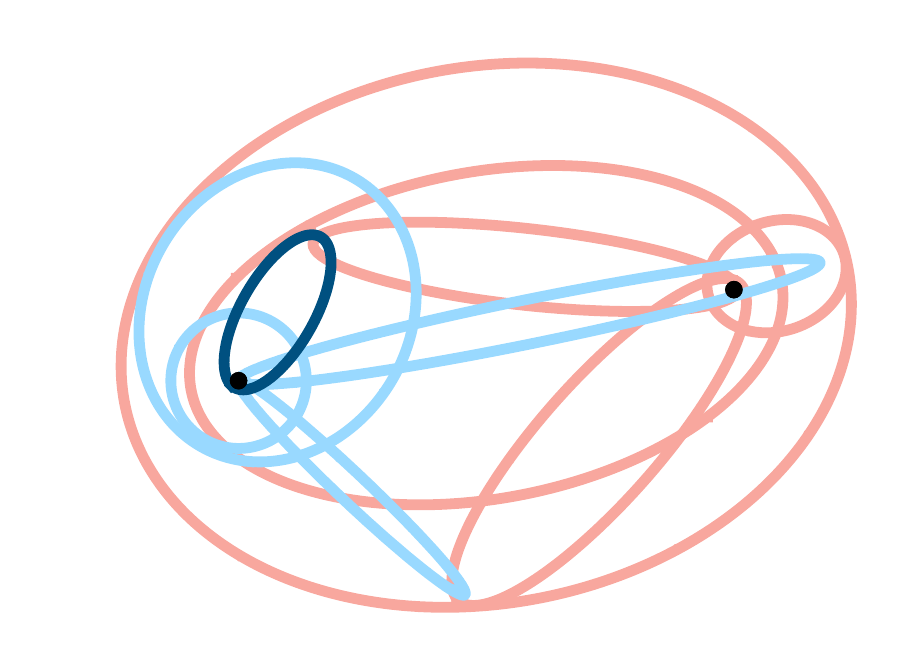}
        \caption{\label{subfig:hexagon_boosted_groups}
    }
    \end{subfigure}
    \caption{\label{fig:hexagon_groups}
    The ten E-surfaces of the hexagon configuration \texttt{1L6P.VIII} (cf. sect.~\ref{sec:multiple_groups}) in spatial momentum space in two Lorentz frames.
    Only the intersections of ellipses in the same color result in poles in the E-surface residues.
    The groups of each color can therefore be parameterised independently.
    Here, we can choose the groups' sources to be the foci, which for massless propagators are simply given by $-p_i$ (cf. appendix~\ref{app:dual_propagator}).
    The groups and their sources are $\color{myRed} E_{13}^+, E_{23}^+, E_{43}^+, E_{53}^+, E_{63}^+$ with $-p_3$, $\color{myLightBlue} E_{61}^+, E_{62}^+, E_{64}^+, E_{65}^+$ with $-p_6$ and $\color{myBlue} E_{21}^+$ with $-p_1$.
    In the rest frame of $p_1+p_6$ in (\subref{subfig:hexagon_boosted_groups}) two sources collide and we can merge their respective groups.
    In (\subref{subfig:hexagon_groups}) we use a different parameterisation for each group.
    Curiously, the residues in the red group locally sum up to zero.
    This is not further investigated in this paper.
    }
\end{figure}

In order to use our integral expressions for Monte Carlo numerical integration, it is crucial that the integral is free of poles.
We therefore need to make sure that all poles of E-surface residues locally cancel among themselves.
\par

Ideally, one would determine a coordinate system in which all poles locally cancel.
However, according to sect.~\ref{sec:e_surface_intersections} and as we demonstrated with figs.~\ref{fig:triangle_miscancellations} and \ref{fig:triangle_cancellations}, not every coordinate system, depending on where we place its origin with respect to the E-surfaces, realises local cancellations.
It may not even be possible to find such a spherical coordinate system.
%
%
In this case one can try to group E-surfaces in a suitable way and treat each group separately, such that each group has its own associated coordinate system.
\par

Before we consider the subtracted LTD integral of the real part, we first investigate the E-surface residues of the imaginary part.
Note that an E-surface residue without poles can be integrated on its own, independently of other E-surface residues.
Only when two E-surface residues share locally cancelling poles, they have to be integrated in the same coordinate system in order not to break the cancellation.
This applies to any group of E-surface residues that realise local cancellations among themselves.
We can therefore choose a parameterisation for each such group.
In particular, this may allow us to avoid miscancellations that were unavoidable if all E-surfaces had to be parameterised in the same frame.
\par

Luckily, the same grouping can be applied to the counterterms in the subtracted integral.
First, note that in general, the counterterm $\CT_{i_\sigma j_{-\sigma}}^\rho$ is not only singular at $r=r_{i_\sigma j_{-\sigma}}^\rho$ but it is plagued by the same poles in $\hat k$ as the residue $R_{i_\sigma j_{-\sigma}}^\rho$ for all values of $r$.
However, if the residues realise local cancellations, then all poles in $\hat k$ are spurious in the sum of counterterms, such that only the ones in $r$ remain.
This applies for any such group of residues and their counterterms.
Nevertheless, the poles in $r$ still have to be cancelled locally with the corresponding ones in the LTD integrand.
One therefore has to make sure to shift back the groups' coordinate system \emph{after} parameterisation of its E-surfaces to align it with the integration frame.
\par


A group is characterised by E-surfaces $\mathcal{E}$ and a parameterisation coordinate system, identified by its origin\footnote{
The frame's orientation has no effect on the realisation of local cancellations.
}, which we call the group's \emph{source} $\vec{s}$, a spatial vector with respect to e.g. the integration frame.
Suitable groups are easily identified.
If $E_{ij}^\sigma$ and $E_{ab}^\sigma$ intersect and their intersection corresponds to a double pole in the LTD integrand, i.e. $(a,b)=(i,l)$ or $(a,b)=(l,j)$, then they belong to the same group.
Note that if they do not intersect but they have group members that do, they are also in the same group.
An example is given in fig.~\ref{fig:hexagon_groups}.
\par

If we can identify a source that allows for local cancellations within each group, we can replace the expressions in eqs.~\eqref{eq:subtracted_ltd} and \eqref{eq:fract_residues_ltd} with their generalisation
\begin{align}
\label{eq:subtracted_ltd_grouped}
	\Re I
	&= -\int_{H^2} \frac{\dd^2\hat k}{(2\pi)^3}
	\int_\mathbb{R}\dd r r^2
	\Bigg(\mathcal{I}_{\rm LTD}(r \hat k) - 
	\sum_{(\vec{s},\mathcal{E}) \in \mathcal{G}}
	\sum_{(i,j)\in\mathcal{E}} \sum_{\rho\in\{\pm1\}}
	\frac{\mathrm{CT}_{i_\rho j_{-\sigma}}^{\sigma}(r_{\vec{s}},\hat k_{\vec{s}};\vec{s})}{r_{\vec{s}}^2}\Bigg) \\
\label{eq:fract_residues_ltd_grouped}
	-\Im I
	&= \frac12 \int_{H^2} \frac{\dd^2\hat k}{(2\pi)^2}
	\sum_{(\vec{s},\mathcal{E}) \in \mathcal{G}}
	\sum_{(i,j)\in\mathcal{E}} \sum_{\rho\in\{\pm1\}}
	\sigma R_{i_\sigma j_{-\sigma}}^{\rho}(\hat k;\vec{s}),
\end{align}
where $r_{\vec{s}}$ and $\hat k_{\vec{s}}$ are spherical coordinates of the frame with origin at $\vec{s}$ and functions of the integration variables $r$ and $\hat k$.
The residue and counterterm parameterised with respect to that frame are labelled with the source as their last argument and can be calculated as before from eqs.~\eqref{eq:e_surface_residue} and \eqref{eq:ltd_counterterm} but with $\vec{p}_i$ replaced by $\vec{p}_i+\vec{s}$.
\par

The factor $r_{\vec{s}}^{-2}$ that accompanies the counterterms corrects for the Jacobian determinant in \eqref{eq:ltd_counterterm}.
It leads to an integrable singularity.
Whereas this does not directly spoil numerical integration, it however increases the variance of the integrand, which results in a larger Monte Carlo error on the estimate.
It is therefore of advantage to have as few groups as possible.
This can be achieved by merging groups and parameterising their E-surfaces with respect to the same coordinates.
\par

\subsection{Divergent loop integrals}
\label{sec:divergent_loop_integrals}
Divergent loop integrals have ultraviolet or/and infrared singularities.
The former arise when the integrand does not fall off fast enough for increasing magnitude of loop momenta.
The latter are attributed to \emph{pinch surfaces} (cf. \cite{Collins:1989gx}), where, as with the pinched poles we encountered, two or more singularities of the integrand pinch the integration contour, however, not only in one dimension of the integral but in all (usually four)\footnote{
In dimensional regularisation in $D$ dimensions, these singular surfaces are not pinched anymore, such that the integral converges but with poles at $D=4$.
This is analogous to pinched poles in one dimension, as in sect.~\ref{sec:degenerate_case} that are only locally pinched in multi-dimensional integrals (see sect.~\ref{sec:multidimensional_integrals}) unless they pinch in all integration dimensions.
} spacetime dimensions.
This forces the integration contour to go through the singularity, resulting in a divergent integral.
Pinch surfaces are found in regions where the loop momenta vanish, i.e. are soft, or collinear to external momenta.
\par

In the framework of LTD, infrared singularities correspond to pinched (squeezed) E-surfaces, whose endpoints, the soft singularities, are connected by a line, the collinear singularity (cf. app.~\ref{app:dual_propagator}).
\par

Divergent loop integrals can be regulated by subtracting suitable counterterms that locally approximate the integrand in the problematic regions.
This has been first studied at one-loop in \cite{Nagy_2003,Assadsolimani_2010,Becker_2010}.
Ultraviolet counterterms at two loops were investigated in \cite{Driencourt_Mangin_2019}.
Moreover, using a local representation of the BPHZ forest formula \cite{Bogoliubov:1957gp,Hepp:1966eg,Zimmermann:1969jj} the ultraviolet subtraction can in principle be extended to all loop orders.
Infrared counterterms for two-loop amplitudes based on their factorization properties were devised in \cite{Anastasiou_2019,Anastasiou_2021}.
\par

The regulated integral is finite in $D=4$ dimensions and can in principle be integrated with the method presented in this article.
To obtain the result of the divergent integral however, one has to add back the counterterms.
Fortunately, these counterterms are in general much simpler than the original integral and can usually be evaluated analytically in dimensional regularisation.
\par

\section{Numerical Results}
\label{sec:results}
In this section, we will apply our method in sect.~\ref{sec:feynman_integrals} in combination with Monte Carlo numerical integration to a range of examples and demonstrate validity and performance.
\par

More explicitly, we will use our eqs.~\eqref{eq:subtracted_ltd_grouped} and \eqref{eq:fract_residues_ltd_grouped} to evaluate a range of scalar one-loop configurations that have previously been calculated in spatial loop momentum space making use of contour deformation \cite{buchta2015numerical, Capatti_2020}.

First, in sect.~\ref{sec:single_group} we will apply eqs.~\eqref{eq:subtracted_ltd} and \eqref{eq:fract_residues_ltd} to a subset of loop integrals, where a single residue group can be identified with source in the interior of all E-surfaces. 
Secondly, in sect.~\ref{sec:multiple_groups} we will look at more complicated one-loop integrals, that do not allow for a single source but are still such that each of their residue groups allow for a single source.
Thirdly, in sect.~\ref{sec:tringigons} with configurations of a triacontagon with 30 external legs and up to over 400 E-surfaces we push our threshold subtraction method into a regime, in which a calculation with contour deformation would be very challenging.
Even in this case, the Monte Carlo integration is stable and yields precise results.
\par

In sect.~\ref{sec:exterior_sources} we demonstrate with an example, where no source can be found in the interior of every E-surface of a residue group, that external sources are permitted as long as they do not introduce miscancelling pinched poles, as described in sect.~\ref{sec:e_surface_intersections}.
\par

Moreover, in sect.~\ref{sec:unitarity_raised_props_ot} we illustrates with an example of forward scattering how eq.~\eqref{eq:fract_residues_ltd} provides a local representation of the optical theorem.
\par

Last but not least, in sect.~\ref{sec:numerator_ir_singularities} we apply our method to infrared and ultraviolet divergent scalar integrals as well as one-loop amplitudes, whose singularities are subtracted with local counterterms.
\par

\subsection{Implementation}

The numerical results presented in this section were obtained using a \textsc{Rust} implementation of the expressions in sect.~\ref{sec:feynman_integrals} and using the \textsc{Vegas} adaptive Monte-Carlo algorithm \cite{Lepage:1977sw,Lepage_2021} of the \textsc{Cuba} library \cite{Hahn_2005} for multidimensional numerical integration.
The preprocessing of topologies, including the identification of residue groups and a heuristic algorithm to find suitable sources, is performed by a \textsc{Python} routine.
The implementation is automated for one-loop integrals without raised propagators, inasmuch that only the numerical values of the external momenta and the masses of the internal propagators must be provided, and supports arbitrary numerators.
If the heuristic source finder fails, one then has to, at the current stage of development, inspect manually whether or not suitable sources can be found and where to place them.
The problem of identifying sources can be improved on in the future, e.g. by tackling it algorithmically by solving a convex problem.
This is similar to the identification of the overlap structure of ellipsoids for the construction of a contour deformation in \cite{Capatti_2020}, although generally less restrictive.
\par

The evaluation of the integrand relies on cancellations between many orders of magnitude, potentially in the vicinity of integrable singularities.
This could cause numerical instabilities as floating point arithmetic breaks down.
This instability threshold can be shifted by raising the machine precision from double \texttt{f64} to quadruple \texttt{f128}.
For the one-loop integrals shown below this has however not made a significant difference and the results below were therefore generated with much faster \texttt{f64} precision.
\par

However, the particular LTD representation one uses has a large impact on numerical stability \cite{Capatti_2020,capatti2020manifestly,Ram_rez_Uribe_2021}.
In principle, the representation of dual integrands as in eq.~\eqref{eq:one_loop_ltd_averaged} should be avoided as it contains spurious H-surface singularities.
This is especially critical in the ultraviolet region, where individual dual integrands are generally divergent.
Instead one should employ the \emph{causal} LTD \cite{capatti2020manifestly,Aguilera_Verdugo_2021,Ram_rez_Uribe_2021,bobadilla2021lotty,Torres_Bobadilla_2021,Sborlini_2021} representation, where the spurious singularities of dual integrands were algebraically removed.
Nevertheless, for the results presented in this article, we stuck to the representation of dual integrands.
In fact, numerical instabilities were negligible for all integrals we considered except for the amplitudes in sect.~\ref{sec:numerator_ir_singularities}, where it was necessary to suppress probing the unstable ultraviolet region.
This was achieved through importance sampling of the Monte~Carlo samples in the hypercube such that the spatial loop momenta grow logarithmically. 
\par

To improve the convergence of the Monte Carlo integration, we can reduce the variance of the integrand.
Apart from \textsc{Vegas}' importance sampling and adaptive stratified sampling, no specially designed flattening of the integrand is employed.
Since the imaginary part of the integral is given by a sum of phase space integrals, it is conceivable to profit from the common techniques of tree-level phase space generators, such as specialised propagator mappings \cite{Roth:1999kk,Dittmaier:2002ap,Denner_1999} to further flatten the integrand through importance sampling combined with multi-channelling using adaptive weight optimization \cite{Kleiss_1994}.
An implementation of these optimisations is left for future work.
\par

We emphasise that the placement of the sources can have a direct impact on the variance of the integrands of both the real and the imaginary part of the loop integral.
In particular, there are two ways integrable singularities are introduced to the integrand:
1. a source outside the maximal overlap of a residue group gives rise to integrable singualities at tangents of E-surfaces and 
2. each residue group has an integrable singularity at the source (applies to the real part only).
Note that the Jacobian of the spherical parameterisation can remove one of the integrable singularities of the second type.
It is therefore of advantage to place the source in the maximal overlap wherever possible and to merge groups if they allow a common source in their maximal overlap.
Since the integrable singularities at the sources of a residue group are located at single points, they could be easily removed with a multi-channelling approach.
\par

Furthermore, as we will see below the Monte~Carlo estimates are generally more accurate for the imaginary part than the real part of the integral.
This is mainly attributed to the smaller dimension of the imaginary integral.
Note also that the integral for the real part is more susceptible to numerical instabilities than the one for the imaginary part.
This is because the subtracted integrand not only realises cancellations between the LTD integrand and the counterterms but also carries over the cancelling regions among the residues to the counterterms.
\par

Finally, one may want to explore other numerical integration rules with our subtraction method, such as the deterministic \textsc{Cuhre} algorithm \cite{10.1145/210232.210233}, which for low dimensions usually outperforms Monte~Carlo based methods.
Indeed, for our one-loop integrals in two and three dimensions we generally found much faster convergence with \textsc{Cuhre} than with \textsc{Vegas}.
Nevertheless, we decided to present \textsc{Vegas} results only to enable a future comparison with multi-loop results.
\par

\subsection{Single residue groups with interior sources}
\label{sec:single_group}

In tab.~\ref{tab:single_group} we give the numerical results of various scalar one-loop topologies, whose kinematic configurations allow for a single group of residues, whose E-surfaces are all parameterised in the same reference frame.
Furthermore, the kinematics are such that a source can be found in the interior of all E-surfaces, such that local cancellation of all threshold singularities is guaranteed.
Integrals of this type define the simplest use case of our method.
\par

The phase space points in this and the next section starting with \texttt{P} and \texttt{1L} are taken from \cite{buchta2015numerical} and \cite{Capatti_2020}, respectively, where the corresponding integrals were evaluated using contour deformation
The respective reference results were obtained with \textsc{LoopTools} \cite{Hahn_1999} and \textsc{MadLoop} \cite{Hirschi_2011,Alwall_2014}.
\par

\begin{table}[H]
\centering
\resizebox{\columnwidth}{!}{%
\texttt{%
\begin{tabular}{clrrrrcllcrrr}
\hline
Topology & Kin. & $\mathtt{N_{\text{E}}}$ & $\mathtt{N_{\text{G}}}$ & $\mathtt{N_{\text{G}}^{\text{max}}}$ & $\mathtt{N_{\text{P}}}$ & Phase & Exp. & Reference & Numerical & $\mathtt{\Delta \ [\sigma]}$ & $\mathtt{\Delta \ [\%]}$ & $\mathtt{\Delta \ [\%] |\cdot|}$ \\
\hline
\multirow{4}{*}{%
\begin{tabular}{@{}c@{}}
\begin{tikzpicture}[scale=1]
	\pgfmathsetmacro\n{3}
	\pgfmathsetmacro\radius{0.4}
	\foreach \i in {0,...,\n} {
		\pgfmathsetmacro\r{(\i-0.25)*(360/\n)}
		\fill (\r:\radius) circle (1.5pt);
		\draw[thick] (\radius,0) arc (0:\r:\radius);
		\draw[thick] (\r:\radius) -- (\r:\radius*3/2);
}
\end{tikzpicture} \\
Triangle\end{tabular}}
& \multirow{2}{*}{P3}& \multirow{2}{*}{1}& \multirow{2}{*}{1}& \multirow{2}{*}{1}& $\mathtt{10^9}$& Re& \multirow{2}{*}{-04}& \texttt{ 5.372715}& \texttt{~5.372733~+/-~0.000378}& 0.048& 3e-04& \multirow{2}{*}{2e-04}\\
& & & & & $\mathtt{10^9}$& Im& & \texttt{-6.681071}& \texttt{-6.681071~+/-~0.000103}& 5e-04& 7e-07& \\
\cline{2-13}
& \multirow{2}{*}{P4}& \multirow{2}{*}{1}& \multirow{2}{*}{1}& \multirow{2}{*}{1}& $\mathtt{10^9}$& Re& \multirow{2}{*}{-06}& \texttt{-0.561372}& \texttt{-0.561372~+/-~0.000014}& 0.014& 3e-05& \multirow{2}{*}{2e-05}\\
& & & & & $\mathtt{10^9}$& Im& & \texttt{-1.016658}& \texttt{-1.016658~+/-~0.000006}& 6e-05& 4e-08& \\
\hline
\multirow{16}{*}{%
\begin{tabular}{@{}c@{}}
\begin{tikzpicture}[scale=1]
	\pgfmathsetmacro\n{4}
	\pgfmathsetmacro\radius{0.4}
	\foreach \i in {0,...,\n} {
		\pgfmathsetmacro\r{(\i+0.5)*(360/\n)}
		\fill (\r:\radius) circle (1.5pt);
		\draw[thick] (\radius,0) arc (0:\r:\radius);
		\draw[thick] (\r:\radius) -- (\r:\radius*3/2);
}
\end{tikzpicture} \\
Box\end{tabular}}
& \multirow{2}{*}{P7}& \multirow{2}{*}{4}& \multirow{2}{*}{1}& \multirow{2}{*}{2}& $\mathtt{10^9}$& Re& \multirow{2}{*}{-10}& \texttt{-2.387669}& \texttt{-2.387670~+/-~0.000087}& 0.018& 7e-05& \multirow{2}{*}{5e-05}\\
& & & & & $\mathtt{10^9}$& Im& & \texttt{-3.030811}& \texttt{-3.030812~+/-~0.000022}& 0.048& 4e-05& \\
\cline{2-13}
& \multirow{2}{*}{P8}& \multirow{2}{*}{2}& \multirow{2}{*}{1}& \multirow{2}{*}{1}& $\mathtt{10^9}$& Re& \multirow{2}{*}{-11}& \texttt{-4.271184}& \texttt{-4.271188~+/-~0.000117}& 0.039& 1e-04& \multirow{2}{*}{1e-04}\\
& & & & & $\mathtt{10^9}$& Im& & \texttt{ 4.493049}& \texttt{~4.493044~+/-~0.000065}& 0.078& 1e-04& \\
\cline{2-13}
& \multirow{2}{*}{P9}& \multirow{2}{*}{3}& \multirow{2}{*}{1}& \multirow{2}{*}{2}& $\mathtt{10^9}$& Re& \multirow{2}{*}{-10}& \texttt{-0.737898}& \texttt{-0.737896~+/-~0.000027}& 0.065& 2e-04& \multirow{2}{*}{1e-04}\\
& & & & & $\mathtt{10^9}$& Im& & \texttt{-1.196570}& \texttt{-1.196570~+/-~0.000003}& 0.018& 4e-06& \\
\cline{2-13}
& \multirow{2}{*}{P10}& \multirow{2}{*}{2}& \multirow{2}{*}{1}& \multirow{2}{*}{1}& $\mathtt{10^9}$& Re& \multirow{2}{*}{-10}& \texttt{-1.855449}& \texttt{-1.855447~+/-~0.000093}& 0.018& 9e-05& \multirow{2}{*}{8e-05}\\
& & & & & $\mathtt{10^9}$& Im& & \texttt{ 2.135544}& \texttt{~2.135545~+/-~0.000037}& 0.039& 7e-05& \\
\cline{2-13}
& \multirow{2}{*}{1L4P.K1}& \multirow{2}{*}{5}& \multirow{2}{*}{1}& \multirow{2}{*}{3}& $\mathtt{10^9}$& Re& \multirow{2}{*}{-03}& \texttt{-0.554879}& \texttt{-0.554876~+/-~0.000055}& 0.050& 5e-04& \multirow{2}{*}{2e-04}\\
& & & & & $\mathtt{10^9}$& Im& & \texttt{-1.131167}& \texttt{-1.131167~+/-~0.000017}& 0.010& 2e-05& \\
\cline{2-13}
& \multirow{2}{*}{1L4P.K2}& \multirow{2}{*}{5}& \multirow{2}{*}{1}& \multirow{2}{*}{3}& $\mathtt{10^9}$& Re& \multirow{2}{*}{-05}& \texttt{-7.240038}& \texttt{-7.239970~+/-~0.000596}& 0.114& 0.001& \multirow{2}{*}{0.001}\\
& & & & & $\mathtt{10^9}$& Im& & \texttt{-5.719278}& \texttt{-5.719281~+/-~0.000148}& 0.020& 5e-05& \\
\cline{2-13}
& \multirow{2}{*}{1L4P.K3}& \multirow{2}{*}{5}& \multirow{2}{*}{1}& \multirow{2}{*}{3}& $\mathtt{10^9}$& Re& \multirow{2}{*}{-06}& \texttt{-2.069838}& \texttt{-2.069871~+/-~0.000168}& 0.194& 0.002& \multirow{2}{*}{0.001}\\
& & & & & $\mathtt{10^9}$& Im& & \texttt{-1.553783}& \texttt{-1.553779~+/-~0.000035}& 0.116& 3e-04& \\
\cline{2-13}
& \multirow{2}{*}{1L4P.K3*}& \multirow{2}{*}{3}& \multirow{2}{*}{1}& \multirow{2}{*}{3}& $\mathtt{10^9}$& Re& \multirow{2}{*}{-06}& \texttt{-2.276333}& \texttt{-2.276333~+/-~0.000102}& 0.002& 1e-05& \multirow{2}{*}{1e-05}\\
& & & & & $\mathtt{10^9}$& Im& & \texttt{ 0.179910}& \texttt{~0.179910~+/-~0.000024}& 2e-05& 3e-07& \\
\hline
\multirow{10}{*}{%
\begin{tabular}{@{}c@{}}
\begin{tikzpicture}[scale=1]
	\pgfmathsetmacro\n{5}
	\pgfmathsetmacro\radius{0.4}
	\foreach \i in {0,...,\n} {
		\pgfmathsetmacro\r{(\i+0.25)*(360/\n)}
		\fill (\r:\radius) circle (1.5pt);
		\draw[thick] (\radius,0) arc (0:\r:\radius);
		\draw[thick] (\r:\radius) -- (\r:\radius*3/2);
}
\end{tikzpicture} \\
Pentagon\end{tabular}}
& \multirow{2}{*}{P13}& \multirow{2}{*}{4}& \multirow{2}{*}{1}& \multirow{2}{*}{1}& $\mathtt{10^9}$& Re& \multirow{2}{*}{-11}& \texttt{ 1.023505}& \texttt{~1.023505~+/-~0.000052}& 0.001& 8e-06& \multirow{2}{*}{9e-05}\\
& & & & & $\mathtt{10^9}$& Im& & \texttt{ 1.403822}& \texttt{~1.403823~+/-~0.000011}& 0.152& 1e-04& \\
\cline{2-13}
& \multirow{2}{*}{P14}& \multirow{2}{*}{7}& \multirow{2}{*}{1}& \multirow{2}{*}{2}& $\mathtt{10^9}$& Re& \multirow{2}{*}{-14}& \texttt{-0.153898}& \texttt{-0.153897~+/-~0.000010}& 0.086& 0.001& \multirow{2}{*}{9e-05}\\
& & & & & $\mathtt{10^9}$& Im& & \texttt{-1.037575}& \texttt{-1.037575~+/-~0.000008}& 0.016& 1e-05& \\
\cline{2-13}
& \multirow{2}{*}{P15}& \multirow{2}{*}{2}& \multirow{2}{*}{1}& \multirow{2}{*}{1}& $\mathtt{10^9}$& Re& \multirow{2}{*}{-14}& \texttt{-0.429464}& \texttt{-0.429465~+/-~0.000492}& 0.002& 2e-04& \multirow{2}{*}{2e-04}\\
& & & & & $\mathtt{10^9}$& Im& & \texttt{-6.554400}& \texttt{-6.554412~+/-~0.000051}& 0.229& 2e-04& \\
\cline{2-13}
& \multirow{2}{*}{1L5P.I}& \multirow{2}{*}{8}& \multirow{2}{*}{1}& \multirow{2}{*}{4}& $\mathtt{10^9}$& Re& \multirow{2}{*}{-12}& \texttt{-2.564923}& \texttt{-2.564918~+/-~0.000489}& 0.009& 2e-04& \multirow{2}{*}{0.001}\\
& & & & & $\mathtt{10^9}$& Im& & \texttt{ 3.443330}& \texttt{~3.443352~+/-~0.000239}& 0.094& 0.001& \\
\cline{2-13}
& \multirow{2}{*}{1L5P.II}& \multirow{2}{*}{10}& \multirow{2}{*}{1}& \multirow{2}{*}{10}& $\mathtt{10^9}$& Re& \multirow{2}{*}{-13}& \texttt{ 5.971432}& \texttt{~5.971441~+/-~0.000223}& 0.039& 1e-04& \multirow{2}{*}{1e-04}\\
& & & & & $\mathtt{10^9}$& Im& & \phantom{+}\texttt{0}\phantom{.00000}& \texttt{~0.000000~+/-~0.000137}& 1e-04&  & \\
\hline
\multirow{2}{*}{%
\begin{tabular}{@{}c@{}}
Hexagon\end{tabular}}
& \multirow{2}{*}{1L6P.I}& \multirow{2}{*}{12}& \multirow{2}{*}{1}& \multirow{2}{*}{6}& $\mathtt{10^9}$& Re& \multirow{2}{*}{-13}& \texttt{-1.176800}& \texttt{-1.176797~+/-~0.000080}& 0.028& 2e-04& \multirow{2}{*}{2e-04}\\
& & & & & $\mathtt{10^9}$& Im& & \texttt{-0.030396}& \texttt{-0.030397~+/-~0.000022}& 0.055& 0.004& \\
\end{tabular}%
}%
}%
\caption{
\label{tab:single_group}
Scalar one-loop integrals for various kinematic configurations that allow for the direct application of eqs.~\eqref{eq:subtracted_ltd} and \eqref{eq:fract_residues_ltd}, without the need of grouping residues and counterterms.
Here, it was possible to place the origin in the interior of all E-surfaces such that no integrable singularities were introduced to the integrand.
}%
\end{table}

The topologies are ordered in complexity, starting with scalar triangles up to a hexagon, with increasing number of E-surfaces $N_E$.
For the topologies in this section a single residue group $N_G=1$ could be found, although the kinematics may allow for more $N_G^{\rm max}$.
The numerical estimate after $N_P=10^9$ Monte Carlo samples agrees with the reference values with a relative accuracy of at least $\Delta\, [\%] |\cdot| \leq 0.01$\textperthousand~on the magnitude.
With $\Delta \, [\sigma]$ we denote the deviation in units of the Monte~Carlo error between the numerical estimate and the reference and with $\Delta \, [\%]$ their relative error.
\par

\subsection{Multiple residue groups with interior sources}
\label{sec:multiple_groups}

In general, the threshold structure of one-loop topologies is such that several groups of independent E-surface residues exist.
This can be exploited to parameterise each group of E-surfaces in their own coorinate system, as described in sect.~\ref{sec:e_surface_grouping}.
Thanks to this grouping we can avoid miscancelling poles that would be unavoidable if all E-surfaces were parameterised with respect to a single source.
\par

In tab.~\ref{tab:multiple_groups} we present scalar integrals with up to eight external legs that require such a grouping.
Moreover, they correspond to kinematic configurations, whose residue groups are such that each group's source is in the interior of all its E-surfaces.
This guarantees that all remaining threshold singularities of E-surface residues and corresponding counterterms locally cancel.
\par

Most of the topologies in \cite{Capatti_2020}, starting with \texttt{1L}, fall into that category.
However, for the topologies \texttt{Box4E}, \texttt{1L5P.V}, \texttt{1L6P.V} and \texttt{1L6P.VI} no such sources exist.
The topology \texttt{Box4E} allows an exterior sources, as shown in sect.~\ref{sec:exterior_sources}.
However, the three other topologies always have locally pinched poles with our radial E-surface parameterisation and therefore cannot be directly integrated with our framework.
Possible solutions are discussed in the outlook in sect.~\ref{sec:conclusion}.
\par

In principle, the kinematic configurations allow for $N_G^{\text{max}}$ independent groups.
We however merged groups whenever possible, since each residue group introduces an integrable singularity at the source in the real part of the integral\footnote{
One integrable singularity can be removed by the Jacobian of the spherical coordinate system with origin at the source.
Moreover, by placing the origin at the source of the smallest E-surface one can avoid integrable singularities in vicinity of locally cancelling regions.
}.
This improves the variance of the integrand and therefore the Monte~Carlo estimate.
\par

The topologies are roughly ordered in complexity, starting with scalar boxes up to octagons, with increasing number of E-surfaces $N_E$ and groups $N_G$.
The numerical estimate after $N_P=10^9$ Monte~Carlo samples agrees with the reference values with a relative accuracy $\Delta\, [\%] |\cdot|$ of below one percent on the magnitude.
The deviation $\Delta \, [\sigma]$ of all numerical estimates remains below one sigma.
There is a tendency of increasing relative Monte~Carlo error for increasing number of E-surfaces and groups.
\par

\begin{table}[H]
\centering
\resizebox{\columnwidth}{!}{%
\texttt{%
\begin{tabular}{clrrrrcllcrrr}
\hline
Topology & Kin. & $\mathtt{N_{\text{E}}}$ & $\mathtt{N_{\text{G}}}$ & $\mathtt{N_{\text{G}}^{\text{max}}}$ & $\mathtt{N_{\text{P}}}$ & Phase & Exp. & Reference & Numerical & $\mathtt{\Delta \ [\sigma]}$ & $\mathtt{\Delta \ [\%]}$ & $\mathtt{\Delta \ [\%] |\cdot|}$ \\
\hline
\multirow{4}{*}{%
\begin{tabular}{@{}c@{}}
\begin{tikzpicture}[scale=1]
	\pgfmathsetmacro\n{4}
	\pgfmathsetmacro\radius{0.4}
	\foreach \i in {0,...,\n} {
		\pgfmathsetmacro\r{(\i+0.5)*(360/\n)}
		\fill (\r:\radius) circle (1.5pt);
		\draw[thick] (\radius,0) arc (0:\r:\radius);
		\draw[thick] (\r:\radius) -- (\r:\radius*3/2);
}
\end{tikzpicture} \\
Box\end{tabular}}
& \multirow{2}{*}{1L4P.K1*}& \multirow{2}{*}{5}& \multirow{2}{*}{2}& \multirow{2}{*}{5}& $\mathtt{10^9}$& Re& \multirow{2}{*}{-03}& \texttt{-2.184002}& \texttt{-2.183949~+/-~0.000102}& 0.525& 0.002& \multirow{2}{*}{0.002}\\
& & & & & $\mathtt{10^9}$& Im& & \texttt{-1.852246}& \texttt{-1.852246~+/-~0.000016}& 3e-04& 3e-07& \\
\cline{2-13}
& \multirow{2}{*}{1L4P.K2*}& \multirow{2}{*}{5}& \multirow{2}{*}{2}& \multirow{2}{*}{5}& $\mathtt{10^9}$& Re& \multirow{2}{*}{-04}& \texttt{-1.081252}& \texttt{-1.081228~+/-~0.000052}& 0.459& 0.002& \multirow{2}{*}{0.002}\\
& & & & & $\mathtt{10^9}$& Im& & \texttt{-0.302700}& \texttt{-0.302700~+/-~0.000009}& 3e-04& 9e-07& \\
\hline
\multirow{18}{*}{%
\begin{tabular}{@{}c@{}}
\begin{tikzpicture}[scale=1]
	\pgfmathsetmacro\n{5}
	\pgfmathsetmacro\radius{0.4}
	\foreach \i in {0,...,\n} {
		\pgfmathsetmacro\r{(\i+0.25)*(360/\n)}
		\fill (\r:\radius) circle (1.5pt);
		\draw[thick] (\radius,0) arc (0:\r:\radius);
		\draw[thick] (\r:\radius) -- (\r:\radius*3/2);
}
\end{tikzpicture} \\
Pentagon\end{tabular}}
& \multirow{2}{*}{1L5P.III}& \multirow{2}{*}{8}& \multirow{2}{*}{2}& \multirow{2}{*}{4}& $\mathtt{10^9}$& Re& \multirow{2}{*}{-12}& \texttt{-1.713406}& \texttt{-1.713344~+/-~0.000218}& 0.286& 0.004& \multirow{2}{*}{0.003}\\
& & & & & $\mathtt{10^9}$& Im& & \texttt{ 0.839048}& \texttt{~0.839054~+/-~0.000051}& 0.114& 0.001& \\
\cline{2-13}
& \multirow{2}{*}{1L5P.IV}& \multirow{2}{*}{8}& \multirow{2}{*}{2}& \multirow{2}{*}{4}& $\mathtt{10^9}$& Re& \multirow{2}{*}{-12}& \texttt{-3.900140}& \texttt{-3.899798~+/-~0.000921}& 0.371& 0.009& \multirow{2}{*}{0.007}\\
& & & & & $\mathtt{10^9}$& Im& & \texttt{ 3.489942}& \texttt{~3.489974~+/-~0.000222}& 0.142& 0.001& \\
\cline{2-13}
& \multirow{2}{*}{1L5P.VI}& \multirow{2}{*}{8}& \multirow{2}{*}{3}& \multirow{2}{*}{4}& $\mathtt{10^9}$& Re& \multirow{2}{*}{-13}& \texttt{-2.180572}& \texttt{-2.180520~+/-~0.000242}& 0.214& 0.002& \multirow{2}{*}{0.002}\\
& & & & & $\mathtt{10^9}$& Im& & \texttt{-0.041187}& \texttt{-0.041197~+/-~0.000070}& 0.145& 0.025& \\
\cline{2-13}
& \multirow{2}{*}{1L5P.K1}& \multirow{2}{*}{8}& \multirow{2}{*}{2}& \multirow{2}{*}{4}& $\mathtt{10^9}$& Re& \multirow{2}{*}{-05}& \texttt{-6.453455}& \texttt{-6.453245~+/-~0.001131}& 0.186& 0.003& \multirow{2}{*}{0.003}\\
& & & & & $\mathtt{10^9}$& Im& & \texttt{ 1.908472}& \texttt{~1.908451~+/-~0.000325}& 0.063& 0.001& \\
\cline{2-13}
& \multirow{2}{*}{1L5P.K2}& \multirow{2}{*}{8}& \multirow{2}{*}{2}& \multirow{2}{*}{4}& $\mathtt{10^9}$& Re& \multirow{2}{*}{-06}& \texttt{-1.806787}& \texttt{-1.806439~+/-~0.000669}& 0.520& 0.019& \multirow{2}{*}{0.019}\\
& & & & & $\mathtt{10^9}$& Im& & \texttt{ 0.151074}& \texttt{~0.151080~+/-~0.000106}& 0.054& 0.004& \\
\cline{2-13}
& \multirow{2}{*}{1L5P.K3}& \multirow{2}{*}{8}& \multirow{2}{*}{2}& \multirow{2}{*}{4}& $\mathtt{10^9}$& Re& \multirow{2}{*}{-09}& \texttt{-1.235314}& \texttt{-1.235031~+/-~0.000516}& 0.549& 0.023& \multirow{2}{*}{0.020}\\
& & & & & $\mathtt{10^9}$& Im& & \texttt{ 0.662398}& \texttt{~0.662402~+/-~0.000106}& 0.032& 0.001& \\
\cline{2-13}
& \multirow{2}{*}{1L5P.K1*}& \multirow{2}{*}{8}& \multirow{2}{*}{3}& \multirow{2}{*}{5}& $\mathtt{10^9}$& Re& \multirow{2}{*}{-05}& \texttt{-7.949164}& \texttt{-7.948492~+/-~0.000765}& 0.879& 0.008& \multirow{2}{*}{0.008}\\
& & & & & $\mathtt{10^9}$& Im& & \texttt{-2.603962}& \texttt{-2.603963~+/-~0.000226}& 0.008& 7e-05& \\
\cline{2-13}
& \multirow{2}{*}{1L5P.K2*}& \multirow{2}{*}{8}& \multirow{2}{*}{4}& \multirow{2}{*}{6}& $\mathtt{10^9}$& Re& \multirow{2}{*}{-06}& \texttt{-3.276952}& \texttt{-3.276448~+/-~0.000546}& 0.923& 0.015& \multirow{2}{*}{0.015}\\
& & & & & $\mathtt{10^9}$& Im& & \texttt{ 0.483021}& \texttt{~0.483015~+/-~0.000104}& 0.060& 0.001& \\
\cline{2-13}
& \multirow{2}{*}{1L5P.K3*}& \multirow{2}{*}{6}& \multirow{2}{*}{2}& \multirow{2}{*}{6}& $\mathtt{10^9}$& Re& \multirow{2}{*}{-09}& \texttt{-1.531295}& \texttt{-1.531188~+/-~0.000365}& 0.294& 0.007& \multirow{2}{*}{0.005}\\
& & & & & $\mathtt{10^9}$& Im& & \texttt{ 1.214974}& \texttt{~1.214974~+/-~0.000093}& 1e-04& 8e-07& \\
\hline
\multirow{26}{*}{%
\begin{tabular}{@{}c@{}}
\begin{tikzpicture}[scale=1]
	\pgfmathsetmacro\n{6}
	\pgfmathsetmacro\radius{0.4}
	\foreach \i in {0,...,\n} {
		\pgfmathsetmacro\r{\i*(360/\n)}
		\fill (\r:\radius) circle (1.5pt);
		\draw[thick] (\radius,0) arc (0:\r:\radius);
		\draw[thick] (\r:\radius) -- (\r:\radius*3/2);
}
\end{tikzpicture} \\
Hexagon\end{tabular}}
& \multirow{2}{*}{1L6P.II}& \multirow{2}{*}{6}& \multirow{2}{*}{2}& \multirow{2}{*}{4}& $\mathtt{10^9}$& Re& \multirow{2}{*}{+01}& \texttt{ 0.423433}& \texttt{~0.423433~+/-~0.000082}& 0.003& 6e-05& \multirow{2}{*}{2e-04}\\
& & & & & $\mathtt{10^9}$& Im& & \texttt{ 2.070140}& \texttt{~2.070144~+/-~0.000078}& 0.055& 2e-04& \\
\cline{2-13}
& \multirow{2}{*}{1L6P.III}& \multirow{2}{*}{12}& \multirow{2}{*}{2}& \multirow{2}{*}{6}& $\mathtt{10^9}$& Re& \multirow{2}{*}{-15}& \texttt{-2.259010}& \texttt{-2.258576~+/-~0.000623}& 0.698& 0.019& \multirow{2}{*}{0.016}\\
& & & & & $\mathtt{10^9}$& Im& & \texttt{-1.369185}& \texttt{-1.369196~+/-~0.000186}& 0.062& 0.001& \\
\cline{2-13}
& \multirow{2}{*}{1L6P.IV}& \multirow{2}{*}{12}& \multirow{2}{*}{2}& \multirow{2}{*}{6}& $\mathtt{10^9}$& Re& \multirow{2}{*}{-15}& \texttt{-2.165894}& \texttt{-2.170801~+/-~0.012222}& 0.401& 0.227& \multirow{2}{*}{0.194}\\
& & & & & $\mathtt{10^9}$& Im& & \texttt{-1.297703}& \texttt{-1.297724~+/-~0.001807}& 0.011& 0.002& \\
\cline{2-13}
& \multirow{2}{*}{1L6P.VII}& \multirow{2}{*}{10}& \multirow{2}{*}{2}& \multirow{2}{*}{2}& $\mathtt{10^9}$& Re& \multirow{2}{*}{-17}& \texttt{-7.733373}& \texttt{-7.733390~+/-~0.000743}& 0.023& 2e-04& \multirow{2}{*}{2e-04}\\
& & & & & $\mathtt{10^9}$& Im& & \texttt{ 3.019394}& \texttt{~3.019388~+/-~0.000711}& 0.008& 2e-04& \\
\cline{2-13}
& \multirow{2}{*}{1L6P.VIII}& \multirow{2}{*}{10}& \multirow{2}{*}{2}& \multirow{2}{*}{4}& $\mathtt{10^9}$& Re& \multirow{2}{*}{-02}& \texttt{ 0.640326}& \texttt{~0.643323~+/-~0.006342}& 0.473& 0.468& \multirow{2}{*}{0.143}\\
& & & & & $\mathtt{10^9}$& Im& & \texttt{-2.119278}& \texttt{-2.120308~+/-~0.001200}& 0.859& 0.049& \\
\cline{2-13}
& \multirow{2}{*}{1L6P.IX}& \multirow{2}{*}{12}& \multirow{2}{*}{3}& \multirow{2}{*}{6}& $\mathtt{10^9}$& Re& \multirow{2}{*}{-14}& \texttt{-1.152818}& \texttt{-1.152711~+/-~0.000139}& 0.768& 0.009& \multirow{2}{*}{0.009}\\
& & & & & $\mathtt{10^9}$& Im& & \texttt{-0.007940}& \texttt{-0.007937~+/-~0.000034}& 0.083& 0.036& \\
\cline{2-13}
& \multirow{2}{*}{1L6P.X}& \multirow{2}{*}{10}& \multirow{2}{*}{3}& \multirow{2}{*}{4}& $\mathtt{10^9}$& Re& \multirow{2}{*}{+00}& \texttt{ 2.473271}& \texttt{~2.473398~+/-~0.000228}& 0.558& 0.005& \multirow{2}{*}{0.003}\\
& & & & & $\mathtt{10^9}$& Im& & \texttt{ 2.814754}& \texttt{~2.814756~+/-~0.000032}& 0.060& 7e-05& \\
\cline{2-13}
& \multirow{2}{*}{1L6P.K1}& \multirow{2}{*}{12}& \multirow{2}{*}{2}& \multirow{2}{*}{6}& $\mathtt{10^9}$& Re& \multirow{2}{*}{-06}& \texttt{-1.547556}& \texttt{-1.547394~+/-~0.000255}& 0.634& 0.010& \multirow{2}{*}{0.010}\\
& & & & & $\mathtt{10^9}$& Im& & \texttt{-0.510252}& \texttt{-0.510248~+/-~0.000069}& 0.067& 0.001& \\
\cline{2-13}
& \multirow{2}{*}{1L6P.K2}& \multirow{2}{*}{12}& \multirow{2}{*}{2}& \multirow{2}{*}{6}& $\mathtt{10^9}$& Re& \multirow{2}{*}{-08}& \texttt{-6.963373}& \texttt{-6.960169~+/-~0.007156}& 0.448& 0.046& \multirow{2}{*}{0.046}\\
& & & & & $\mathtt{10^9}$& Im& & \texttt{-0.604400}& \texttt{-0.604415~+/-~0.000446}& 0.033& 0.002& \\
\cline{2-13}
& \multirow{2}{*}{1L6P.K3}& \multirow{2}{*}{12}& \multirow{2}{*}{2}& \multirow{2}{*}{6}& $\mathtt{10^9}$& Re& \multirow{2}{*}{-12}& \texttt{-2.519558}& \texttt{-2.519262~+/-~0.002864}& 0.103& 0.012& \multirow{2}{*}{0.012}\\
& & & & & $\mathtt{10^9}$& Im& & \texttt{-0.406595}& \texttt{-0.406598~+/-~0.000198}& 0.013& 0.001& \\
\cline{2-13}
& \multirow{2}{*}{1L6P.K1*}& \multirow{2}{*}{12}& \multirow{2}{*}{4}& \multirow{2}{*}{8}& $\mathtt{10^9}$& Re& \multirow{2}{*}{-06}& \texttt{-2.273543}& \texttt{-2.273682~+/-~0.000416}& 0.334& 0.006& \multirow{2}{*}{0.005}\\
& & & & & $\mathtt{10^9}$& Im& & \texttt{-1.302132}& \texttt{-1.302148~+/-~0.000118}& 0.130& 0.001& \\
\cline{2-13}
& \multirow{2}{*}{1L6P.K2*}& \multirow{2}{*}{12}& \multirow{2}{*}{3}& \multirow{2}{*}{8}& $\mathtt{10^9}$& Re& \multirow{2}{*}{-08}& \texttt{-6.379295}& \texttt{-6.378242~+/-~0.003305}& 0.319& 0.017& \multirow{2}{*}{0.016}\\
& & & & & $\mathtt{10^9}$& Im& & \texttt{ 2.199344}& \texttt{~2.199348~+/-~0.000544}& 0.008& 2e-04& \\
\cline{2-13}
& \multirow{2}{*}{1L6P.K3*}& \multirow{2}{*}{10}& \multirow{2}{*}{4}& \multirow{2}{*}{10}& $\mathtt{10^9}$& Re& \multirow{2}{*}{-12}& \texttt{-2.216660}& \texttt{-2.216821~+/-~0.005975}& 0.027& 0.007& \multirow{2}{*}{0.006}\\
& & & & & $\mathtt{10^9}$& Im& & \texttt{ 1.281615}& \texttt{~1.281615~+/-~0.000671}& 2e-04& 9e-06& \\
\hline
\multirow{12}{*}{%
\begin{tabular}{@{}c@{}}
\begin{tikzpicture}[scale=1]
	\pgfmathsetmacro\n{8}
	\pgfmathsetmacro\radius{0.4}
	\foreach \i in {0,...,\n} {
		\pgfmathsetmacro\r{(\i+0.5)*(360/\n)}
		\fill (\r:\radius) circle (1.5pt);
		\draw[thick] (\radius,0) arc (0:\r:\radius);
		\draw[thick] (\r:\radius) -- (\r:\radius*3/2);
}
\end{tikzpicture} \\
Octagon\end{tabular}}
& \multirow{2}{*}{1L8P.K1}& \multirow{2}{*}{23}& \multirow{2}{*}{3}& \multirow{2}{*}{13}& $\mathtt{10^9}$& Re& \multirow{2}{*}{-10}& \texttt{-1.627993}& \texttt{-1.624523~+/-~0.005629}& 0.616& 0.213& \multirow{2}{*}{0.065}\\
& & & & & $\mathtt{10^9}$& Im& & \texttt{-5.099169}& \texttt{-5.098951~+/-~0.001718}& 0.127& 0.004& \\
\cline{2-13}
& \multirow{2}{*}{1L8P.K2}& \multirow{2}{*}{23}& \multirow{2}{*}{4}& \multirow{2}{*}{13}& $\mathtt{10^9}$& Re& \multirow{2}{*}{-12}& \texttt{-1.952888}& \texttt{-1.941883~+/-~0.018752}& 0.587& 0.563& \multirow{2}{*}{0.237}\\
& & & & & $\mathtt{10^9}$& Im& & \texttt{-4.209151}& \texttt{-4.209013~+/-~0.002236}& 0.062& 0.003& \\
\cline{2-13}
& \multirow{2}{*}{1L8P.K3}& \multirow{2}{*}{23}& \multirow{2}{*}{4}& \multirow{2}{*}{13}& $\mathtt{10^9}$& Re& \multirow{2}{*}{-19}& \texttt{-0.825671}& \texttt{-0.824363~+/-~0.003329}& 0.393& {0.159}& \multirow{2}{*}{0.087}\\
& & & & & $\mathtt{10^9}$& Im& & \texttt{-1.273792}& \texttt{-1.273663~+/-~0.000348}& 0.369& 0.010& \\
\cline{2-13}
& \multirow{2}{*}{1L8P.K1*}& \multirow{2}{*}{23}& \multirow{2}{*}{4}& \multirow{2}{*}{16}& $\mathtt{10^9}$& Re& \multirow{2}{*}{-09}& \texttt{-1.468062}& \texttt{-1.467656~+/-~0.000472}& 0.860& 0.028& \multirow{2}{*}{0.027}\\
& & & & & $\mathtt{10^9}$& Im& & \texttt{ 0.356928}& \texttt{~0.356952~+/-~0.000141}& 0.170& 0.007& \\
\cline{2-13}
& \multirow{2}{*}{1L8P.K2*}& \multirow{2}{*}{23}& \multirow{2}{*}{5}& \multirow{2}{*}{16}& $\mathtt{10^9}$& Re& \multirow{2}{*}{-12}& \texttt{-2.705869}& \texttt{-2.699826~+/-~0.010242}& 0.590& 0.223& \multirow{2}{*}{0.206}\\
& & & & & $\mathtt{10^9}$& Im& & \texttt{ 1.147180}& \texttt{~1.147210~+/-~0.001610}& 0.019& 0.003& \\
\cline{2-13}
& \multirow{2}{*}{1L8P.K3*}& \multirow{2}{*}{21}& \multirow{2}{*}{4}& \multirow{2}{*}{21}& $\mathtt{10^9}$& Re& \multirow{2}{*}{-08}& \texttt{-4.042207}& \texttt{-4.034787~+/-~0.007951}& 0.933& 0.184& \multirow{2}{*}{0.182}\\
& & & & & $\mathtt{10^9}$& Im& & \texttt{ 0.575146}& \texttt{~0.575152~+/-~0.001044}& 0.005& 0.001& \\
\end{tabular}%
}%
}%
\caption{\label{tab:multiple_groups}
Scalar one-loop integrals for various kinematic configurations that require grouping of counterterms and residues, according to eqs.~\eqref{eq:subtracted_ltd_grouped}  and \eqref{eq:fract_residues_ltd_grouped}, to fully realise local cancellations.
Here, it was possible to place each group's source in the interior of its E-surfaces such that the only integrable singularities are in the subtracted integrand located at isolated points, the sources.
}%
\end{table}

\subsection{Triacontagons}
\label{sec:tringigons}

To test the stability of our threshold subtraction method in the presence of a large number of threshold singularities, we apply it to a few triacontagons, i.e. one-loop integrals with 30 external legs, with physical kinematics in tab.~\ref{tab:triacontagons}.
Such kinematic configurations are very challenging for numerical integration based on contour deformation, in particular because the contour is likely to be in the close vicinity of singularities.
The example with over 400 E-surface and over 350 groups, which can be merged to 15 groups, show that in practice many E-surfaces are mostly contained in one another.
Despite the large multiplicity of external legs, the numerical estimate after $N_P=10^9$ Monte~Carlo samples agrees with the reference values with a relative accuracy $\Delta \, [\%]|\cdot|$ of below one percent on the magnitude.

\begin{table}[H]
\centering
\resizebox{\columnwidth}{!}{%
\texttt{%
\begin{tabular}{clrrrrcllcrrr}
\hline
Topology & Kin. & $\mathtt{N_{\text{E}}}$ & $\mathtt{N_{\text{G}}}$ & $\mathtt{N_{\text{G}}^{\text{max}}}$ & $\mathtt{N_{\text{P}}}$ & Phase & Exp. & Reference & Numerical & $\mathtt{\Delta \ [\sigma]}$ & $\mathtt{\Delta \ [\%]}$ & $\mathtt{\Delta \ [\%] |\cdot|}$ \\
\hline
\multirow{8}{*}{%
\begin{tabular}{@{}c@{}}
\begin{tikzpicture}[scale=1]
	\pgfmathsetmacro\n{30}
	\pgfmathsetmacro\radius{0.4}
	\foreach \i in {0,...,\n} {
		\pgfmathsetmacro\r{\i*(360/\n)}
		\fill (\r:\radius) circle (1.5pt);
		\draw[thick] (\radius,0) arc (0:\r:\radius);
		\draw[thick] (\r:\radius) -- (\r:\radius*3/2);
}
\end{tikzpicture} \\
Triacontagon\end{tabular}}
& \multirow{2}{*}{1L30P.I}& \multirow{2}{*}{5}& \multirow{2}{*}{1}& \multirow{2}{*}{1}& $\mathtt{10^9}$& Re& \multirow{2}{*}{-02}& \texttt{-1.007398}& \texttt{-1.007449~+/-~0.001467}& 0.035& 0.005& \multirow{2}{*}{0.002}\\
& & & & & $\mathtt{10^9}$& Im& & \texttt{ 3.175180}& \texttt{~3.175183~+/-~0.000085}& 0.030& 8e-05& \\
\cline{2-13}
& \multirow{2}{*}{1L30P.II}& \multirow{2}{*}{6}& \multirow{2}{*}{1}& \multirow{2}{*}{1}& $\mathtt{10^9}$& Re& \multirow{2}{*}{-12}& \texttt{-4.166377}& \texttt{-4.165527~+/-~0.006697}& 0.127& 0.020& \multirow{2}{*}{0.016}\\
& & & & & $\mathtt{10^9}$& Im& & \texttt{ 3.413930}& \texttt{~3.413917~+/-~0.000075}& 0.182& 4e-04& \\
\cline{2-13}
& \multirow{2}{*}{1L30P.III}& \multirow{2}{*}{408}& \multirow{2}{*}{15}& \multirow{2}{*}{354}& $\mathtt{10^9}$& Re& \multirow{2}{*}{-09}& \texttt{-2.991654}& \texttt{-2.984733~+/-~0.026977}& 0.257& 0.231& \multirow{2}{*}{0.231}\\
& & & & & $\mathtt{10^9}$& Im& & \texttt{-0.000000}& \texttt{-0.000001~+/-~0.003831}& 3e-04& & \\
\cline{2-13}
& \multirow{2}{*}{1L30P.IV}& \multirow{2}{*}{408}& \multirow{2}{*}{15}& \multirow{2}{*}{354}& $\mathtt{10^9}$& Re& \multirow{2}{*}{-07}& \texttt{-1.757748}& \texttt{-1.757913~+/-~0.002169}& 0.076& 0.009& \multirow{2}{*}{0.009}\\
& & & & & $\mathtt{10^9}$& Im& & \texttt{-0.000000}& \texttt{~0.000001~+/-~0.000199}& 0.007& & \\
\end{tabular}%
}%
}%
\caption{\label{tab:triacontagons}
Four kinematic configurations of scalar one-loop integrals with thirty external legs. The reference results were calculated with \textsc{MadLoop} \cite{Hirschi_2011,Alwall_2014}.}%
\end{table}

\subsection{Exterior sources}
\label{sec:exterior_sources}

The limitations of our method become manifest when no suitable source can be found for a residue group, in which case not all of the remaining threshold singularities locally cancel between the E-surface residues.
\par

\begin{figure}
    \centering
    \includegraphics[scale=1]{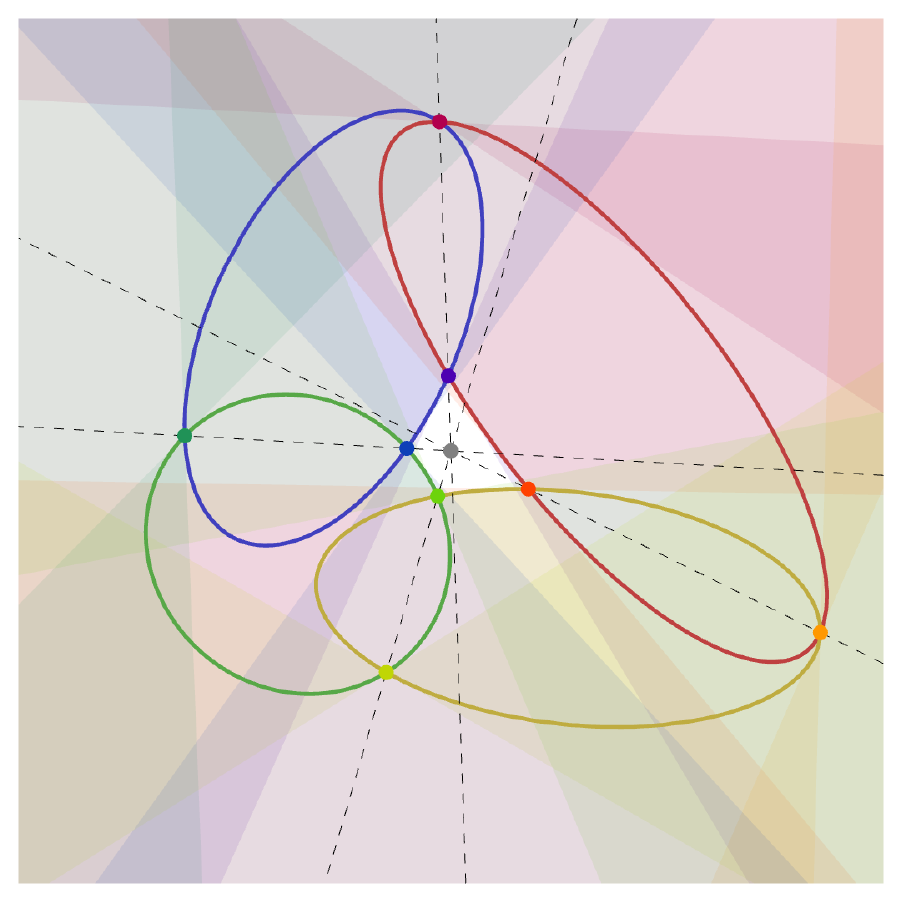}
    \caption{\label{fig:box_4e}
    The threshold singularities of \texttt{Box\_4E}.
    All E-surfaces belong to the same group, such that a common source has to be found.
    The tangents at each intersection define an $X$-shaped region.
    If the source were located within this region, the corresponding E-surface intersection would result in a miscancelling pole in the respective residues.
    We therefore exlcude the colorful regions.
    Only a source in the white region allows for the local cancellation of all poles among the E-surface residues.
    Curiously, the (dashed) lines connecting the two intersections of two ellipses all intersect in a single point (gray).
    }
\end{figure}

We demonstrate this issue with the example of a specific configuration of external momenta of a box diagram, the \texttt{Box\_4E} from \cite{Capatti_2019}.
Its E-surfaces all belong to the same group and have therefore to be integrated in the same reference frame.
As we can see in fig.~\ref{fig:box_4e}, it is not possible to find a source that is in the interior of all E-surfaces simultaneously.
This is in contrast to the groups in sect.~\ref{sec:single_group} and sect.~\ref{sec:multiple_groups}, where this was possible.
We are therefore forced to choose a source in the exterior of (at least some) E-surfaces.
Note however that this does not immediately imply that our method is not applicable.
Nevertheless, one has to search more carefully for a source, where only non-pinched higher-order poles exist, i.e. where plus-poles (i.e. the poles at $r_{i_\sigma j_{-\sigma}}^+$ with $\rho=+1$) only intersect with plus-poles (and analogously for $\rho=-1$).
\par

In this specific example, we can geometrically construct such a point in fig.~\ref{fig:box_4e}, and confirm that it integrates to the right value in tab.~\ref{tab:box_4e}.
However, the numerical error is larger since the integrand now features a range of integrable singularities at angles where the line through the source is tangent to an E-surface.
Note that these tangent regions can be in vicinity of locally cancelling regions (intersections of E-surfaces), such that integrable singularities and poles are close.
In this case, the individual E-surface residues diverge strongly, resulting in large numerical cancellations that may render the integrand numerically unstable.
\par

Note that in this particular case of the \texttt{Box\_4E} it was possible to find a global parameterisation that does not lead to pinched poles.
In general however, it is \emph{not} possible to always find a suitable source in our particular framework, where we integrated out the magnitude of the spatial loop momentum in spherical coordinates.
Examples are  the topologies \texttt{1L5P.V}, \texttt{1L6P.V} and \texttt{1L6P.VI} from \cite{Capatti_2020}.
To tackle such integrals, one may explore different parameterisations or think of some sort of slicing of the three-dimensional integration space, where each sector has its own parameterisation. 

\begin{table}[H]
\centering
\resizebox{\columnwidth}{!}{%
\texttt{%
\begin{tabular}{clrrrrcllcrrr}
\hline
Topology & Kin. & $\mathtt{N_{\text{E}}}$ & $\mathtt{N_{\text{G}}}$ & $\mathtt{N_{\text{G}}^{\text{max}}}$ & $\mathtt{N_{\text{p}}}$ & Phase & Exp. & Reference & Numerical & $\mathtt{\Delta \ [\sigma]}$ & $\mathtt{\Delta \ [\%]}$ & $\mathtt{\Delta \ [\%] |\cdot|}$ \\
\hline
\multirow{2}{*}{%
\begin{tabular}{@{}c@{}}
Box\end{tabular}}
& \multirow{2}{*}{Box\_4E}& \multirow{2}{*}{4}& \multirow{2}{*}{1}& \multirow{2}{*}{1}& $\mathtt{10^9}$& Re& \multirow{2}{*}{-08}& \texttt{-7.437071}& \texttt{-7.430810~+/-~0.017054}& 0.367& 0.084& \multirow{2}{*}{0.101}\\
& & & & & $\mathtt{10^9}$& Im& & \texttt{ 6.578304}& \texttt{~6.570476~+/-~0.005288}& {1.480}& 0.119& \\
\end{tabular}%
}%
}%
\caption{\label{tab:box_4e}
The scalar box topology \texttt{Box\_4E}, whose kinematic configuration does not allow for a source in the interior of all E-surfaces (see fig.~\ref{fig:box_4e}).
However, an exterior source can be found that enables local cancellations of all poles of E-surface residues.
The subtracted integrand (real part) and the sum of residues (imaginary part) have integrable singularities at the tangents of the E-surfaces through the source.
}%
\end{table}

\subsection{Optical theorem and raised propagators}
\label{sec:unitarity_raised_props_ot}
In this section, we explore the connection of our expression in eq.~\eqref{eq:fract_residues_ltd} with the optical theorem with a simple forward scattering process.
Moreover, we see a first application of raised propagators in our framework.
\par

We consider a forward scattering box diagram, such that the incoming momenta $p_1$ and $p_2$ with $p_1^0>0$ and $p_2^0>0$ are the same as the outgoing ones.
Furthermore, we assume that the particle exchanged between the two incoming legs has the same mass as one between the two outgoing legs, i.e. $m_1=m_3$.
We therefore have a propagator that is raised to the second power.
The integral can then be written as
\begin{align}
    \ii I
    &=
    \begin{gathered}
	    \vcenter{\hbox{\includegraphics[page=7,scale=1]{img/diagrams.pdf}}}
	\end{gathered}
	=
    \int \frac{\dd^4 k}{(2\pi)^4} \mathcal{I},
\end{align}
where the integrand is given by
\begin{align}
	\mathcal{I} = 
	\frac{1}{((k^0+p_1^0)^2-E_1^2)^2}
	\frac{1}{(k^0+p_1^0+p_2^0)^2-E_2^2}
	\frac{1}{(k^0)^2-E_4^2},
\end{align}
with on-shell energies
\begin{align}
    E_1=\sqrt{(\vec{k}+\vec{p}_1)^2+m_1^2-\ii \epsilon}, \qquad
    E_2=\sqrt{(\vec{k}+\vec{p}_1+\vec{p}_2)^2+m_2^2-\ii \epsilon}
\end{align}
and $E_4=\sqrt{\vec{k}^2+m_4^2-\ii \epsilon}$.
The LTD expression is simply given by the sum of residues
\begin{align}
    R_{i_+} = \Res\left[\mathcal{I}(k^0),k^0=E_i-p_i^0\right].
\end{align}
where we chose $\sigma=+1$.
Note that since the first propagator has a second order pole in $k^0$, the residue is given by the limit
\begin{align}
    R_{1_+} = \lim_{k^0\to E_1-p_1^0}
    \frac{\dd }{\dd k^0}
    \bigg( (k^0-E_1+p_1^0)^2 \, \mathcal{I}(k^0)\bigg).
\end{align}
$R_{2_+}$ and $R_{4_+}$ are residues of single poles as in eq.~\eqref{eq:dual_integrand}.
Before we can apply eq.~\eqref{eq:subtracted_ltd} and eq.~\eqref{eq:fract_residues_ltd}, we have to determine the threshold singularities of our integrand.
Given that $p_1$ and $p_2$ both have positive energy components, we can infer that
\begin{align}
    &E_{21} \text{ exists} \quad
    \Longleftrightarrow \quad 
    p_2^2>(m_1+m_2)^2, \\
    &E_{14} \text{ exists} \quad
    \Longleftrightarrow \quad 
    p_1^2>(m_1+m_4)^2, \\
    &E_{24} \text{ exists} \quad
    \Longleftrightarrow \quad
    (p_1+p_2)^2>(m_2+m_4)^2.
\end{align}
Typically one would consider kinematics with large enough center of mass energy, such that $E_{24}$ exists.
Furthermore, one may want only one additional scale, i.e. $p_1^2=p_2^2=m_1^2=m_2^2=m_4^2=m^2>0$, such that the thresholds $E_{21}$ and $E_{14}$ do not exist.
It is however more interesting to consider the kinematic region, where all three E-surfaces exist.
Their residues are then given by
\begin{align}
    R_{i_+j_-}^\rho = \Res\left[
        r^2 R_{i_+}(r \hat k),r=r^\rho_{i_+j_-}(\hat k)
        \right],
\end{align}
where we emphasise that both $R_{2_+1_-}$ and $R_{1_+4_-}$ are residues of second order poles, which have to be calculated accordingly, with e.g. eq.~\eqref{app:residue_second_order_pole}.
For $R_{2_+4_-}$ we can however directly use eq.~\eqref{eq:e_surface_residue}.
Diagrammatically, we can therefore write
\begin{align}
    2 \Im \ii
    \!\!
    \begin{gathered}
	    \vcenter{\hbox{\includegraphics[page=13,scale=0.7]{img/diagrams.pdf}}}
	\end{gathered}
	=
	\!\!\!
	\begin{gathered}
	    \vcenter{\hbox{\includegraphics[page=9,scale=0.7]{img/diagrams.pdf}}}
	\end{gathered}
	+
	\!\!\!
	\begin{gathered}
	    \vcenter{\hbox{\includegraphics[page=10,scale=0.7]{img/diagrams.pdf}}}
	\end{gathered}
	+
	\!\!\!
	\begin{gathered}
	    \vcenter{\hbox{\includegraphics[page=12,scale=0.7]{img/diagrams.pdf}}}
	\end{gathered}
	+
	\!\!\!
	\begin{gathered}
	    \vcenter{\hbox{\includegraphics[page=11,scale=0.7]{img/diagrams.pdf}}}
	\end{gathered}
	+
	\!\!\!
	\begin{gathered}
	    \vcenter{\hbox{\includegraphics[page=8,scale=0.7]{img/diagrams.pdf}}}
	\end{gathered}
\end{align}
where we identify the diagrams with the residues as in eq.~\eqref{eq:e_surface_residue_diagram}, with the subtlety that the residues of the raised E-surfaces are each sums of two diagrams.
More preceisely, the sum of the first two diagrams is $R_{2_+1_-}$, the sum of diagram three and four is $R_{1_+4_-}$, and the last one is $R_{2_+4_-}$.

The counterterm corresponding to $R_{2_+4_-}$ is therefore simply given by eq.~\eqref{eq:counterterm}.
For the other two thresholds of second order one has to be more careful.
Since the counterterm not only has to cancel the first but also the second order pole one therefore has to construct it according to eq.~\eqref{eq:appendix_counterterms_local_cancellations}, which guarantees that its Cauchy principal value still vanishes.
\par

In tab.~\ref{tab:ot}, we evaluated the box integral for two kinematic configurations.
In the second, we set $m_1=m_3=0$.
In this case the integral has an infrared divergence arising from the pole in the raised propagator.
It can be regulated with an infrared counterterm, a tadpole diagram with a raised massless propagator.
This scaleless integral however has a ultraviolet divergence that needs to be regulated with another tadpole diagram with a massive propagator.
These counterterms only affect the real part of the integral.
This implies that the infrared pole at $D=4$ spacetime dimensions is purely real and the imaginary part remains finite.

\begin{table}[H]
\centering
\resizebox{\columnwidth}{!}{%
\texttt{%
\begin{tabular}{clrrrrcllcrrr}
\hline
Topology & Kin. & $\mathtt{N_{\text{E}}}$ & $\mathtt{N_{\text{G}}}$ & $\mathtt{N_{\text{G}}^{\text{max}}}$ & $\mathtt{N_{\text{p}}}$ & Phase & Exp. & Reference & Numerical & $\mathtt{\Delta \ [\sigma]}$ & $\mathtt{\Delta \ [\%]}$ & $\mathtt{\Delta \ [\%] |\cdot|}$ \\
\hline
\multirow{4}{*}{%
\begin{tabular}{@{}c@{}}
\begin{tikzpicture}[scale=1]
	\pgfmathsetmacro\n{4}
	\pgfmathsetmacro\radius{0.4}
	\foreach \i in {0,...,\n} {
		\pgfmathsetmacro\r{(\i+0.5)*(360/\n)}
		\fill (\r:\radius) circle (1.5pt);
		\draw[thick] (\radius,0) arc (0:\r:\radius);
		\draw[thick] (\r:\radius) -- (\r:\radius*3/2);
}
\end{tikzpicture} \\
Box\end{tabular}}
& \multirow{2}{*}{1L4P.OT}& \multirow{2}{*}{5}& \multirow{2}{*}{1}& \multirow{2}{*}{3}& $\mathtt{10^9}$& Re& \multirow{2}{*}{-04}& \texttt{-0.959216}& \texttt{-0.959437~+/-~0.001444}& 0.153& 0.023& \multirow{2}{*}{0.005}\\
& & & & & $\mathtt{10^9}$& Im& & \texttt{-4.416317}& \texttt{-4.416317~+/-~0.000002}& 3e-05& 1e-09& \\
\cline{2-13}
& \multirow{2}{*}{1L4P.OT\_IR}& \multirow{2}{*}{5}& \multirow{2}{*}{1}& \multirow{2}{*}{3}& $\mathtt{10^9}$& Re& \multirow{2}{*}{-03}& \texttt{\phantom{-0.}inf.}& \texttt{}&  &  & \multirow{2}{*}{}\\
& & & & & $\mathtt{10^9}$& Im& & \texttt{-0.894078}& \texttt{-0.894078~+/-~0.000001}& 1e-04& 7e-09& \\
\end{tabular}%
}%
}%
\caption{\label{tab:ot}
Two forward scattering configurations of a one-loop box integral featuring raised threshold singularities.
The latter features a soft singularity, which results in a divergence in the real part.
The reference results were calulated with \textsc{Package-X} \cite{Patel_2015}.
}%
\end{table}

We can compare our expression for the imaginary part with the optical theorem (see eq.~\eqref{eq:generalised_optical_theorem} with $f=i$) at fixed order.
For example, for $m_2=m_4$, the interferences of the tree level diagrams can be identified directly with the cut diagrams as
\begin{align}
    \!\!\!
	\begin{gathered}
	    \vcenter{\hbox{\includegraphics[page=9,scale=0.7]{img/diagrams.pdf}}}
	\end{gathered}
	&=
	-\int \dd \Pi_3
	\bigg(
	\!\!\!
	\begin{gathered}
	    \vcenter{\hbox{\includegraphics[page=15,scale=0.7]{img/diagrams.pdf}}}
	\end{gathered}
	\bigg)
	\bigg(
	\!\!\!
	\begin{gathered}
	    \vcenter{\hbox{\includegraphics[page=16,scale=0.7]{img/diagrams.pdf}}}
	\end{gathered}
	\bigg)^*
	\\
	\!\!\!
	\begin{gathered}
	    \vcenter{\hbox{\includegraphics[page=10,scale=0.7]{img/diagrams.pdf}}}
	\end{gathered}
	&=
	-\int \dd \Pi_3
	\bigg(
	\!\!\!
	\begin{gathered}
	    \vcenter{\hbox{\includegraphics[page=16,scale=0.7]{img/diagrams.pdf}}}
	\end{gathered}
	\bigg)
	\bigg(
	\!\!\!
	\begin{gathered}
	    \vcenter{\hbox{\includegraphics[page=15,scale=0.7]{img/diagrams.pdf}}}
	\end{gathered}
	\bigg)^*
	\\
	\!\!\!
	\begin{gathered}
	    \vcenter{\hbox{\includegraphics[page=12,scale=0.7]{img/diagrams.pdf}}}
	\end{gathered}
	&=
	-\int \dd \Pi_3
	\bigg(
	\!\!\!
	\begin{gathered}
	    \vcenter{\hbox{\includegraphics[page=14,scale=0.7]{img/diagrams.pdf}}}
	\end{gathered}
	\bigg)
	\bigg(
	\!\!\!
	\begin{gathered}
	    \vcenter{\hbox{\includegraphics[page=17,scale=0.7]{img/diagrams.pdf}}}
	\end{gathered}
	\bigg)^*
	\\
	\!\!\!
	\begin{gathered}
	    \vcenter{\hbox{\includegraphics[page=11,scale=0.7]{img/diagrams.pdf}}}
	\end{gathered}
	&=
	-\int \dd \Pi_3
	\bigg(
	\!\!\!
	\begin{gathered}
	    \vcenter{\hbox{\includegraphics[page=17,scale=0.7]{img/diagrams.pdf}}}
	\end{gathered}
	\bigg)
	\bigg(
	\!\!\!
	\begin{gathered}
	    \vcenter{\hbox{\includegraphics[page=14,scale=0.7]{img/diagrams.pdf}}}
	\end{gathered}
	\bigg)^*
	\\
	\!\!\!
	\begin{gathered}
	    \vcenter{\hbox{\includegraphics[page=8,scale=0.7]{img/diagrams.pdf}}}
	\end{gathered}
	&=
	-\int \dd \Pi_2
	\bigg(
	\!\!\!
	\begin{gathered}
	    \vcenter{\hbox{\includegraphics[page=18,scale=0.7]{img/diagrams.pdf}}}
	\end{gathered}
	\bigg)
	\bigg(
	\!\!\!
	\begin{gathered}
	    \vcenter{\hbox{\includegraphics[page=18,scale=0.7]{img/diagrams.pdf}}}
	\end{gathered}
	\bigg)^*
\end{align}
This identification shows that
the tree level interferences with three final states, corresponding to the diagonal cuts, also contribute to the imaginary part of the box diagram.
Furthermore, it clarifies that indeed eq.~\eqref{eq:fract_residues_ltd} is a local representation of the optical theorem.
\par

\subsection{UV \& IR subtraction, numerators and amplitudes}
\label{sec:numerator_ir_singularities}

In this section we apply our method to one-loop integrals and amplitudes, whose infrared and ultraviolet singularities were subtracted with local counterterms.
\par

As a first example we consider a scalar box diagram with massless on-shell external legs, $(p_i^{\rm ext})^2=0$, and massless propagators, $m_i=0$.
A suitable counterterm accounting for all four soft and four collinear singular regions was constructed in \cite{Anastasiou_2019} and is given by a sum of triangle diagrams
\begin{align}
    \CT^{\mathrm{IR}}_{\rm box} = \frac{1}{t}\left(
        \frac{1}{D_1 D_3 D_4 }
        +
        \frac{1}{D_1 D_2 D_3}
        \right)
        +
        \frac{1}{s}\left(
        \frac{1}{D_2 D_3 D_4}
        +
        \frac{1}{D_1 D_2 D_4}
        \right),
\end{align}
with $s=(p_3-p_1)^2$ and $t=(p_4-p_2)^2$ and $D_i = k + p_i$, where the $p_i=\sum_{j=1}^i p_i^{\rm ext}$ are the sum of the incoming external momenta, arranged clockwise.
Momentum conservation imposes $p_4=0$.
Equivalently, one can introduce a numerator
\begin{align}
    N^{\rm IR}_{\rm box} = 1-\frac{D_2+D_4}{t}-\frac{D_1+D_3}{s}
\end{align}
to the box integral to regulate the infrared divergences.
\par
We can do a similar exercise for an infrared divergent scalar pentagon integral with two adjacent massless on-shell external particles, $(p^{\rm ext}_1)^2=(p^{\rm ext}_5)^2=0$, and massless internal particles, $m_i=0$.
It has two collinear and one soft limit that can be regulated by introducing the numerator
\begin{align}
    \begin{split}
    N^{\rm IR}_{\rm pentagon} = 1
    - \frac{D_2 D_3}{[D_2 D_3 ]_{k=0}}
    &- \frac{D_3 D_4}{[D_3 D_4 ]_{k=x_{21}p_1}}
    - \frac{D_2 D_4}{[D_2 D_4 ]_{k=x_{31}p_1}} \\
    &- \frac{D_1 D_3}{[D_1 D_3 ]_{k=x_{24}p_4}}
    - \frac{D_1 D_2}{[D_1 D_2 ]_{k=x_{34}p_4}},
    \end{split}
\end{align}
where $x_{ij}=-\frac{p_i^2}{2 p_i \cdot p_j}$.
The remaining definitions are the same as above. 
In particular, we have that $p_1=p_1^{\rm ext}$, $p_4=-p^{\rm ext}_5$ and $p_5=0$.
Again, the counterterm is a sum of triangle diagrams and follows the construction of \cite{Capatti_2020}.
\par

\par

The infrared regulated box and pentagon integrals can directly be integrated with our method.
In tab.~\ref{tab:ir_box_penta} we present our results for two physical phase space points.
Despite the infrared counterterms both integrals converge similar to comparable phase space configurations in sect.~\ref{sec:single_group} and \ref{sec:multiple_groups}.
Note that the full result of the infrared divergent box and pentagon diagram is not presented here but can simply be obtained by adding the respective integrated counterterm (in dimensional regularisation) to the subtracted integrals in tab.~\ref{tab:ir_box_penta}.
\par

\begin{table}[H]
\centering
\resizebox{\columnwidth}{!}{%
\texttt{%
\begin{tabular}{llrrrrcllcrrr}
\hline
Topology & Kin. & $\mathtt{N_{\text{E}}}$ & $\mathtt{N_{\text{G}}}$ & $\mathtt{N_{\text{G}}^{\text{max}}}$ & $\mathtt{N_{\text{p}}}$ & Phase & Exp. & Reference & Numerical & $\mathtt{\Delta \ [\sigma]}$ & $\mathtt{\Delta \ [\%]}$ & $\mathtt{\Delta \ [\%] |\cdot|}$ \\
\hline
\multirow{2}{*}{%
\begin{tabular}{@{}c@{}}
Box\end{tabular}}
& \multirow{2}{*}{1L4P.4S4C}& \multirow{2}{*}{1}& \multirow{2}{*}{1}& \multirow{2}{*}{1}& $\mathtt{10^9}$& Re& \multirow{2}{*}{-03}& \texttt{ 0.380313}& \texttt{~0.380322~+/-~0.000552}& 0.016& 0.002& \multirow{2}{*}{0.001}\\
& & & & & $\mathtt{10^9}$& Im& & \texttt{-3.447431}& \texttt{-3.447458~+/-~0.000248}& 0.106& 0.001& \\
\hline
\multirow{2}{*}{%
\begin{tabular}{@{}c@{}}
Pentagon\end{tabular}}
& \multirow{2}{*}{1L5P.1S2C}& \multirow{2}{*}{6}& \multirow{2}{*}{2}& \multirow{2}{*}{6}& $\mathtt{10^9}$& Re& \multirow{2}{*}{+00}& \texttt{-4.801140}& \texttt{-4.800655~+/-~0.000736}& 0.658& 0.010& \multirow{2}{*}{0.010}\\
& & & & & $\mathtt{10^9}$& Im& & \texttt{-0.414486}& \texttt{-0.414435~+/-~0.000209}& 0.243& 0.012& \\
\end{tabular}%
}%
}%
\caption{\label{tab:ir_box_penta}
Two scalar one-loop integrals whose infrared singularities were subtracted with local counterterms.
The reference result for the box is from \cite{Anastasiou_2019} and the pentagon result was obtained with \textsc{MadLoop} \cite{Hirschi_2011,Alwall_2014}.
}%
\end{table}

As a last test, we apply our method to one-loop amplitudes in QED of two processes $e^+ e^- \to n \gamma^*$ for $n=2$ and $n=3$.
Apart from infrared singularities due to the massless initial states, the amplitudes also feature ultraviolet singularities arising from bubble and triangle contributions.
Both infrared and ultraviolet singularities were subtracted following the scheme described in \cite{Anastasiou_2021}.
\par

The subtracted amplitudes are finite in four dimensions and can be integrated numerically after regulating the remaining threshold singularities.
For simplicity, we performed the interference of the one-loop with the Born contributions to get rid of uncontracted spinor indices and summed over all four polarisations, including the unphysical ones.
The numerical results obtained with our method for the interfered amplitudes at two physical phase space points are presented in tab.~\ref{tab:amplitudes}.
Note that the respective integrated counterterms (in dimensional regularisation) have to be added to the results in tab.~\ref{tab:amplitudes} to get the full amplitude.
This calculation is independent of our method and therefore not presented here.
\par

\begin{table}[H]
\centering
\resizebox{\columnwidth}{!}{%
\texttt{%
\begin{tabular}{llrrrrcllcrrr}
\hline
Amplitude & Kin. & $\mathtt{N_{\text{E}}}$ & $\mathtt{N_{\text{G}}}$ & $\mathtt{N_{\text{G}}^{\text{max}}}$ & $\mathtt{N_{\text{P}}}$ & Phase & Exp. & Reference & Numerical & $\mathtt{\Delta \ [\sigma]}$ & $\mathtt{\Delta \ [\%]}$ & $\mathtt{\Delta \ [\%] |\cdot|}$ \\
\hline
\multirow{2}{*}{%
\begin{tabular}{@{}c@{}}
$e^+e^-\to \gamma^*\gamma^*$
\end{tabular}}
& \multirow{2}{*}{PS1}& \multirow{2}{*}{3}& \multirow{2}{*}{1}& \multirow{2}{*}{3}& $\mathtt{10^9}$& Re& \multirow{2}{*}{+00}& \texttt{-1.596105}& \texttt{-1.596350~+/-~0.000682}& 0.360 & 0.015 & \multirow{2}{*}{0.015}\\
& & & & & $\mathtt{10^9}$& Im& & \texttt{-0.059046}& \texttt{-0.059046~+/-~0.000065}&0.008 & 0.001 & \\
\hline
\multirow{2}{*}{%
\begin{tabular}{@{}c@{}}
$e^+e^-\to \gamma^*\gamma^*\gamma^*$
\end{tabular}}
& \multirow{2}{*}{PS2}& \multirow{2}{*}{4}& \multirow{2}{*}{1}& \multirow{2}{*}{2}& $\mathtt{10^9}$& Re& \multirow{2}{*}{-04}& \phantom{+}\texttt{1.004337}& \texttt{~1.004428~+/-~0.000529}& 0.172& 0.009 & \multirow{2}{*}{0.004}\\
& & & & & $\mathtt{10^9}$& Im& & \phantom{+}\texttt{1.878322}& \texttt{~1.878321~+/-~0.000090}& 0.014 & 1e-04 & \\
\end{tabular}%
}%
}%
\caption{\label{tab:amplitudes}
Two one-loop amplitudes at physical kinematic points, whose ultraviolet and infrared singularities were subtracted with local counterterms.
The reference results were calculated using \textsc{Form} \cite{Vermaseren:2000nd} and \textsc{Package-X} \cite{Patel_2015}.
}%
\end{table}

Despite the increased complexity from fermionic propagators and infrared and ultraviolet counterterms, our threshold subtraction scheme performs similarly to the 4- and 5-point scalar integrals presented in sect.~\ref{sec:multiple_groups}.
Note however that numerical instabilities arise in the ultraviolet region due to the large cancellations between dual integrands, which are individually ultraviolet divergent and only finite if summed up.
This effect is enhanced by the presence of fermionic propagators.
We found that we can suppress probing the unstable ultraviolet region by adjusting the scaling of Monte~Carlo samples in the hypercube to be logarithmic for large spatial loop momenta.

\section{Conclusion}
\label{sec:conclusion}

In this article, we derived a new representation of one-loop integrals that is locally free of poles in the integration domain.
This representation is particularly well suited for numerical integration with a Monte Carlo event generator and does not rely on a contour deformation.
Its real (dispersive) part of the integral is given by the Loop-Tree Duality representation supplied with a simple counterterm for each existing threshold singularity.
Its imaginary (absorptive) part of the integral is given by a sum of residues, which correspond to tree-level interferences integrated over the two-body phase space.
The latter serves as a locally finite representation of the generalised optical theorem at one-loop.
\par

To determine the boundary conditions of our method, a careful investigation of intersecting threshold singularities was in order.
We distinguished between locally pinched and non-pinched poles, where the latter result in local cancellations between residues and counterterms and the former introduce poles in the residues.
Our method is applicable whenever locally pinched poles are absent.
By means of grouping thresholds and appropriately choosing the parameterisation frame, we could easily avoid locally pinched poles for a wide range of one-loop configurations with various multiplicities of threshold singularities, external legs and mass scales.
In this case we found that the Monte Carlo integration is accurate, stable and efficient, outperforming existing methods, e.g. contour deformation, that tackle the numerical integration of one-loop integrals directly in momentum space.
\par

The described threshold subtraction method cannot be applied to one-loop integrals with momentum configurations, which feature locally pinched poles that cannot be avoided through grouping thresholds and/or suitable parameterisation.
More concretely, this applies to topologies, whose groups of intersecting thresholds cannot be parameterised in the radial coordinate with a matching orientation on all intersections\footnote{
This is similar to limiting ourselves to (the groups of thresholds of) topologies which can be integrated using a single radial deformation field but not entirely the same since, as we showed in sect.~\ref{sec:exterior_sources}, a radial parameterisation may work in cases where a radial deformation does not.
}.
Whether this constraint can be satisfied or not can be probed geometrically (c.f. fig.~\ref{fig:box_4e}).
This limitation can in principle be overcome, since locally pinched poles can be avoided for example by slicing the phase space into regions around intersections that each are parameterised independently.
Another approach is to employ a different parameterisation of spatial loop momentum space and to integrate out a different variable than the magnitude of the spatial loop momentum.
Such a parameterisation must exists, since we know that a global deformation vector field can always be found.
Explicitly applying such a parameterisation is related to solving for the causal flow as introduced in \cite{Capatti_2021}.
\par

The particular contour deformation beyond one-loop developed in \cite{Capatti_2020} may serve as a guide to construct a threshold subtraction scheme for multi-loop diagrams.
Furthermore, since the general E-surface is parameterised by the $n$-body phase space, we can also draw inspiration from the algorithms of tree-level phase space generators, such as phase space factorisation, parameterisation by kinematic invariants and common optimisation strategies \cite{Roth:1999kk,Dittmaier:2002ap,Kleiss_1994,Denner_1999} using importance sampling and multi-channelling.
Moreover, in the framework of Local Unitarity \cite{Capatti_2021}, where infrared singularities are locally cancelled between real and virtual contributions, our approach has the potential not only to replace the contour deformation of subgraphs but also to serve as an alternative to Soper's trick (of putting intermediate particles on-shell), which conveniently also reduces the integration dimension by one.
\par

For future work, its is planned to extend our subtraction-based treatment of threshold singularities to generic one-loop and multi-loop integrals.
We believe that this generalisation is crucial for the development of an efficient numerical calculation of higher-order corrections to observables.
In fact, regulating a loop integral through subtraction rather than deformation has the potential to drastically reduce the integrand's variance, directly improving the convergence of the Monte~Carlo integration, as we have shown with many examples at one loop.
Moreover, the fact that our approach results in a locally finite representation of the generalised optical theorem suggests that it can play an important role in methods aiming at the \emph{local} realisation of infrared cancellations between real radiation and virtual loop corrections in momentum space.

\acknowledgments
We are grateful to Babis Anastasiou and Valentin Hirschi for their guidance and support, Nikos Kalntis, Andrea Pelloni and Armin Schweitzer for sharing their insights on infrared subtraction, Ben Ruijl for his help with the computer implementation and Zeno Capatti, Tim Engel and Rayan Haindl for many fruitful and inspiring discussions.
This research was supported by the European Research Council (ERC) under grand agreement ID 694712 (PertQCD) and by the Swiss National Science Foundation (SNSF) grant number 179016.

\appendix

\section{Local cancellations due to merging poles}
\label{app:residue_cancellations}
In this section, we derive the results in sect.~\ref{sec:local_cancellations}, namely eqs.~\eqref{eq:limit_sum_residues}, \eqref{eq:limit_sum_counterterms} and \eqref{eq:limit_subtracted_integrand}.
We show explicitly that the residues and counterterms of two separate single poles, although divergent individually, combine into the residue and a valid counterterm of a double pole in the limit, where the two single poles merge.
\par

Note that if the integrand has a global double pole (e.g. a propagator raised to the second power) one shall directly construct the corresponding residue and counterterm according to eqs.~\eqref{app:residue_second_order_pole} and \eqref{eq:appendix_counterterms_local_cancellations}, respectively (cf. sect.~\ref{sec:unitarity_raised_props_ot}).
\par

We consider the function $\mathcal{I}(x,y,z)=f(x)/g(x,y,z)$ as defined in eq.~\eqref{eq:integrand_local_cancellations}, where $g$ has two single poles at $x=y$ and $x=z$ if $y\neq z$ that merge into a double pole if $y=z$.
Under the assumption $y\neq z$, the residues $R_y(y,z)$ and $R_z(y,z)$ at $x=y$ and $x=z$, respectively, are given in eq.~\eqref{eq:residues_local_cancellations}.
Their corresponding counterterms $\CT_y(x,y,z)$ and $\CT_z(x,y,z)$ are given in eq.~\eqref{eq:counterterms_local_cancellations}.

\subsection{Residues}
The sum of the two residues in the limit $z\to y$ can simply be computed by Taylor expanding the residues in $z$ around $y$.
Therefore, we first expand the denominators of $R_y$ and $R_z$ to second order,
\begin{align}
    g^{(1,0,0)}(y,y,z)
    &= 
    a(y) (z-y)
    + b(y) (z-y)^2
    + \order((z-y)^3)
    \\
    g^{(1,0,0)}(z,y,z)
    &=
    c(y) (z-y)
    + d(y) (z-y)^2
    + \order((z-y)^3),
\end{align}
with coefficients $a$, $b$, $c$ and $d$ given by partial derivatives of $g$ at $x=z=y$, i.e.
\begin{align}
\begin{gathered}
    a(y) = g^{(1,0,1)}(\mathbf{y}),
    \qquad
    b(y) = \frac{1}{2} g^{(1,0,2)}(\mathbf{y}), \\
    c(y) = g^{(2,0,0)}(\mathbf{y}) + g^{(1,0,1)}(\mathbf{y}),
    \qquad
    d(y) =\frac12 g^{(3,0,0)}(\mathbf{y})+  g^{(2,0,1)}(\mathbf{y})+ \frac12 g^{(1,0,2)}(\mathbf{y}),
\end{gathered}
\end{align}
where $\mathbf{y}=(y,y,y)$. 
One can now exploit the assumptions on $g$ to determine relations between its partial derivatives.
More explicitly, the total derivatives w.r.t $y$ and $z$ of our initial assumptions $g(y,y,z)=0$, $g(z,y,z)=0$ and $g^{(1,0,0)}(y,y,y)=0$ imply that
\begin{align}
    c(y) = -a(y) = \frac{1}{2!} \tilde g''(y),
    \quad
    b(y) + d(y) = \frac{1}{3!} \tilde g'''(y),
\end{align}
with $\tilde g(x)=g(x,y,y)$.
The sum of residues therefore converges to
\begin{align} 
    \lim_{z\to y}\left(R_y(y,z)+R_z(y,z)\right)
    &=
    \lim_{y\to z}
        \frac{
            (z-y)(a+c)f + (z-y)^2(b+d)f
            + (z-y)^2 a f'
            }{(z-y)^2 ac}
    \\
    &=
    \frac{
        6f' \tilde g''
        -2f \tilde g'''
        }{
        3(\tilde g'')^2},
\end{align}
where all functions with omitted arguments are evaluated at $y$.
This is precisely the residue of a second order pole, i.e.
\begin{align}
    R(y) \coloneqq
    \Res[\mathcal{I}(x,y,y),x=y]
    &=
    \lim_{x\to y}\frac{\dd}{\dd x}\left((x-y)^2\frac{f(x)}{\tilde g(x)}\right) \\
    &=
    \lim_{x\to y}\frac{\dd}{\dd x}\left(
        \frac{f(y)+f'(y)(x-y)+\order((x-y)^2)}{
            \frac{1}{2} \tilde g''(y)
            +\frac{1}{3!} \tilde g'''(y)(x-y)
            +\order((x-y)^2)}
    \right) \\
    &=
    \lim_{x\to y}
    \frac{
        f'(y)\frac{1}{2} \tilde g''(y)
        -
        f(y)\frac{1}{3!} \tilde g'''(y)
        +\order((x-y))
    }{
        \left(\frac{1}{2} \tilde g''(y)\right)^2
        +\order((x-y))
    } \\
    &= 
    \frac{
        6f'(y)\tilde g''(y)
        -
        2f(y) \tilde g'''(y)
    }{
        3 (\tilde g''(y))^2
    }.
\label{app:residue_second_order_pole}
\end{align}
It therefore follows that the relation
\begin{align}
\label{eq:appendix_limit_sum_residues}
    \lim_{z\to y}
    \left(
        R_y(y,z)
        +
        R_z(y,z)
    \right)
    = 
    R(y)
\end{align}
holds true.
In the limit $z\to y$ the residues of two single poles at $x=y$ and $x=z$ therefore combine into the residue of the double pole at $x=y=z$.


\subsection{Counterterms}
Each counterterm is proportional to its corresponding residue and since the residue diverges for $z\to y$, so will the counterterm, independent of the value of $x$.
This means that each of the two counterterms introduces new singularities.
First, we show that these new singularities cancel locally between the two counterterms.
For this pairwise cancellation to happen, it is crucial that the UV suppression functions are chosen to be equal, i.e. $\chi\equiv\chi_y\equiv\chi_z$.
Secondly, we show that the double pole of $\mathcal{I}$ at $x=y=z$ is indeed removed completely by the sum of counterterms $\CT_y+\CT_z$.

\paragraph{Cancellations among counterterms ($z\to y$, $x\neq y$)}
The sum of the two counterterms in the limit $z\to y$ for $x\neq y$ can simply be computed by Taylor expanding the counterterms in $z$ around $y$.
We first expand the $x$-dependent factor of $\CT_z$ to first order
\begin{align}
    \frac{\chi(x-z)}{x-z}
    =
    \frac{\chi(x-y)}{x-y}
    +
    \left(
        \frac{\chi(x-y)}{(x-y)^2}
        -
        \frac{\chi'(x-y)}{(x-y)}
    \right)
    (z-y)
    +
    \order((z-y)^2).
\end{align}
When plugging this expansion into the sum of counterterms, we can make use of the previous result in eq.~\eqref{eq:appendix_limit_sum_residues}, as well as the fact that $R_z$ has a single pole at $z=y$ with residue
\begin{align}
    \Res[R_z(y,z),z=y] = \frac{2f(y)}{\tilde g''(y)},
\end{align}
where we used that $g^{(1,0,1)}(y,y,y)=-\frac{1}{2} g^{(2,0,0)}(y,y,y)$.
It follows that the sum of counterterms converges for $x\neq y$ to
\begin{align}
\label{eq:appendix_counterterms_local_cancellations}
    \lim_{z\to y} \left( \CT_y(x,y,z) + \CT_z(x,y,z) \right)
    &=
    \frac{\chi(x-y)}{x-y}
    R(y)
    +
    \left(
        \frac{\chi(x-y)}{(x-y)^2}
        -
        \frac{\chi'(x-y)}{(x-y)}
    \right)
    \frac{2f(y)}{\tilde g''(y)}.
\end{align}
By assumption also $\mathcal{I}(x,y,y)=f(x)/g(x,y,y)=f(x)/\tilde g(x)$ is finite for $x\neq y$, such that $\mathcal{I}-\CT_y-\CT_z$ converges for $z\to y$.

\paragraph{Cancellation of double pole ($z\to y,x\to y$)}
To demonstrate the local cancellations of the double pole of $\mathcal{I}$ at $x=y=z$ with the counterterms $\CT_y$ and $\CT_z$, we first expand the above expression in eq.~\eqref{eq:appendix_counterterms_local_cancellations} in $x$ around $y$, yielding
\begin{align}
    \left(
        \frac{1}{x-y}
        +
        \chi'(0)
    \right)
    R(y)
    +
    \left(
        \frac{1}{(x-y)^2}
        -
        \frac{1}{2}\chi''(0)
    \right)
    \frac{2f(y)}{\tilde g''(y)}
    + \order((x-y)^1),
\end{align}
where we used that $\chi(0)=1$.
Similarly, we then expand $\mathcal{I}(x,y,y)=f(x)/\tilde g(x)$ in $x$ around $y$, i.e.
\begin{align}
    \mathcal{I}(x,y,y)
    &=
    \frac{1}{(x-y)^2}
    \frac{f(x)
        }{
        \frac{1}{2!}\tilde g''(y)
        +\frac{1}{3!}(x-y)\tilde g'''(y)
        +\frac{1}{4!}(x-y)^2\tilde g''''(y)
        +\order((x-y)^3)} \\
    &=
    \frac{1}{(x-y)^2}
    \frac{2f(y)}{\tilde g''(y)}
    +
    \frac{R(y)}{(x-y)}
    +
    C(y)
    + \order((x-y)^1),
\end{align}
where
\begin{align}
    C \coloneqq \frac{
        18 \tilde g''^2 f''-12 \tilde g'' \tilde g''' f'-3 f \tilde g'' \tilde g''''+4 f \tilde g'''^2
        }{
        18 \tilde g''^3}.
\end{align}
Note that the above expansions of counterterms and $\mathcal{I}$ have the exact same poles.
Their difference is therefore free of poles, implying that the limits $z\to y$ and $x\to y$ exist and are given by \footnote{For this limit to exist, it is not strictly necessary that $\chi(x)$ is differentiable at $x=0$ as long as the one-sided derivatives exist.}
\begin{align}
    \lim_{x\to y}
    \lim_{z\to y}
    \left(
        \mathcal{I}(x,y,z)-\CT_y(x,y,z)-\CT_z(x,y,z)
    \right)
    &=
    C(y) - \chi'(0) R(y)
    + \chi''(0) \frac{f(y)}{\tilde g''(y)}.
\end{align}
In other words, the counterterms $\CT_y$ and $\CT_z$ locally cancel the poles of $\mathcal{I}$ also in the case where the two single poles at $x=y$ and $x=z$ merge into a double pole at $x=y=z$.
We have therefore shown that $\mathcal{I}-\CT_y-\CT_z$ converges for $z\to y$.

\section{Singular surfaces of the dual propagator}
\label{app:dual_propagator}

The singularities of loop integrands in the LTD representation are given by the singular surfaces of the dual propagator.
Some singularities, namely the H-surfaces, are spurious in the sum of dual integrands as they are subject to dual cancellations.
The others, the E-surfaces, remain singular.
It is useful to investigate these surfaces in more detail.
\par

The general one-loop dual propagator takes the form
\begin{align}
\label{eq:app_dual_prop}
D_j\vert_{i_\sigma} &= (\sigma E_i-p_i^0+p_j^0)^2-E_j^2 \\
&=
\left(\sigma \sqrt{(\vec{k}+\vec{p}_i)^2+m_i^2-\ii\epsilon} - p_i^0 + p_j^0\right)^2 - \left(\sqrt{(\vec{k}+\vec{p}_j)^2+m_j^2-\ii\epsilon}\right)^2.
\end{align}
The surfaces described by $D_j\vert_{i_\sigma}=0$ are the intersections of the hyperboloid
\begin{align}
    (k+p_j)^0=(\vec{k}+\vec{p_j})^2+m_j^2-\ii\epsilon
    \quad
    \text{with}
    \quad
    k^0=\sigma \sqrt{(\vec{k}+\vec{p}_i)^2+m_i^2-\ii\epsilon},
\end{align}
the forward ($\sigma=1$) or backward ($\sigma=-1$) hyperboloid in four dimensions.
The solutions can be hyperboloids, paraboloids, ellipsoids, lines or single points in three-dimensional spatial momentum space.
We can see this more clearly in the principal axis coordinate system.
\par

First we perform a shift, followed by a rotation\footnote{
A compact way to explicitly write down the matrix that rotates a vector $\vec{e}$ along the $x$-axis is given by
\begin{align}
	R_{\hat e} = \mathbb{I}_3 + C(\vec{s}_{\hat e}) +\frac{C(\vec{s}_{\hat e})^2}{1+c_{\hat e}},
	\quad
	\begin{array}{ll}
	\vec{s}_{\hat e} &= \hat e \wedge \left(1,0,0\right) \\
	c_{\hat e} &= \hat e \cdot \left(1,0,0\right),
	\end{array},
	\quad
	C(\vec{s}) =
	\begin{pmatrix}
		0 & -s_3 & s_2 \\
		s_3 & 0 & -s_1 \\
		-s_2	& s_1 & 0 \\
	\end{pmatrix}.
\end{align}}
$(x,y,z)=R_{\hat p_{ij}}(\vec{k}+\vec{p}_i)$, such that $(p,0,0)=R_{\hat p_{ij}}\vec{p}_{ij}$, where $p_{ij} = p_i-p_j$.
The dual propagator then reads
\begin{align}
D_j\vert_{i_\sigma} &= \left(\sigma \sqrt{x^2+y^2+z^2+m_i^2-\ii\epsilon} - p_{ij}^0\right)^2 - \left(\sqrt{(x-p)^2+y^2+z^2+m_j^2-\ii\epsilon}\right)^2.
\end{align}
To obtain a quadratic form, we need to remove the square roots.
This can be achieved either by squaring the equation appropriately or by multiplying with the factor $D_j\vert_{i_{-\sigma}}$, the dual propagator with on-shell energy of opposite sign.
In either case, this operation discards the sign of $\sigma p_{ij}^0$.
If $p_{ij}^2\neq 0$, we can complete the square by shifting $x=\tilde x+ \frac{p}{2} \chi$, with $\chi = 1+(m_i^2-m_j^2)/p_{ij}^2$ and find the standard form of a two dimensional quadric surface
\begin{align}
\label{eq:quadric}
    \frac{\tilde x^2}{a^2}\pm\frac{y^2}{ b^2}\pm\frac{z^2}{b^2} = 1,
\end{align}
where
\begin{align}
\label{eq:quadric_parameters}
    a^2 &= \frac{(p_{ij}^0)^2}{p_{ij}^2} Q,
    \quad
    c^2 = \frac{|\vec{p}_{ij}|^2}{p_{ij}^2} Q,
    \quad
    \pm b^2 = a^2 -c^2 = Q,
    \quad
    Q=\frac{\lambda(p_{ij}^2,m_i^2,m_j^2)}{4p_{ij}^2}+\ii\epsilon.
\end{align}
In general, eq.~\eqref{eq:quadric} describes a rotational surface around the $x$-axis.
However, it has no real solutions if $\lambda(p_{ij}^2,m_i^2,m_j^2)<0$.
If $\lambda(p_{ij}^2,m_i^2,m_j^2)>0$, the solutions either define an ellipsoid ($p_{ij}^2>0$) or a hyperboloid of two sheets ($p_{ij}^2<0$) with focal points at $(\tilde x, y, z) = (\pm c,0,0)$, or in $\vec{k}$-space at
\begin{align}
    \label{eq:quadric_focus}
    \vec{k}_\pm
    &= -\vec{p}_i + \vec{p}_{ij}\left( \frac{\chi}{2} \pm \sqrt{\frac{Q}{p_{ij}^2}}\right)
    ,
\end{align}
In the massless case, the foci simplify to $\vec{k}_+ = -\vec{p}_j$ and $\vec{k}_- = -\vec{p}_i$.
Furthermore, the quadric becomes a sphere if $\vec{p}_{ij}=0$.
If $\lambda(p_{ij}^2,m_i^2,m_j^2)=0$ the ellipsoid shrinks into a point at $\vec{k}_+=\vec{k}_-$.
Moreover, if $p_{ij}^0=0$ the hyperboloid flattens into the plane through $(k_++k_-)/2$ with normal vector $\vec{p}_{ij}$.
\par

If $p_{ij}^2=0$ and $m_i\neq m_j$ the surface describes a paraboloid
\begin{align}
\label{eq:paraboloid}
    d(y^2+z^2)
    = x+\left(\frac{1}{4d}-d(m_i^2-\ii\epsilon)\right),
\end{align}
where $d= p/(m_i^2-m_j^2)$ with focus at $(x,y,z)=(d(m_i^2-\ii\epsilon),0,0)$, or in $\vec{k}$-space at
\begin{align}
    \vec{k}_* = -\vec{p}_i+\vec{p}_{ij}\frac{m_i^2-\ii \epsilon}{m_i^2-m_j^2},
\end{align}
which is the limit from below of $k_{\sigma \sgn p_{ij}^0}$ for $p_{ij}^2\to 0$.
For $p_{ij}^2=0$ and $m_i=m_j\neq 0$ the dual propagator cannot vanish for real momenta.
\par

Finally, if $p_{ij}^2=0$ and $m_i=m_j=0$ the dual propagator vanishes on the ray $\vec{k}=-\vec{p}_i+x \sigma\sgn(p_{ij}^0) \vec{p}_{ij}$ for $x\in(0,\infty)$.
\par

We can further categorise the poles of the dual propagator.
Recall that it factorises into an E- and an H-surface, $D_{j}\vert_{i_\sigma}=H_{ij}^\sigma E_{ij}^\sigma$, where
\begin{align}
    E_{ij}^\sigma\equiv \sigma(E_i+E_j)-p_{ij}^0,
    \qquad
    H_{ij}^\sigma\equiv \sigma(E_i-E_j)-p_{ij}^0.
\end{align}
The E-surface is the intersection of a forward an a backward hyperboloid in four dimensions and the H-surface is the intersection of two forward ($\sigma=1$) or two backward ($\sigma=-1$) hyperboloids.
\par

Note that for $p_{ij}^2\neq 0$ the E-surface $E_{ij}^\sigma=0$ can only describe an ellipsoid.
The H-surface $H_{ij}^\sigma=0$ however can either be a single sheet of a hyperboloid or an ellipsoid.
Note that whereas solutions of eqs.~\eqref{eq:quadric} and \eqref{eq:paraboloid} are insensitive to the sign of $\sigma p_{ij}^0$, the existence of the E- and H-surface will in general depend on it.
\par

For $p_{ij}^2=(m_i+m_j)^2=0$ and if $\sigma p_{ij}^0>0$  the E-surface gets squeezed and becomes the line segment between the two foci $\vec{k}_+=-\vec{p}_j$ and $\vec{k}_-=-\vec{p}_i$, parameterised by $\vec{k}=-\vec{p}_i+x\vec{p}_{ij}$ for $x\in[0,1]$.
Note that the endpoints of the squeezed E-surface $x=0$ and $x=1$ are regions where the propagator momenta $q_i=k+p_i$ and $q_j=k+p_j$ become soft, respectively. 
In the region between, the momenta $q_i=x p_{ij}$ and $q_j=(x-1)p_{ij}$ are collinear momentum fractions of $p_{ij}=q_i-q_j$.
Therefore, the presence of pinched (squeezed) E-surfaces in the loop integrand results in infrared singularities of the integral.
\par

For $p_{ij}^2=0$ the H-surface either describes a paraboloid if $m_i\neq m_j$ or get squeezed into a ray if $m_i=m_j=0$
The latter, if $\sigma p_{ij}^0<0$ is the ray $\vec{k}=-\vec{p}_i-x \vec{p}_{ij}$ for $x\in[0,\infty)$ from $-\vec{p}_i$ along $-\vec{p}_{ij}$ to $-\infty$.
If $\sigma p_{ij}^0>0$ the ray goes into the opposite direction and starts from the other focus $\vec{k}=-\vec{p}_j+x \vec{p}_{ij}$.
\par

Note that in general the existence of an E-surface for some values of spatial momenta excludes the existence of the H-surface corresponding to the same dual propagator.
This however does not apply for the pinched E-surface.
If it exists for some spatial momenta, then the pinched H-surface also exists in other regions of spatial momentum space.
\par

A summary of the various existence conditions for the singular surfaces of the dual propagator is given in tab.~\ref{tab:surfaces_pneq0} for $p_{ij}^2\neq 0$ and in tab.~\ref{tab:surfaces_peq0} for $p_{ij}^2=0$.

\begin{landscape}
\renewcommand{\arraystretch}{1.8}

\begin{table}[t]
\begin{center}
\footnotesize
\begin{tabularx}{\textwidth}{XXXXXX}
\hline
\multicolumn{3}{|c|}{$\lambda(p_{ij}^2,m_i^2,m_j^2)\geq0$}
	& \multicolumn{1}{c|}{$\lambda(p_{ij}^2,m_i^2,m_j^2)<0$}
	& \multicolumn{2}{c|}{$\lambda(p_{ij}^2,m_i^2,m_j^2)\geq0$}
	\\\hline
\multicolumn{1}{|c|}{\multirow{2}{*}{$p_{ij}^2<0$}}
	& \multicolumn{2}{c|}{$0 < p_{ij}^2 \leq (m_i-m_j)^2$}
	& \multicolumn{1}{c|}{$(m_i-m_j)^2 < p_{ij}^2 < (m_i+m_j)^2$}
	& \multicolumn{2}{c|}{$(m_i+m_j)^2 \leq p_{ij}^2$}
	\\\cline{2-6}
\multicolumn{1}{|c|}{~}
	& \multicolumn{1}{c|}{$\sgn(m_i-m_j)=\sigma\sgn(p_{ij}^0)$}
	& \multicolumn{1}{c|}{$\sgn(m_i-m_j)\neq\sigma\sgn(p_{ij}^0)$}
	& \multicolumn{1}{c|}{\multirow{4}{*}{no solutions}}
	& \multicolumn{1}{c}{$\sigma p_{ij}^0<0$}
	& \multicolumn{1}{|c|}{$\sigma p_{ij}^0>0$}
	\\\cline{1-3}\cline{5-6}
\multicolumn{1}{|c|}{one sheet of hyperboloid}
	& \multicolumn{1}{c|}{ellipsoid}
	& ~
	& ~
	& ~
	& \multicolumn{1}{|c|}{ellipsoid}
	\\\cline{1-2}\cline{6-6}
\multicolumn{1}{|c|}{$\vec{k}_{\sigma \sgn p_{ij}^0}$}
	& \multicolumn{1}{c|}{$\vec{k}_\pm$}
	& ~
	& ~
	& ~
	& \multicolumn{1}{|c|}{$\vec{k}_\pm$}
	\\\cline{1-2}\cline{6-6}
\multicolumn{2}{|c|}{H-surface}
	& ~
	& ~
	& ~
	& \multicolumn{1}{|c|}{E-surface}
	\\\hline
\end{tabularx}
\end{center}
\caption{
\label{tab:surfaces_pneq0}
The singular surfaces of the dual propagator in eq.~\eqref{eq:app_dual_prop} if $p_{ij}^2\neq 0$.
}
\end{table}%

\begin{table}[b]
\begin{center}
\footnotesize
\begin{tabularx}{\textwidth}{XXXXXX}
\hline
\multicolumn{2}{|c|}{$\lambda(0,m_i^2,m_j^2)>0$}
	& \multicolumn{4}{c|}{$\lambda(0,m_i^2,m_j^2)=0$}
	\\\hline
\multicolumn{2}{|c|}{$m_i\neq m_j$}
	& \multicolumn{1}{c}{\multirow{2}{*}{$m_i=m_j\neq 0$}}
	& \multicolumn{2}{|c}{\multirow{2}{*}{$m_i=m_j=0$}}
	& \multicolumn{1}{c|}{\multirow{1}{*}{$m_i=m_j=0$}}
	\\\cline{1-2}\cline{6-6}
\multicolumn{1}{|c|}{$\sgn(m_i-m_j) = \sigma\sgn(p_{ij}^0)$}
	& \multicolumn{1}{c|}{$\sgn(m_i-m_j)\neq\sigma\sgn(p_{ij}^0)$}
	& ~
	& \multicolumn{2}{|c|}{~}
	& \multicolumn{1}{c|}{$\sigma p_{ij}^0>0$}
	\\\hline
\multicolumn{1}{|c|}{paraboloid}
	& \multicolumn{2}{c}{\multirow{3}{*}{no solutions}}
	& \multicolumn{2}{|c|}{ray}
	& \multicolumn{1}{c|}{line segment}
	\\\cline{1-1}\cline{4-6}
\multicolumn{1}{|c|}{$\vec{k}_{\sigma\sgn p_{ij}^0}$}
	& ~
	& ~
	& \multicolumn{2}{|c|}{from $\vec{k}_{\sigma \sgn p_{ij}^0}$ along $-\sigma \sgn(p_{ij}^0)\vec{p}_{ij}$}
	& \multicolumn{1}{c|}{between $\vec{k}_+=-\vec{p}_j$ and $\vec{k}_-=-\vec{p}_i$}
	\\\cline{1-1}\cline{4-6}
\multicolumn{1}{|c|}{H-surface}
	& ~
	& ~
	& \multicolumn{2}{|c|}{pinched H-surface}
	& \multicolumn{1}{c|}{pinched E-surface}
	\\\hline
\end{tabularx}
\end{center}
\caption{
\label{tab:surfaces_peq0}
The singular surfaces of the dual propagator in eq.~\eqref{eq:app_dual_prop} if $p_{ij}^2= 0$.
}
\end{table}%
\end{landscape}


\bibliographystyle{JHEP}
\bibliography{biblio}

\end{document}